\documentclass[utf8]{FrontiersinVancouver} 

\usepackage{url,hyperref,lineno,microtype,subcaption}
\usepackage[onehalfspacing]{setspace}
\usepackage{comment}
\usepackage{xcolor}
\usepackage{lipsum}
\usepackage{todonotes}
\usepackage{multirow}
\usepackage{subfig}
\usepackage{siunitx}
\usepackage{graphicx}
\usepackage{csquotes}
\usepackage{soul}
\usepackage[capitalise,noabbrev]{cleveref}
\usepackage[most]{tcolorbox}

\makeatletter
\AtBeginDocument
 {
   \def\ltx@label#1{\cref@label{#1}}
   \def\label@in@display@noarg#1{\cref@old@label@in@display{#1}}
 } %
\makeatother

\usepackage[nolist,nohyperlinks]{acronym}
\acrodef{AI}{Artificial Intelligence}
\acrodef{AMT}{Asynchronous Many-Task}
\acrodef{AOD}{Analysis Object Data}
\acrodef{APGAS}{Asynchronous PGAS}
\acrodef{API}{Application Programming Interface}
\acrodef{CG}{Conjugate Gradient}
\acrodef{CI/CD}{Continuous Integration and Continuous Deployment}
\acrodef{CPU}{Central Processing Unit}
\acrodef{DAG}{Directed Acyclic Graph}
\acrodef{DAOD}{Derived Analysis Object Data}
\acrodef{DLC}{Direct Liquid Cooling}
\acrodef{DSL}{Domain Specific Language}
\acrodef{DVFS}{Dynamic Voltage and Frequency Scaling}
\acrodef{FAIR}{Findable, Accessible, Interoperable, Reusable}
\acrodef{FGMRES}{Flexible Generalised Minimal Residual}
\acrodef{FITS}{Flexible Image Transport System}
\acrodef{FP64}{double precision, or 64-bit floating point performance}
\acrodef{FPGA}{Field Programmable Gate Array}
\acrodef{FLOP}{Floating Point Operation}
\acrodef{GCR}{Generalised Conjugate Residual}
\acrodef{GPU}{Graphics Processing Unit}
\acrodef{HEP}{High Energy Physics}
\acrodef{HL-LHC}{High Luminosity LHC}
\acrodef{HMC}{Hamiltonian/Hybrid Monte Carlo}
\acrodef{HPC}{High Performance Computing}
\acrodef{HTC}{High Throughput Computing}
\acrodef{LFT}{Lattice Field Theory}
\acrodef{LHC}{Large Hadron Collider}
\acrodef{LLM}{Large Language Model}
\acrodef{LQCD}{Lattice Quantum Chromodynamics}
\acrodef{MG}{multigrid}
\acrodef{ML}{Machine Learning}
\acrodef{MCMC}{Markov Chain Monte Carlo}
\acrodef{MPI}{Message Passing Interface}
\acrodef{PCIe}{Peripheral Component Interconnect Express}
\acrodef{PGAS}{Partitioned Global Address Space}
\acrodef{R2O}{Research-to-Operations}
\acrodef{RSE}{Research Software Engineer}
\acrodef{SKA}{Square Kilometer Array}
\acrodef{SM}{Standard Model}
\acrodef{WLCG}{Worldwide LHC Computing Grid}

\newcommand{\revision}{}

\ifx\myfixmestatus\undefined
\usepackage[status=draft]{fixme}
\else
\usepackage[status=\myfixmestatus]{fixme}
\fi

\fxsetup{%
  multiuser,
  layout=inline,
  theme=color,
  inlineface=\bfseries,
  envface=\bfseries
}
\FXRegisterAuthor{es}{aes}{ES} 
\FXRegisterAuthor{mf}{amf}{MF} 
\FXRegisterAuthor{mg}{amg}{MG} 
\FXRegisterAuthor{ja}{aja}{JA} 
\FXRegisterAuthor{sp}{asp}{SP} 
\FXRegisterAuthor{bk}{abk}{BK} 

\def\keyFont{\fontsize{8}{11}\helveticabold }
\def\firstAuthorLast{Suarez {et~al.}} 
\def\Authors{Estela Suarez\,$^{1,2,3,*}$, Jorge Amaya\,$^{4}$, Martin Frank\,$^{5}$, Oliver Freyermuth\,$^{6}$, Maria Girone\,$^{7}$, Bartosz Kostrzewa\,$^{8}$, and Susanne Pfalzner\,$^{1}$}
\def\Address{$^{1}$J\"{u}lich Supercomputing Centre, Forschungszentrum Juelich GmbH, J\"{u}lich, Germany \\
$^{2}$ SiPEARL, Maisons-Laffitte, France \\
$^{3}$Institute of Computer Science, Rheinische Friedrich-Wilhelms-Universit\"{a}t Bonn, Bonn, Germany \\
$^{4}$European Space Agency, Space Weather Office, Space Safety Programme, ESA/ESOC, Darmstadt, Germany \\
$^{5}$Karlsruhe Institute of Technology~(KIT), Karlsruhe, Germany \\
$^{6}$Physikalisches Institut, Rheinische Friedrich-Wilhelms-Universit\"{a}t Bonn, Bonn, Germany \\
$^{7}$CERN openlab, CERN, Geneva, Switzerland \\
$^{8}$Helmholtz-Institut für Strahlen- und Kernphysik \revision{ (Theorie)}, Rheinische Friedrich-Wilhelms-Universit\"{a}t Bonn, Bonn, Germany
}

\def\corrAuthor{Estela Suarez, J\"{u}lich Supercomputing Centre, Forschungszentrum J\"{u}lich GmbH, Leo Brandt Str, 52428 J\"{u}lich, Germany}

\def\corrEmail{e.suarez@fz-juelich.de}

\begin{document}
\onecolumn
\firstpage{1}

\title[Energy efficiency in HPC]{Energy Efficiency trends in HPC: what high-energy and astrophysicists need to know} 

\author[\firstAuthorLast ]{\Authors} 
\address{} 
\correspondance{} 

\extraAuth{}

\maketitle

\begin{abstract}
The growing energy demands of \ac{HPC} systems have made energy efficiency a critical concern for system developers and operators. However, \ac{HPC} users are generally less aware of how these energy concerns influence the design, deployment, and operation of supercomputers even though they experience the consequences. This paper examines the \revision{implications} of \ac{HPC}'s energy consumption, providing an overview of current trends aimed at improving energy efficiency. We describe how hardware innovations such as energy-efficient processors, novel system architectures, power management techniques, and advanced scheduling policies do have a direct impact on how applications need to be programmed and executed on \ac{HPC} systems. For application developers, understanding how these new systems work and how to analyse and report the performances of their own software is critical in the dialog with \ac{HPC} system designers and administrators. The paper aims to raise awareness about energy efficiency among users, particularly in the high energy physics and astrophysics domains, offering practical advice on how to analyse and optimise applications to reduce their \revision{energy consumption} without compromising \revision{on performance}.


\tiny
 \keyFont{ \section{Keywords:} High Performance Computing, HPC, energy efficiency, monitoring, programming, application optimisation} 
\end{abstract}

\section{Introduction} 
\label{sec:intro}

Computational methods are considered the third pillar of science, together with theory and experiments~\cite{Weinzierl:2021, Skuse:2019}. As science advances, problem complexity grows and the volumes of data necessary to extract scientific conclusions as well, increasing the demand for \revision{computing} and data management capabilities. In consequence, nowadays all scientific fields, and especially Astrophysics and \ac{HEP}, heavily rely on the use of supercomputers. \acf{HPC} infrastructures are absolutely necessary to run computer simulations that test theories explaining the fundamental nature of matter and the forces that rule its behavior --~all the way from femtometer scales up to the size of the Universe~--, as well as to analyse the data generated by physical experiments and observations. This makes \ac{HPC} infrastructures as important for the physical sciences as  experimental instrumentation, such as telescopes, satellites, or particle accelerators. \revision{While} relatively few of the latter kind of instruments \revision{are} deployed worldwide, the number of \ac{HPC} systems and their sizes are steadily growing, driven by the \revision{increasing} computational demands from all research and engineering fields, further enhanced by the relatively recent exponential growth in training \ac{AI} models~\citep{sevilla2022}. Therefore, even if all science infrastructures have a high energy consumption and in consequence a large carbon footprint, increasing energy efficiency in \ac{HPC} bears the highest potential to reduce the environmental impact of \revision{science} overall. Therefore, many scientific communities have started to report on energy consumption in scientific work, and some journals start demanding this information.

\revision{Considering one individual run of an application on an \ac{HPC} system, if its energy consumption diminishes, also its carbon or environmental footprint become smaller. However, it shall be mentioned that the gain is often overcompensated by either executing more runs of the same application, running more applications, or even installing larger \ac{HPC} systems for the same costs, so that the overall environmental impact is not reduced. This is generally referred to as the \emph{Jevons paradox}~\cite{Jevons:1865} or \emph{rebound effect}, which can only be solved by decision makers (i.e, funding agencies or governments) setting up strict upper limits on the overall consumption. Therefore, we focus in this paper strictly on energy efficiency and refrain from discussing carbon or environmental impact. However, this caveat does neither diminish the importance of striving for the maximum energy efficiency in \ac{HPC}, nor the need of \ac{HPC} users to be aware of and contribute to these efforts. Because while the gain in energy efficiency of individual jobs do not necessarily materialise in overall producing less CO$_2$, it does lead to a higher \emph{scientific throughput}: more scientific results are produced using the \ac{HPC} system for the same amount of time and energy.}

\ac{HPC} providers and operators are applying a variety of measures to maximise the energy efficiency of \ac{HPC} infrastructures. Energy savings can be achieved by selecting energy efficient hardware in the first place, by operating it with system software that ensure maximum utilisation of resources, and by running optimised applications. While \ac{HPC} sites can directly impact the former aspects, the latter is in the hands of application developers alone. Strategies such as reporting the energy \revision{consumption} per job back to users, or awarding more compute time to applications that are more energy efficient, are envisioned. This makes it therefore very important for \textit{domain scientists} \revision{(i.e, researchers \revision{in} specific scientific areas, such as biology, material science, astrophysics, high energy physics, etc.)} to be aware of the main factors contributing to the energy consumption of their codes. Furthermore, it is important that \ac{HPC} users \revision{understand} the impact that some of the energy-saving measures applied by hardware developers and system operators have on the way supercomputers are exposed to the end users.

\begin{figure}
    \includegraphics[width=\linewidth]{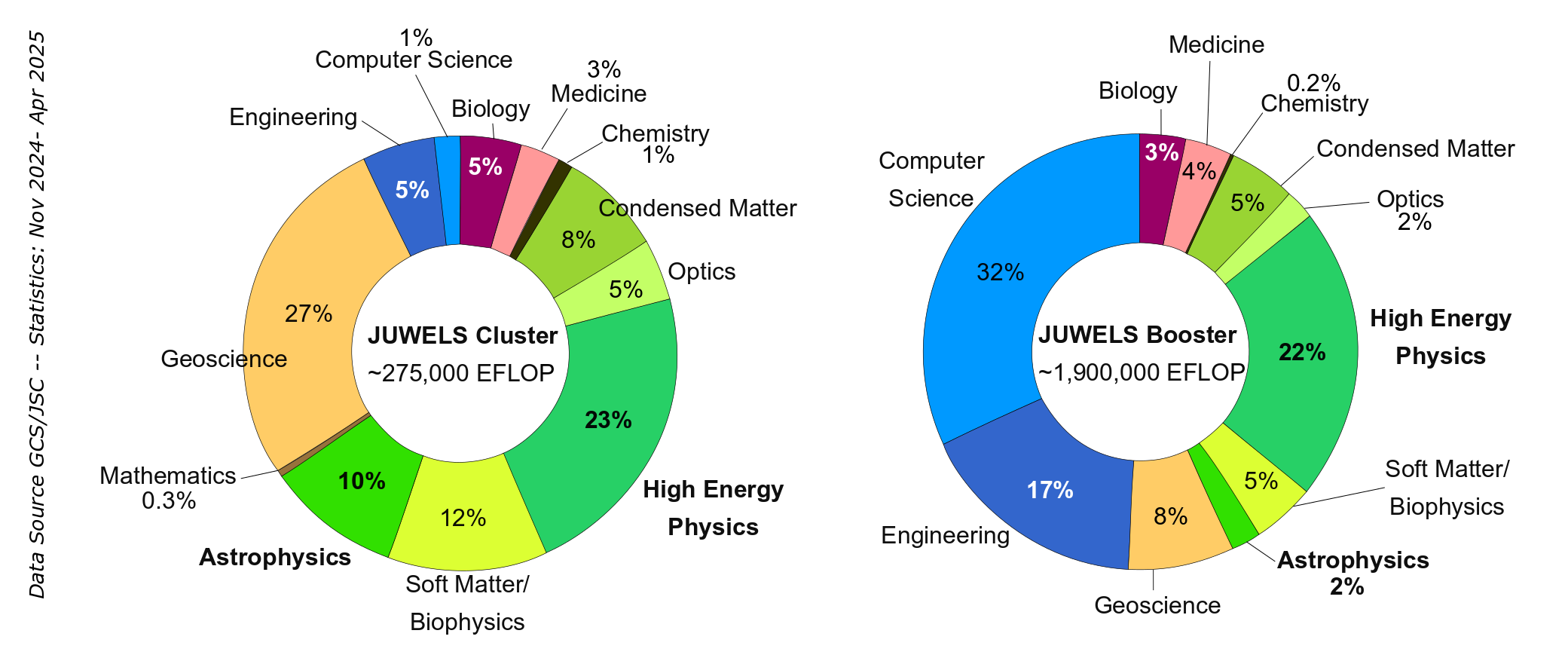}
    \caption{Statistics of allocated compute time per scientific domain in the period November 2024 to April 2025, on the JUWELS supercomputer~\cite{Alvarez2021}. Left shows the CPU module, aka \emph{JUWELS Cluster}. Right shows the GPU module, aka \emph{JUWELS Booster}. Astrophysics and \ac{HEP} (the latter being to a large extent \ac{LQCD} applications) account together for between 1/4 and 1/3 of the available compute time. Data Source: GCS/JSC. \revision{Data available in~\cite{plots_data_zenodo}}.}
    \label{fig:usestats}
\end{figure}

Astrophysics and \ac{HEP} are amongst the strongest consumers \revision{of compute time (core and node-hours) on public \ac{HPC} systems} (see \cref{fig:usestats} and, e.g.~\cite{nersc_stats:2023, cscs_stats:2023, hlrs_stats:2023, eurohpc_stats:2022}), and therefore also in terms of energy. It is therefore especially important that application developers and users in these areas know about the techniques and strategies that they can apply to minimise the energy footprint of their \ac{HPC} workloads. This paper gives an overview of what is done in \ac{HPC} at the hardware and software level, with specific recommendations for application developers and users. It aims at raising their awareness, motivating them to work on improving not only the performance of their applications but also the scientific output per Watt they are able to produce.

Contributions of this paper are:
\begin{itemize}
    \item Summarise the energy-saving techniques applied in \ac{HPC}, describing them from the user perspective and highlighting their impact on the user-space. \revision{\cref{sec:hpc} also reports general recommendations for users and application developers, which apply in principle to any scientific domain}.
    \item Review the historical evolution and specific concerns of applications in computational astrophysics, particularly focusing in the areas of planetary and solar astrophysics, space weather, and space physics. \revision{\cref{sec:astro} gives specific advice for the astrophysics users and application developers, which go beyond the general comments stated in \cref{sec:hpc}}.
    \item Review the historical evolution and specific concerns of computational high-energy physics, particularly in the subareas of experimental collider physics (\cref{sec:hep}) and \revision{lattice quantum} field theory (\cref{sec:lqcd}). \revision{These two sections focus on aspects particularly relevant for high-energy physicists}.
    \item Summarise the main \revision{observations or} recommendations to application developers and users \revision{to facilitate readers identifying the} take-away messages.
\end{itemize}

\section{Energy efficiency trends in High Performance Computing}
\label{sec:hpc}

Energy efficiency is a major factor in the design and operation of today's \ac{HPC} systems. Hosting sites build up data centres --~the building and physical infrastructure in which the \ac{HPC} systems are physically located~-- so that the energy used for cooling and operation of a supercomputer is minimised, and the unavoidable waste heat can be reused. Because water can absorb more heat than air, water cooling systems are more efficient than air cooling ones. Compute nodes are integrated in blades and chassis that include \ac{DLC} pipes, in direct contact with the hottest electronic components on the motherboard (\acp{CPU}, \acp{GPU}, and memory). Water circulates from one blade to the next, absorbing the heat in the process, and creating an internal water loop that reaches all elements within one computer rack. The entry temperature of water in this internal loop is in the range of 35-45~$^\circ$C, gaining $\sim$5$^\circ$C during its pass through the computer. A heat exchanger transfers the gained heat from the internal water loop into a secondary one, which runs across the whole data \revision{centre}. 
At most \ac{HPC} sites, particularly in central and northern Europe where the outside air temperature is \revision{all year round} well below the \revision{aforementioned} values, the secondary water loop simply runs out of the building and is cooled down via dry coolers (basically ventilators in contact with outside air). A further optimisation is to utilise the warm water from the secondary loop to heat neighboring buildings, e.g., offices or living spaces, saving money and energy on other heating infrastructures, and therefore maximising energy efficiency. 

While \ac{HPC} centres are already applying energy efficiency measures voluntarily, policies mandating efficient use of computational resources are being put into place. For example, the European Union regulates data \revision{centre}s through its Energy Efficiency Directive~\cite{EUenergyefficiency}, last updated in 2023. The directive contains energy efficiency targets, energy savings obligations, requirements for the establishment of energy management systems and reporting duties specifically for data \revision{centre}s. 


\subsection{System Hardware}
\label{sec:hpc:hw}

\begin{figure}[!ht]
    \includegraphics[width=\linewidth]{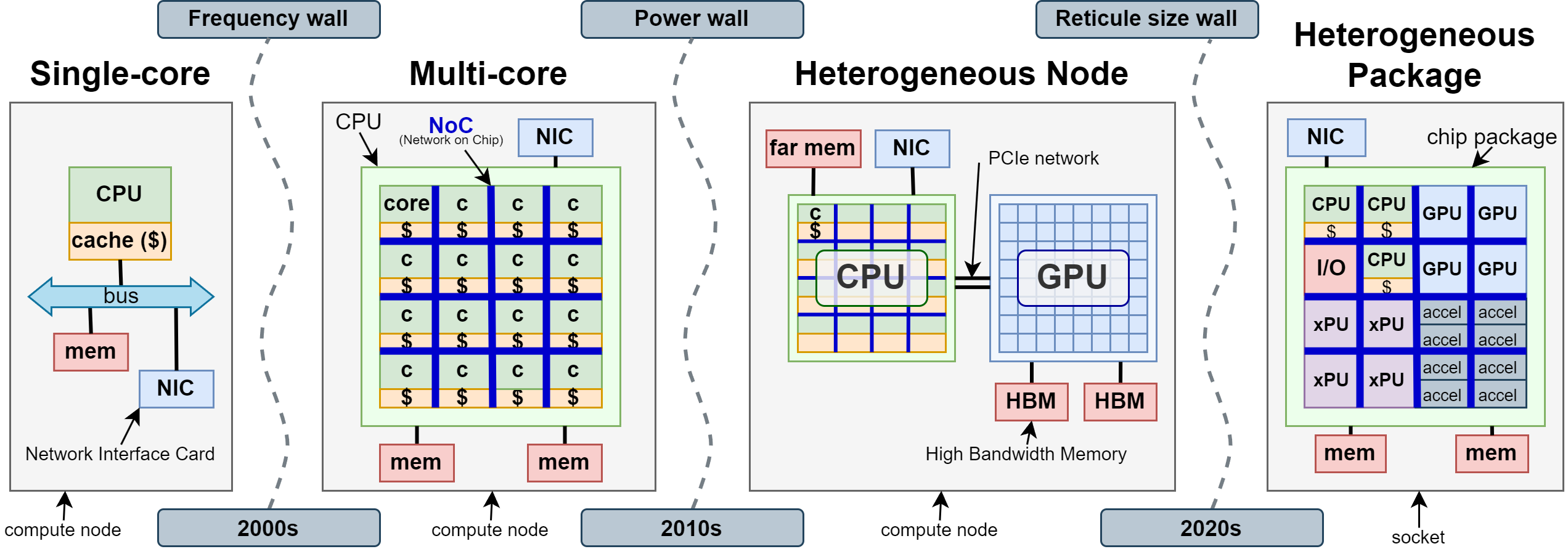}
    \caption{Evolution of processor architectures over time: from single-core \acp{CPU}, to many-core \acp{CPU}, to heterogeneous compute nodes with \ac{CPU} and \acp{GPU}. \revision{The most recent trend goes towards specialised chip designs that are} internally heterogeneous, combining different processing technologies in the form of chiplets.}
    \label{fig:chips}
\end{figure}

Thanks to miniaturisation in microelectronics, historically more and more transistors \revision{have been} included in the same area of silicon. Smaller structures also mean that electrical signals travel shorter distances across the chip, which implies lower energy consumption. This phenomenon has allowed processors in the last decades to become more computationally powerful and energy efficient from generation to generation. The exponential increase of computing power over time has been formulated in \emph{Moore's law}~\citep{moore2006}. Until around 2004 miniaturisation was accompanied by a steady increase of operational frequency, following the \emph{Dennard scaling}: faster \ac{CPU} clocks mean faster operations and therefore more computing capacity. But miniaturisation and high frequency have as consequence higher power density (more power per area) due to leakage currents, and more generated heat, to a point where electronics would get damaged. The end of Dennard scaling marked the end of single-core \acp{CPU}\revision{. After that moment in time} more compute power could only be added by increasing the number of cores in the \ac{CPU}, and parallelising tasks across them (see \cref{fig:chips}). 

When endless miniaturisation of \acp{CPU} is impossible and operational frequency cannot grow any further, \emph{specialisation} might at least bring some added performance. While \acp{CPU} are \emph{general purpose} processors aiming at solving all possible tasks, compute \emph{accelerators} are devices designed to solve \emph{a limited class} of tasks in the most efficient way possible. Some accelerators used in \ac{HPC} are \acp{GPU}, many-core processors, \acp{FPGA}, and \ac{AI}-accelerators. While \acp{CPU} are still needed to perform some tasks, one can obtain the computational results faster and consuming less energy by employing accelerators to execute exactly the operations at which they excel. 

\begin{figure}[!ht]
        \includegraphics[width=0.8\linewidth]{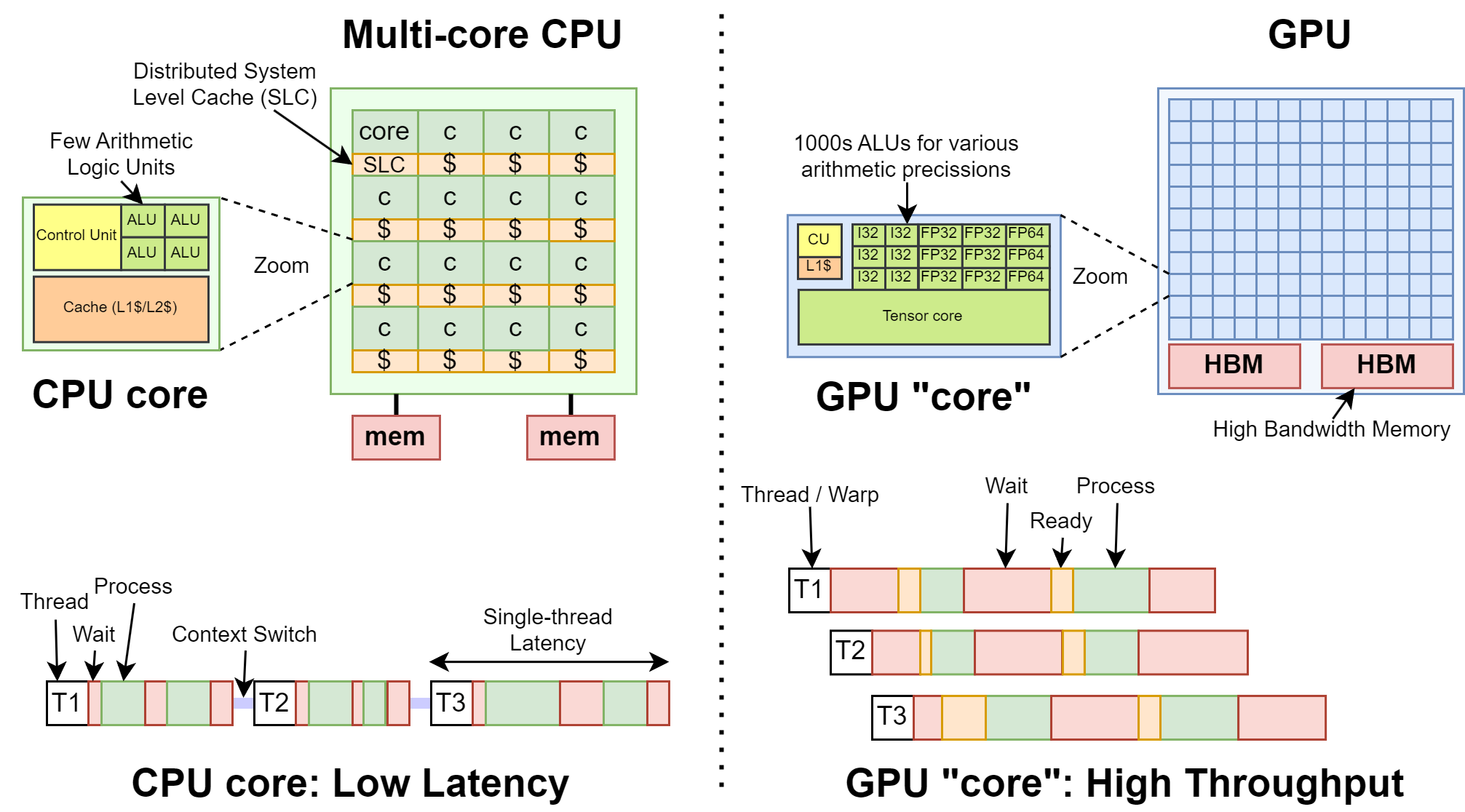}
        \caption{Schematic description of the architectures of a \ac{CPU} (left) and a \ac{GPU} (right). \acp{CPU} typically contain a limited amount of large, complex, highly capable processing cores, while \acp{GPU} contain thousands of simple arithmetic execution units (increasingly for low-precision operations). \acp{CPU} are therefore good for latency limited applications, while \acp{GPU} serve best highly parallel workloads}
        \label{fig:cpuVSgpu}
\end{figure}

The principle behind \emph{heterogeneous architectures} is to combine \acp{CPU} with accelerators (most frequently \acp{GPU}) to build supercomputers and deliver the highest performance in an energy efficient way. The high-level architecture differences between \acp{CPU} and \acp{GPU} are schematically described in~\cref{fig:cpuVSgpu}. \acp{GPU} are equipped with a very large number of execution units for different types of arithmetic operations. Lacking advanced \ac{CPU} features such as out-of-order execution, \acp{GPU} have higher latency for each individual operation, but they enable very high throughput through massive parallelism. Because their operating clock frequency is lower, the overall power consumption per operation (Watt per \ac{FLOP}) is lower (better) on \acp{GPU} than on \acp{CPU}.

Typically, one, two or more \acp{GPU} are connected to a \ac{CPU} via a network interface called \ac{PCIe}. Historically, the \ac{CPU} had to actively participate in inter-device and inter-network data transfers, resulting in a communications bottleneck \revision{across} \ac{PCIe} as its bandwidth is one or more orders of magnitude lower than \ac{CPU}-to-memory or \ac{GPU}-to-memory links. Inter-\ac{GPU} network links such as NVlink and advanced communication features such as \ac{GPU}-aware \ac{MPI} have strongly improved the situation. The current trend goes even further \revision{(see \cref{fig:chips})}, integrating \ac{CPU} and \ac{GPU} chips in the same package, or even as chiplets on the same substrate~\cite{gh200, mi300A}, improving \ac{CPU}-\ac{GPU} communication and enabling cache coherency between the two.

It is worth noting that at the same time as heterogeneity is increasing at the package level, the trend at the system level is towards \emph{resource disaggregation} or \emph{modularity}: clusters with different node configurations (\ac{CPU}-only, \ac{CPU}+\ac{GPU}, accelerator-only, quantum processors, etc.) are connected to each other via a common high-speed network. The resource scheduler allows users to simultaneously allocate resources on different partitions. In cases where the \ac{MPI} library has been adapted, an application can even communicate across module boundaries via \ac{MPI}. This system design, sometimes called \emph{Modular Supercomputing Architecture (MSA)~\cite{Suarez:2019, Suarez:2022}}, is used in several systems around the world (e.g.~\cite{Alvarez2021, jupiter, leonardo, meluxina, wisteria}). \revision{All these systems} couple \ac{CPU}-only modules with \ac{GPU}-accelerated modules or partitions. The concept itself allows for further heterogeneity, using different interconnect technologies, acceleration devices, and even connecting disruptive modules such as quantum computers to \emph{classic} \ac{HPC} modules. For users, this modularity requires thinking about their application in a coarse-granular way\footnote{If the code is partitioned between two modules, the \emph{cut} should be done so that most communication happens within the modules, and not in between.}. If the code \revision{simulates} multi-physics or multi-scale problems, which are often tackled by coupling different physical models, it is worth considering whether they could benefit from running on different hardware, i.e, across modules. If the code is monolithic (e.g., a dense matrix-matrix operation), it is better to run the whole application on a single module.

\begin{tcolorbox}[colback=green!5!white,colframe=green!75!black]
    \textbf{Observation 1:} \ac{HPC} systems are becoming more heterogeneous, combining \acp{CPU} with \acp{GPU} and other accelerators. Monolithic system architectures combine these devices within massively heterogeneous nodes, but keeping all nodes equal, which requires a fine-granular partition of application \revision{codes}. Modular architectures create partitions or modules each with a different node configuration, which requires a coarse-granular partition of application \revision{codes}, suitable for e.g  multiphysics or multi-scale applications. 
\end{tcolorbox} 

\subsection{System Software}
\label{sec:hpc:sw}

\begin{figure}
    \includegraphics[width=\linewidth]{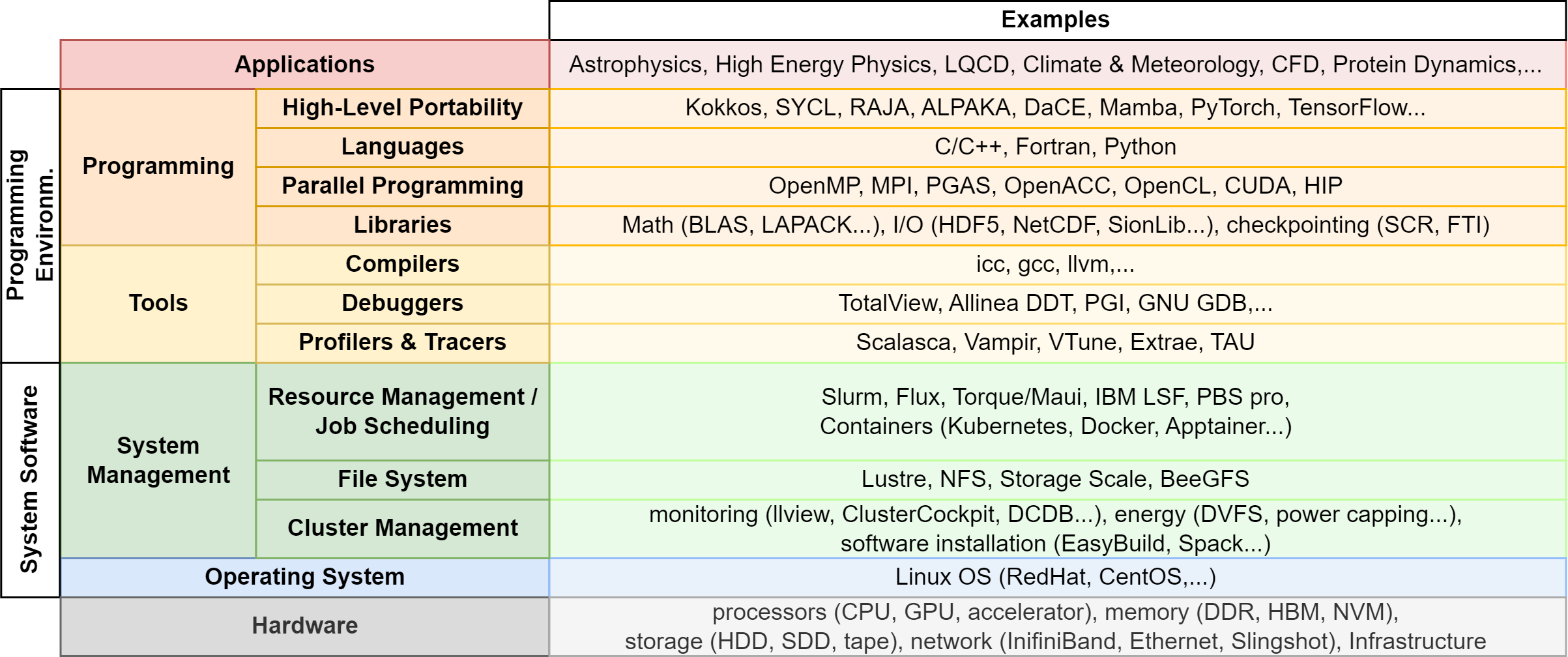}
    \caption{High-level view of the software stack running on \ac{HPC} systems. The product names given as example are merely illustrative, with no intention of giving a comprehensive list of all possible solutions.}
    \label{fig:hpcSW}
\end{figure}

A deep software stack is installed on \ac{HPC} systems to operate them and provide all functionalities needed by the users (see~\cref{fig:hpcSW}). Energy efficiency must be \revision{addressed} at every layer of the stack, ensuring that all hardware resources are employed as efficiently as possible. 

From the administrator perspective, several approaches can be used to optimise power consumption in \ac{HPC} environments. Over-provisioned systems, which have peak power consumption exceeding infrastructure limits, can benefit from dynamic power capping. HPE PowerSched~\cite{powersched}, for example, assigns dynamic power limits to individual jobs to maximise system utilisation. The framework EAR~\cite{EAR}, among other things, adapts frequency to runtime application characteristics and a pre-defined energy policy. Idle power consumption is also a significant factor, especially in modern \acp{CPU} with varying idle power states~\citep{Sundriyal2017, Troepgen_2024_SPEC, ilsche2024}. 

Modern \acp{CPU} and \acp{GPU} offer software interfaces to control power consumption through \ac{DVFS}~\citep{Bhalachandra2015}. While higher frequencies generally improve performance, they can negatively impact energy efficiency. Memory-bound workloads can operate more efficiently at lower frequencies, as their time-to-solution depends more on the speed at which data is retrieved from memory, than the speed at which the processor computes operations, so that their performance does not suffer as much when the processor clock is slowed down~\citep{Bhalachandra2017}. Compute-bound workloads, however, may benefit from higher frequencies, as shorter runtimes can compensate for increased power consumption. Some \ac{HPC} sites offer \emph{energy efficiency knobs}, which users can employ to define per-job system configurations with lower operational frequency~\cite{kodama2020}. Often, \revision{incentives are given} in terms of higher queue priority or additional compute time to users that run their jobs at lower energy consumption~\citep{Auweter14, Solorzano2024}. \revision{But even if users do not employ such knobs, both the hardware itself and the system operators might change the operation mode of individual compute nodes by applying power control techniques (e.g., \ac{DVFS} or power capping) without prior notification, which make it challenging for users to reproduce the same performance numbers across equivalent job runs executed at different points in time.} 


\ac{HPC} systems run comprehensive monitoring tools to optimise energy efficiency. This involves collecting metrics from various components, including compute nodes, infrastructure components, and batch jobs. Data collection is typically handled by node collectors that gather metrics and send them to a central communication component, such as a message broker. The monitoring system must integrate with the batch job scheduler to track resource usage and performance. At many sites, web interfaces provide access to monitoring data, with different levels of access for various user groups~\cite{ClusterCockpit, llview}. In this way, users can read from their job reports \revision{how much energy they consumed}, as well as the amount of resources that they employ. This helps identifying potential improvements for more efficient energy and hardware use. Goal in this optimisation is maximising the scientific throughput.

Most jobs use far less resources than available in the node~\cite{Maloney2024, Li2023}, and minimising idle resources is one of the main objectives of modern scheduling software. \ac{HPC} sites are considering co-scheduling jobs within the same node, especially on \ac{HPC} systems with \emph{fat nodes}, i.e, those containing large numbers of cores and accelerators. Research trends in resource management go also towards dynamic scheduling mechanisms able to react at runtime to variations on both application requirements or overall system workload, or even further, to changes on the electricity network. 
\ac{HPC} users should be therefore prepared for an \ac{HPC} environment in which scheduling decisions might not only depend on their own choices at job submission, but also on energy optimisation strategies applied by the \ac{HPC} centre. Adapting application codes so that they become \emph{malleable}, i.e, able to grow or shrink its hardware needs at runtime, shall bring them the advantage of higher queue priorities~\citep{Tarraf2024}. Furthermore, compute time allocations might start soon to be measured in terms of energy consumed, and not merely core-hours, making it crucial for users to optimise their workloads for maximum efficiency.

\begin{tcolorbox}[colback=green!5!white,colframe=green!75!black]
    \textbf{Observation 2:} \revision{The rising energy demands and cost of energy leads vendors and \ac{HPC} operators to apply numerous techniques to minimise the energy consumption and maximise scientific throughput of \ac{HPC} systems. Some of them are exposed to the users, but not all. Users should be aware that the operational mode of an \ac{HPC} system is not under their control and operational frequencies \revision{of processors} are now more \emph{volatile}. For benchmarking runs in which performance reproducibility is crucial, users are advised to consult job reports and contact the support teams from \ac{HPC} sites to ensure their jobs always run under the same conditions. Furthermore, the allocation of compute resources per job might soon become much more dynamic and be measured in terms of energy consumed than of core-hours. Users are therefore advised to use monitoring data to determine the energy consumption of their jobs and identify the most energy-efficient run configurations before starting production jobs.}
\end{tcolorbox} 

\subsection{Programming Models}
\label{sec:hpc:prog}

\revision{The programming environment in \ac{HPC} abstracts the hardware complexity from the user through general \acp{API}, helping existing code to run on rapidly evolving compute hardware. The following subsections describe state of the art, challenges and recommendations at each of the layers in the programming environment.}

\subsubsection{\revision{Programming languages}} 
\label{sec:hpc:prog:lang}
Many, if not a majority of codes running on \ac{HPC} systems\revision{,} have been developed over many years with their basic architectures designed decades ago relying on relatively low-level approaches (mostly Fortran or C) conceived to run on \acp{CPU} \revision{(e.g.~\cite{Ishiyama2016-qq,Nordlund2018-se,Xia2018-mg,Toth2012-zn,Verbeke2022-ro}). The reasons for these codes to still exist, despite newly available hardware and software solutions, is evident: their goal is to produce data of scientific interest, in a very short time, for a limited number of users. This leads in general to codes that are very difficult to use, not well documented, and not widely distributed. Even codes used to publish results in high impact journals are still considered complex to use and unavailable to the public~\cite{Hotta2021-xi}. In particular, users at the end of the analysis chain employ tools as black boxes to extract scientific insight of their interest. These tools are developed within smaller groups of developers who are often inexperienced with the programming concepts, data structures and performance tuning.} For such legacy codes, it cannot be assumed that more performance can be gained automatically on modern \revision{heterogeneous} \ac{HPC} systems without significant intervention of researchers. 

\revision{With the growth in popularity of data analytics and \ac{ML}, there has been a slow shift in the use of programming languages, moving from the traditional Fortran and C/C++, to high-productivity alternatives like Matlab, Python or Julia~\cite{juliateich}, which offer a simple syntax, a large number of easily accessible libraries, and fast application prototyping. This has lowered the entrance barrier to new computer programmers, at the expense of computing efficiency.} 

\revision{Still, codes optimised for \ac{HPC}, taking advantage of the latest hardware technology, need to be written in traditional languages that are closer to the hardware itself. Unfortunately, new generations of scientists rarely \revision{learn these} (see also \cref{sec:hpc:apps:skills}). Students in many scientific domains lack formal training in software engineering in general, and in \ac{HPC} in particular. Also education in advanced C++ and modern Fortran is extremely rare. Without such prior knowledge, potential new contributors to state-of-the-art software frameworks face a very steep learning curve, often incompatible with their limited time budgets as they begin to establish themselves as domain experts in their academic career paths.}

\subsubsection{\revision{Hardware heterogeneity and portability}}
\label{sec:hpc:prog:portab}

\revision{The increasing heterogeneity of \ac{HPC} systems described in \cref{sec:hpc:hw} and the consequent system complexity make software development more challenging. Transitioning application workflows to energy-efficient platforms, such as \acp{GPU} or \acp{FPGA}, involves more than rewriting codes\revision{. It} requires fundamental redesigns of algorithms to accommodate the distinct memory hierarchies and parallel processing capabilities of these architectures. Each kind of compute device requires different optimisation approaches, e.g., while spatial and temporal locality play a role in the optimisation of access patterns and data layouts on both \acp{CPU} and \acp{GPU}, good performance on \revision{GPUs} can only be reached if threads within a thread team access data in a coalesced fashion and if thread divergence is avoided. Although these efforts promise substantial energy savings, the initial development is effort-intensive, demanding both time and specialised expertise.}

\revision{New hardware also comes with new programming models and libraries that need to be adopted into the software ecosystem. Compilers need to include in their standards additional ways to translate codes into machine binaries and for a new hardware component it can take years before full adoption by the end users. Vendor-specific programming models promote vendor lock-in (e.g., to get optimal performance, NVIDIA \acp{GPU} had to be programmed using CUDA~\cite{cuda}, and AMD \acp{GPU} with HIP~\cite{hip}), which can backfire on application developers as it happened with the Intel Xeon Phi many-core processor launched in 2010:~while many research groups saw this as an opportunity to accelerate their codes without having to deal with complex parallelisation using CUDA on \acp{GPU}, the processor ended up being discontinued in 2020~\cite{Cutress2019} and developments that took years to implement became suddenly dependent on an obsolete technology. Scientists learned that it is important to write codes so that they are portable to different architectures.}

\begin{tcolorbox}[colback=green!5!white,colframe=green!75!black]
\textbf{Observation 3:} 
    Modern compilers are a great source of information to identify low hanging fruit opportunities for improvements in performance. They often provide flags to detect misplaced data accesses, to unroll nested loops, or to replace known inefficient instructions. However, it is our experience that recent compilers remain very unstable, and it is preferable to use reliable older versions of the compilers and miss out on the opportunity to test the \revision{newest} features in the most recent version. It takes from 2 to 5 years for known compiler versions to be fully tested by developers worldwide.
\end{tcolorbox}

\revision{But as hardware becomes more heterogeneous, performance-portability across platforms is harder to attain. Using libraries, especially I/O and numerical linear algebra (e.g., the BLAS \acp{API}), does help because processor vendors provide their own hardware-optimised implementations, reaching highest efficiency. The performance, energy efficiency, and the level of code portability that \ac{HPC} users can achieve depend on the \acp{API} and programming models they choose. Generally, low-level programming models give more control to the users, at the price of lower portability. High-level abstractions (e.g., Python~\cite{python}, OpenMP~\cite{openmp}, SYCL~\cite{sycl}, Kokkos~\cite{kokkos}, ALPAKA~\cite{alpaka}, RAJA~\cite{RAJA}, Mamba~\cite{MAMBA}, Julia~\cite{juliateich}, DaCe~\cite{DaCe}, and many others not named here), on the other hand, are portable and achieve good performances~\cite{Godoy:2023, Vay2021, hunold2020benchmarking, juliateich} by relying on backend optimisations to generate executable codes adapted to different hardware devices}.
However, for these languages to deliver performance in an energy-efficient manner, it is important to call their optimised packages, such as numerical libraries and backends, typically written \revision{in C}. Otherwise applications written in Python can be very power-hungry~\citep{Pereira2017}.

\begin{tcolorbox}[colback=green!5!white,colframe=green!75!black]
    \textbf{Observation 4:} To cope with increasing hardware heterogeneity, application development teams are advised to modularise their codes, decoupling the optimisations for different hardware platforms from the scientific core of the application, e.g., by calling appropriate numerical libraries. Failure to do this can lead to much lower energy and resource usage efficiency. Teams lacking specialised \revision{personnel} with the necessary expertise and effort to optimise code for different hardware devices are advised to employ portable programming models and frameworks that isolate their scientific implementation from the hardware-specific features, which shall be addressed by optimised libraries and backends. 
\end{tcolorbox} 

\subsubsection{\revision{Domain specific languages}}
\label{sec:hpc:prog:dsl}
\acp{DSL} separate user-facing interfaces and performance-critical code altogether. The prime example are \ac{ML} frameworks such as PyTorch and TensorFlow.
Examples in high-energy physics and astrophysics are, in increasing order of abstraction, Grid~\cite{Boyle:2016lbp,Yamaguchi:2022feu}, Chroma~\cite{chroma} and \revision{ROOT~\cite{rene_brun_2020_3895860}} as well as Lyncs~\cite{Bacchio:2022bjk}, PyQUDA~\cite{PyQUDA},  and Grid Python Toolkit (GPT)~\cite{gpt}, where the latter three provide Python interfaces to the highly optimised algorithms implemented in QUDA~\cite{Clark:2009wm,Babich:2011np,Clark:2016rdz} and Grid, respectively. In the domain of space physics, problems that can be approximated by the large-scale approach, where a plasma can be considered as a fluid, can use one of the existing \revision{frameworks (e.g.~\cite{Keppens2021-et, Gombosi2023-rb, Lani2013-jp}) that decouple} numerics, parallelisation, and mathematical equations \revision{from each other}. This approach allows scientists to focus on the equations to be solved, leaving the software heavy-lifting to the underlying code programmed by specialists. 

It is to be seen in what way tools from generative \ac{AI} will impact code generation for \ac{HPC} systems. While using \acp{LLM}/chatbots or GitHub Copilot to generate code for routine tasks is by now common, using generative \ac{AI} also shows promise for HPC code generation~\cite{10.1145/3605731.3605886,10.1145/3625549.3658689}.

\subsubsection{\revision{Parallelisation over multiple compute nodes}}
\label{sec:hpc:prog:parallel}
\revision{Any code or application developer that wants to take advantage of modern \ac{HPC} systems shall start by making sure that: \emph{(i)} their code runs as optimal as possible in a single node, \emph{(ii)} their code is well structured and documented, and \emph{(iii)} it presents an easy to use interface that does not require to recompile the source code for different use cases. Among the applications using the biggest supercomputers in the world it is not uncommon to miss points \emph{(ii)} and \emph{(iii)}.}

While the high-level programming models \revision{described in \cref{sec:hpc:prog:portab}} allow for portability between single nodes, parallel applications need to run across many nodes. The de-facto standard multi-node parallelism framework in \ac{HPC} is \ac{MPI}~\cite{mpi}. This requires explicit transfer of messages between nodes, but is in principle agnostic to the type of node underneath, as long as an \ac{MPI} library is installed and supported. While this is true in most cases, \revision{various} implementations exist, and performance differences are often observed between implementations running on the same hardware. Therefore, users are advised to test their applications against different \ac{MPI} implementations, verify correctness, and identify the most efficient one in each system.

An alternative to message-passing is given by the \ac{PGAS} paradigm in which a global memory space is logically partitioned over many processes and where the various portions of this shared space have an affinity for particular processes.
In \revision{contrast} to message-passing, individual processes may access remote memory directly with the necessary communication taking place either explicitly or implicitly as part of the semantics.
Examples of such \ac{PGAS} systems are Coarray Fortran~\cite{coarray}, Unified Parallel C~\cite{upc} or Global Arrays~\cite{globalarrays}.

For problems subject to load-imbalance, a proliferation of small subtasks, formulations on irregular grids or implementations\revision{,} which aim to make use of highly heterogeneous computational resources, \ac{AMT} systems may provide a more appropriate programming model with the potential of increasing scalability and resource utilisation.
Examples of such \ac{AMT} systems are Uintah~\cite{uintah}, Charm++~\cite{charm++}, ParSEC~\cite{parsec}, 
Legion~\cite{legion} and HPX~\cite{hpx,hpx2}.
A more complete comparison can be found in \revision{reference}~\cite{task_parallel_taxonomy}, but in general it can be said that these \revision{solutions} are either compilers and run-time systems, separate libraries and run-time systems, or systems in which the run-time is linked directly into the application.
Generally, \ac{AMT} systems provide some form of dependency resolution mechanism and a data-flow model based on a \ac{DAG} together with some form of \ac{APGAS} model for representing distributed data.
It is also possible to benefit from the performance-portability of Kokkos and the task-based parallelism of an \ac{AMT} system like HPX to move away from the more traditional fork-join model with good results~\cite{9460406,10029450} by making use of \emph{futures} to launch and synchronise Kokkos kernels.

  \begin{figure}[!ht]
    \begin{center}
    \includegraphics[width=0.6\linewidth]{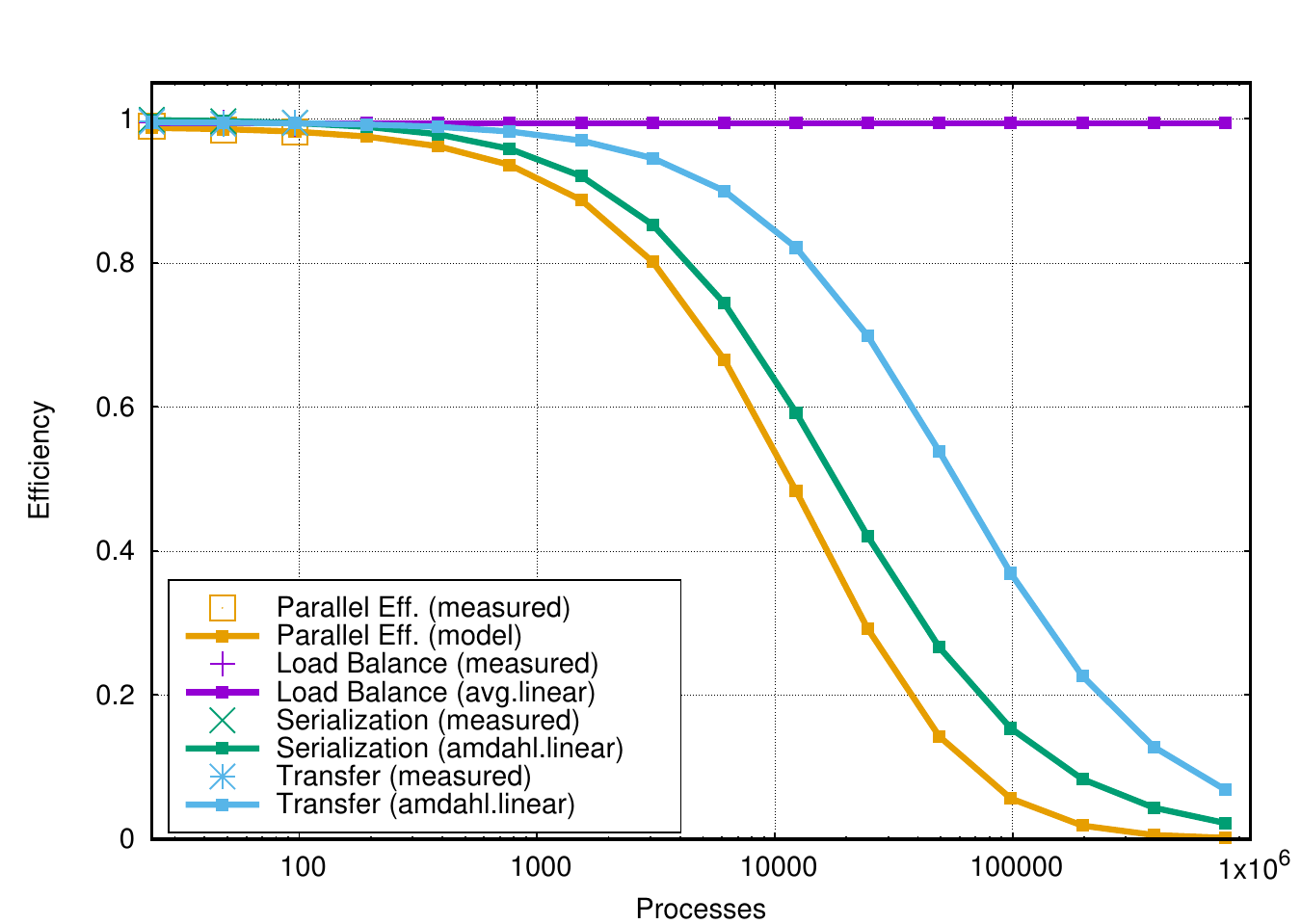}
    \end{center}
    \caption{Measured and extrapolated parallel efficiency of a kinetic space plasma code using DIMEMAS~\cite{dimemas}\revision{, reproduced from Figure~26.a (page~63) in ~\cite{deepest:2021}. The parallel efficiency, representing the time spent on computation (useful work), is composed of the load balance, the serialisation, and the transfer efficiencies, as defined in section 4.1 of~\cite{deepest:2021} and the DIMEMAS~\cite{dimemas} documentation. In this example three measuring points at 1, 24, and 96 processes, were used to extrapolate the parallel efficiency of the code up to one million processes.}}
    \label{fig:dimemas}
\end{figure}
    
\revision{We have observed that computer applications developed to run in parallel architectures with a small number ($<$1000) of cores sometimes struggle with the porting to larger systems, and in particular to Exascale supercomputers. In \cref{fig:dimemas} we present the measured and extrapolated efficiency of a parallel kinetic plasma physics code. Based on three measurement points of the code efficiency, the tool DIMEMAS~\cite{dimemas} can predict the performance of the code on very large systems using Amdahl's law \cite{Ene2022}. Such analysis is an essential step to discover what are the scalability limitations of any software. As the number of processes increase serial sections of the code and parallel transfers become dominant.}

\begin{tcolorbox}[colback=green!5!white,colframe=green!75!black] 
    \textbf{Observation 5:} \revision{The secret to a good parallel application ready for the next generation of supercomputers rely on an early and thorough analysis of single node peak performance, and parallel strategies that reduce data transfers to the minimum.  
    Once this is done, the developer should test as early as possible the scalability of their parallel strategy using performance extrapolation tools. A parallel algorithm is difficult to change when a code is mature with multiple years of development. To develop and test a scalable parallel algorithm it is possible today to use tools like DIMEMAS~\cite{dimemas} that project the potential scalability of a code on any existing or idealised supercomputer using performance traces from experiments run in a few hundred cores}.
\end{tcolorbox}

\subsection{Data management} 
\label{sec:hpc:data}



When discussing energy efficient trends in \ac{HPC}, it is important to include data management and storage. While storage systems may not directly consume as much electricity as \revision{the processors} during peak loads, they can significantly impact overall efficiency. Moreover, storage systems can become persistent power consumers, with data often requiring preservation for decades, leading to high integrated energy costs over time. Fields such as \ac{HEP} and radio astronomy exemplify these challenges\revision{. Constrained} by budget and technological limitations, they cannot store all the raw data generated. Instead, they rely on real-time processing and filtering to reduce storage needs and concentrate on valuable insights. However, this approach introduces its own energy-intensive demands. Advancements in \ac{ML} algorithms offer the potential to enhance data selection accuracy and speed, but their training and optimisation processes come with additional storage and computational requirements, further complicating the energy equation.

High performance data management is an increasingly important consideration as the demand for \ac{HPC} grows across scientific and industrial domains. One major source of energy inefficiency is the movement of data within computing systems. This issue does not stem primarily from the storage or file systems themselves, which are relatively energy-efficient, but from the impact on processing units. \ac{HPC} processors, designed for high throughput, are power-hungry even when idle, consuming almost as much energy as during full operation. These processors often remain underutilised, idling as they wait for data transfers or caching to complete. This mismatch between data availability and processing demands highlights the need for improved data transfer and caching infrastructure. By minimising idle time and ensuring that processors operate at high utilisation rates, significant energy savings can be achieved. Users are therefore advised to consider overlapping communication with computation, to keep the processors busy while they wait for new data. It is also important that they employ optimised I/O libraries (e.g., HDF5~\cite{The_HDF_Group_Hierarchical_Data_Format}, SIONlib~\cite{SIONLib}), which ensure that data is aggregated in a reduced amount of files to avoid saturating the file system. 
\revision{When data transfers between nodes are impossible to avoid, creating a regular pattern for the data transfer is a key factor. Reducing the number of communication steps, and taking advantage of algorithm optimisations by the parallel libraries can increase the scalability of a code.}

\begin{tcolorbox}[colback=green!5!white,colframe=green!75!black]
    \textbf{Observation 6:} Data movement is an important contributor to energy consumption, due to idling processors while they wait for data to be loaded from either disk, memory, or cache. Application developers are advised to employ optimised I/O libraries and overlap computation with communication when possible, to minimise \revision{waiting times} and idling resources. 
\end{tcolorbox} 

Storing data, whether for active use or long-term archival, also contributes substantially to energy consumption. Active storage media must consume power continuously to keep data accessible and to serve it efficiently across distributed systems. This often necessitates maintaining multiple data copies to ensure reliability and availability. Even archival storage, which might seem like a low-energy alternative, requires sophisticated infrastructure to prevent data loss or corruption, such as climate-controlled facilities. Promising advancements in archival technologies aim to create dense, stable media that can preserve data for centuries under normal conditions without requiring power~\cite{Cerabyte}. These innovations could drastically reduce the integrated energy footprint of long-term data storage.

Another critical challenge lies in managing the massive volumes of data generated by instruments and experiments in fields such as physics and astronomy. Projects like the \ac{LHC} and the \ac{SKA} produce data at such scale that it becomes infeasible to store and load all of it. Instead, much of this data must be processed \emph{on the fly}. However, streaming data directly into \ac{HPC} systems is not a traditional use case and poses significant logistical and technical challenges. \ac{HPC} workflows typically rely on pre-scheduled tasks, but the irregular operation of data-generating instruments—affected by maintenance schedules, weather, and other factors—complicates integration. The synchronisation of real-time and scheduled workflows demands innovative solutions, including new scheduling mechanisms and potentially dedicated \ac{HPC} systems optimised for streaming workflows. Nevertheless, gaps in instrument operation could present opportunities for resource sharing, further enhancing efficiency.

\revision{The demand for energy-efficient solutions grows even greater in all scientific areas, as they increasingly rely on \ac{AI} and \ac{ML} techniques\footnote{\revision{For instance, when analysing the user statistics of the JUWELS Booster (right side of \cref{fig:usestats}), we found out that 47\% of the total compute time is used by \ac{AI}-related jobs.}}.} \ac{AI} and \ac{ML} require large datasets for training models, which in turn demand significant computational power. This adds another layer of complexity to the energy equation, underscoring the importance of developing efficient algorithms and hardware optimised for \ac{AI}/\ac{ML} workloads. Innovations in this area can reduce the energy costs of model training and inference, making \ac{AI}-driven data processing more \revision{efficient}.

\revision{Efficiency also means two things: ensuring that scientific results are reliable, and avoiding unnecessary repetition of the same work over and over again. This is why open data and the \ac{FAIR} principles~\cite{Wilkinson:2016} start to play an increasing role. Open data is the essential first step to make simulations reproducible and therefore trustworthy, by providing access to usable software and to raw data from the numerical experiments.}

\begin{tcolorbox}[colback=green!5!white,colframe=green!75!black]
    \textbf{Observation 7:} Every user should strive to make their simulation results \ac{FAIR}~\cite{Wilkinson:2016}, which includes making them reusable. This way, other researchers can avoid running the same simulation again, use the results as input for their own work, and re-analyse the data from a different research perspective. For simulations producing only small statistical samples, reusable data can be accumulated to increase the sample size by pooling the results from different researchers.
\end{tcolorbox} 

The challenges of energy-efficient data management demand a holistic approach that integrates improvements across data movement, storage, and processing. Advancements in transfer and caching infrastructure, archival solutions, and \ac{HPC} systems designed for real-time workflows are critical.

\subsection{Applications and Benchmarks}
\label{sec:hpc:apps}

\revision{Computational sciences create digital representations of scientific problems and in doing so are constantly confronted by a limitation on the available human and computational resources: \emph{(i)} the available budget for a project, the number of people assigned to the development of the code, the availability of the relevant experts, and \emph{(ii)} the available memory, the available disk space, and the available compute time. These constraints are different for different research institutions, and while it is still possible (and elegant) to produce groundbreaking research results \revision{on} a single desktop computer, exploiting the biggest supercomputers can lead to high impact results~\cite{Kawazura2024-tw,Schaye2023-oy,McAlpine2022-uv,Riley2022-rl,Warnecke2023-uj,Vasil2024-zx}. This is the motivation for most computational scientists to approach \ac{HPC}.}

\revision{We give in this section the general recommendations for any developer or user of applications in \ac{HPC}.} The following three sections go into more detail on how scientists specifically in the areas of astrophysics (\cref{sec:astro}), \ac{HEP} (\cref{sec:hep}), and \ac{LFT} (\cref{sec:lqcd}) can address energy efficiency and performance.

\subsubsection{\revision{Skills}}
\label{sec:hpc:apps:skills}

\revision{Up to about the year 2000, in most scientific fields primarily the same researcher developed the code, ran the simulations, and analysed and visualised the results. These researchers knew every line of their codes and understood the numerical challenges. The severe limitations in computing time forced the researchers of this era to spend much effort optimising every aspect of the code for efficiency.}

\revision{Nowadays, scientists starting to use \ac{HPC} typically take their first steps with legacy code inherited from their predecessors, or with large analysis frameworks used as a black box. 
Their tasks often incorporate writing small extensions to study a new aspect of a scientific problem, but there is rarely an immediate motivation or an offer for the necessary education to get accustomed to the existing code base. 
Both time and software engineering training are lacking to perform tasks such as porting code to new environments, refactoring, or performance engineering. 
Furthermore, they implicitly assume that the used code has already been optimised for efficiency, but this assumption is often wrong because the code optimisation, if done at all, happened on an entirely different computer infrastructure.
As a result, the code produced may not fit today's compute hardware, or may perform significantly worse than it could, due to wrong data layouts, inadequate programming models, or outdated libraries and environments. The long term success of a scientific code rely purely on the motivation, training, and acquired expertise of a few experts associated to the institution developing the software, but too few permanent software developer positions are available.} 

\begin{tcolorbox}[colback=green!5!white,colframe=green!75!black]   
    \textbf{Observation 8:} Application developers benefit from working hand in hand with experts in \ac{HPC} software development, usually employed at \ac{HPC} centres. There are multiple means for developers to have access to this expertise (e.g., training courses and direct mail contact with support teams), learning in a very short time how to extract the maximum potential of their codes. 
    However, the best approach for continuous code improvement and adaptation to emerging technology is to include in the development team \ac{RSE} experts with knowledge on the physical domain of interest and on the \ac{HPC} ecosystem. It is expected that the research teams using the latest \ac{HPC} systems to their full potential have in their ranks expert \acp{RSE}, who are commonly postdoctoral researchers with additional specialised training in \ac{HPC} and advanced programming.
\end{tcolorbox}

\revision{In particular, there is a growing risk of losing, over time, the expertise required by research institutions to produce high performance codes, as new generations of researchers are not trained in traditional scientific computing languages (e.g., C/C++ or Fortran). Today, the few research engineers trained with detailed knowledge of traditional programming languages, have a clear advantage in the job market not only in the scientific domain but also in any industry requiring coding (see for example the banking and the video game industries).}

\revision{Application developers can find it difficult to keep up with the very rapid evolution of hardware, and only well-funded research groups with the necessary \ac{HPC} expertise and specialised support can take advantage of emerging new technologies. Given the growing use of \ac{HPC} in science and industry, building the knowledge of the next generations of computational and computer scientists through dedicated courses at universities and training programs becomes crucial. These courses should cover all aspects of \ac{HPC}, including low-level programming languages, performance analysis and engineering, parallelisation and scaling, code optimisation and tuning, and data management. }

\begin{tcolorbox}[colback=green!5!white,colframe=green!75!black]  
\textbf{Observation 9:} \revision{Students and early career researchers in computational sciences should seek training on numerical algorithms, data access patterns, and the frameworks used in their community, as well as on general \ac{HPC} topics such as performance analysis, optimisation, and software engineering. Specific education on energy-efficient simulation techniques is required, including concepts such as the energy footprint of different hardware devices, software tools, and programming models. Similarly, beyond familiarity with a high-level language such as Python or R for data analysis, students should know that some Python packages are more energy efficient than others, and establish a good foundation in modern C++, which is a prerequisite to contribute to the continued development of efficient application software.} 
\end{tcolorbox}

\subsubsection{\revision{Performance analysis}}
\label{sec:hpc:apps:perf}
Generally speaking making an application code more energy efficient is equivalent to ensuring that it reaches the maximum performance on a given compute device, \revision{which itself} is running at the most energy efficient operational configuration. Or said otherwise, the application should not waste the capabilities of the hardware on which it runs. In practice, it means: \revision{\emph{(i)} understanding which is the maximum theoretical performance that the target computer is capable of, \emph{(ii)} measuring the performance that the own code currently attains on this computer, and \emph{(iii)} knowing how to close the gap between \emph{(i)} and \emph{(ii)}}. This can be studied in first approximation with tools like the \emph{roofline} model~\citep{williams2009}, which relates the performance (in \ac{FLOP}/s) with the arithmetic intensity (\revision{in \ac{FLOP}/Byte,} aka operational intensity) of an application with the peak performance of a given computing device, classifying applications into memory and compute bound. Numerous profilers and performance analysis tools integrate the roofline model and \revision{provide} further bottleneck analysis mechanisms for parallel applications (e.g., Vampir~\cite{vampir}, Paraver~\cite{paraver}, Scalasca~\citep{Geimer2010}, Vtune~\cite{vtune}).

\revision{Using these tools, developers have to invest on the analysis of the performances of their code prior to seeking the best technological solutions to remove bottlenecks. Common blocking points include inadequate parallel communications, uneven data transfers, blocking I/O tasks, slow data ingestion, irregular memory access, idling computing resources, and unrealised compiler optimisations. Addressing all these requires access to tutorials, literature, and tools that simplify the analysis of computer software ported to \ac{HPC} systems is needed~\cite{Kreuzer2021}. }


\begin{tcolorbox}[colback=green!5!white,colframe=green!75!black]
    \textbf{Observation 10:} 
    First step to improve the energy efficiency of an application is reducing its \emph{time-to-solution} by investing effort on performance engineering. Performance analysis tools and profilers help identifying communication and computation bottlenecks and sources of performance loss. 
    Small changes in bottlenecks can rapidly provide great gains in performance. \revision{Typical code optimisations include early detection of memory leaks, good alignment of data in memory to take advantage of cache access, use of existing compiler optimisations taking advantage of hardware characteristics, and smart use of multi-threading in all cores in a node.}    
\end{tcolorbox}

\subsubsection{\revision{Artificial intelligence and low precision arithmetics}}
\label{sec:hpc:apps:ai}

\acp{GPU} have been successfully adopted in \ac{HPC}, but even more so in the \ac{AI} market. Model training requires very large amounts of computation, which can be done very efficiently with \acp{GPU}. Unlike most \ac{HPC} applications, \ac{AI} training makes heavy use of reduced precision arithmetic (e.g.~\cite{Doerrich2023}). The large size of the \ac{AI} market motivates \ac{GPU} vendors to devote more silicon area to these arithmetic units at the expense of double precision support. Applications using low-precision arithmetic can achieve higher performance with less power consumption. \ac{HPC} application developers are therefore strongly encouraged to consider whether some of the operations in their codes could be performed in single (or lower) precision. However, it remains to be seen how well the diverse portfolio of \ac{HPC} applications can be successfully ported, taking into account the resulting losses in numerical stability and reproducibility. Intense algorithmic research is being done to enable more \ac{HPC} applications the transition from \ac{FP64} to reduced-precision arithmetic (e.g., FP32, BF16). If the \ac{AI} market continues to grow at its current rate, we can expect to see new accelerators entering the market, particularly for inference operations (e.g., TPU~\cite{TPU}, Graphcore~\cite{graphcore}, Groq~\cite{groq}, etc.), which are likely to support mostly low precision arithmetic, but potentially also new programming models. This again, might require additional adaptations of \ac{HPC} application codes.

\begin{tcolorbox}[colback=green!5!white,colframe=green!75!black] 
    \textbf{Observation 11:} Opportunities to work with reduced arithmetic precision should be exploited, as this translates into a higher performance and energy efficiency of the application when running on the same hardware.
\end{tcolorbox}

\subsubsection{\revision{Benchmarks}}
\label{sec:hpc:apps:bench}
Application-based benchmarks are included in the procurement and acceptance of \ac{HPC} systems to represent the main consumers of computing time~\cite{herten2024}. This allows for choosing a system design and a selection of hardware and software components appropriate for the site's application portfolio, which is critical to maximising energy efficiency in real-world operations. The main difficulty is to accurately predict how the application portfolio at the time of acquisition will evolve over the operational lifetime of the supercomputer. Application developers can contribute to this effort by creating \emph{mini-apps} and application use cases representative of their operational workloads, making their source code and input data publicly available, and keeping them up-to-date with the latest code releases. Benchmarking campaigns are also important for the application developers themselves, as they serve to analyse how \revision{changes in a given code (or environment)} affect performance. Equally important is applying professional software development strategies, with curated code repositories, version control, \ac{CI/CD} pipelines, automated testing, code documentation, and regular software releases. 

\begin{tcolorbox}[colback=green!5!white,colframe=green!75!black] 
    \textbf{Observation 12:} Benchmarking studies should be performed before running production jobs to identify the most energy-efficient job and system configuration for the given application, looking for the lowest possible operational frequency of the hardware without strongly increasing the time-to-solution (e.g., in memory bound applications).
\end{tcolorbox}

\bigskip

\section{Computational Astrophysics}
\label{sec:astro}


Studying the movement of galaxies at the cosmological scale or the motion of a single electron around a magnetic field line, the domain of astrophysics covers a very wide range of scales in time and space. While the equations that govern all the dynamics of the universe are well known, it is still impossible to reproduce in a computer all the different processes at once. Scientists need to make assumptions, neglect terms in equations, reduce the complexity of the environment under study, eliminate non-linearities, and translate mathematical equations into numerical algorithms solved on discrete machines working with ones and zeros. Each one of these \revision{simplifications} introduce an uncertainty to the final result. The role of the \revision{application} developers in computational astrophysics is to minimise such uncertainties and to describe as accurately as possible the real universe into a matrix of discrete values.


Many computer models in the domain of astrophysics have emerged from the work done by scientists over multiple years, even decades. In a traditional academic life-cycle new numerical techniques are typically introduced by senior researchers, which are then extended and refined by their research teams (postdocs, PhDs, master students). Very rarely an academic software is developed by a professional software engineer. Over the past two decades the hardware used to execute such software has become increasingly more complex (see \cref{sec:hpc}). Memory strategies, data traffic, data structures, I/O, accelerators, scalability, interconnections, are some of the terms \revision{that} have come to the attention of research software developers. A new type of expert is emerging from this environment: the \ac{RSE}, who is both \revision{an} expert in physics and computer sciences.

Making simulation results reusable for others is an essential contribution to making astrophysics simulations more energy efficient. In observational astronomy, a large proportion of the data are published. There, the \ac{FITS} format~\cite{Wells:1981} is a widely accepted standard for data publication. It is used for the transport, analysis, and archival storage of scientific data sets. By contrast, open access is much less common in astrophysics simulations, especially if it does not only concern sharing code but also the resulting data. Here, general agreements on used formats and metadata standards are still lacking.  A few large collaborations set good examples of sharing the results data, i.e, the Illustris project~\cite{illustris}, 
the TNG project~\cite{tng}, 
the Eagle project~\cite{eagle}, 
and the Horizon simulations~\cite{horizon}. 
In the context of astronomy and astrophysics, reproduction usually requires re-evaluation of large datasets and the discussion about \ac{FAIR} research data management has just started for astrophysics simulations. 

\revision{It is our goal in this section to present the trends in energy efficient \ac{HPC} to research leaders and research engineers working on astrophysics codes.}

\subsection{Stellar and Planetary Physics}
\label{sec:astro:stellar}

%
            Codes in stellar and planetary physics were initially developed to describe specific physical process, for which a distinct numerical method was used. \revision{For example, the dynamics of particles were calculated using N-body methods and hydrodynamic flows employed numerical fluid dynamics methods}. Over the decades, these codes have developed and nowadays they usually include \revision{multiple} physical processes \revision{simultaneously}. Often, they also allow one to choose which numerical scheme to employ. For example, the newest version of the GADGET code (GADGET 4)~\citep{Springel:2021} also contains a simple model for radiative cooling and star formation, a high dynamic range power spectrum estimator, and an initial conditions generator based on second-order Lagrangian perturbation theory. Besides, it gives several choices of gravity solvers and smooth particle hydrodynamic options.  

            Other approaches are frameworks that allow combining different codes to provide for various physical processes. One example is the Astronomical Multipurpose Software Environment (AMUSE)~\citep{Portegies:2018}, which allows to combine multiple solvers to build new applications, used to study gradually more complex situations. This procedure enables the growth of multi-physics and multi-scale application software hierarchically. Generally, the complexity of the software continues to increase. 
       
        
       
        Most simulation runs are a single realisation of an ensemble of possible states of a system. Thus, it is necessary to perform an entire suite of simulations to increase the statistical significance of the inferred result. However, multi-physics codes, especially, often require such extensive computational resources that only a single simulation \revision{run can be performed}. The statistical significance of a single realisation of a statistical approach is rather limited. 
        Examples of such simulations that went to the limits at the then available computational resources are the Millenium simulation~\citep{Springel:2005} or the simulation of clustered star formation~\cite{Bate:2003}. In principle, this problem should have solved itself with computational \revision{performance} increasing by a factor of several thousands to several ten thousands during the last 20 years. However, this is often not the case. Rather than increasing statistical significance, researchers often prefer to use the increasing resources to boost resolution or include additional physics. 
       
        Due to the modular nature of many physics aspects included in a single code, testing the code to its entire extent becomes increasingly challenging. The researcher can evaluate only \revision{individual} components against a set code test. However, few to no tests exist of how the different parts work together. 
           
        Making use of \ac{GPU}-based \ac{HPC} \revision{systems} can increase energy efficiency \revision{in astrophysics} considerably. For example, the \ac{GPU}-based version of Nbody6 is approximately 10$\times$ faster than its \ac{CPU}-based version~\citep{Nitadori:2012}. 
        The degree of transition to \ac{GPU}-ready codes is mixed in the star and planet formation field. Popular codes for simulating like FLASH~\citep{Fryxell:2000}, GADGET4~\citep{Springel:2021} or REBOUND~\citep{Rein:2012} were not intended for \ac{GPU}-based machines. The user community has become aware, and parts of the codes are gradually changing the algorithms to use \ac{GPU}-based structures. For N-body codes, the situation is better. Codes like Nbody6GPU++ already saw the advantages of using \acp{GPU} about 15 years ago. Other \ac{GPU}-ready N-body codes are, for example, BONSAI~\citep{Bedorf:2020}, GENGA~\cite{Grimm:2014} and PETAR~\cite{Wang:2020}. In summary, the star and planet formation community is currently in the transition phase to \ac{GPU}-ready codes.
           
        Making \revision{open access of results and data,} and ideally fulfilling the \ac{FAIR} principles~\citep{Wilkinson:2016} would make simulations also more energy efficient. 
        However, the community is only starting to evaluate how to adopt \ac{FAIR} principles when dealing with simulation results and data management. 
        Codes are increasingly shared on GitHub or the Astrophysics Source Code Library~\cite{ascl}. However, the shared code versions often differ from those used to \revision{obtain the results reported in publications, because the codes are proprietary or the team simply lacks the resources to properly document new code parts}. 
        Benchmark tests are performed for the \revision{publicly} available code version but are less often performed for the codes with non-public add-ons. Thus, the user is sometimes unaware that these add-ons are less efficient than the main code.
       
        In the past, many codes with similar applications existed in parallel, and there was a constant stream of newly developed codes. While new codes are still being created from scratch, this happens to a lesser degree due to the \revision{growing} code complexity. Nowadays, much of the development effort goes into adapting legacy codes \revision{written in Fortran, C or C++} to new computing structures and extending them by adding new modules \revision{often developed in Python}.    
        Often, \revision{simulation codes} are written by students who are unaware of the energy efficiency issue or lack the knowledge of more energy-efficient programming languages.

        Especially for simulations reaching the limits of computing, replacing part of the code with \ac{AI}-generated information is implemented in some codes. So far, this transformation code approach has only been adopted by a relatively small part of the community. While promising to make the codes more energy-efficient, one loses some of the causal information. Furthermore, the compute and energy cost of training the \ac{AI}-models should not be neglected.

\begin{tcolorbox}[colback=green!5!white,colframe=green!75!black]   
    \textbf{Observation 13:} When an application framework allows \emph{switching on} and \emph{off} different physical processes, it should be ensured that only those necessary to answer the investigated scientific question are activated. As such, during the planning of the simulation approach, one should put emphasis on allocating the minimum necessary amount of computational resources (nodes), while maximising the use of those that are reserved. Users should check after completion whether the code performance stayed high throughout the runs, and learn how to minimise idling resources. 

\end{tcolorbox}

\subsection{Space Weather and Space Physics}
\label{sec:astro:space} 

\begin{figure}[!ht]
    \begin{center}
    \includegraphics[width=0.7\linewidth]{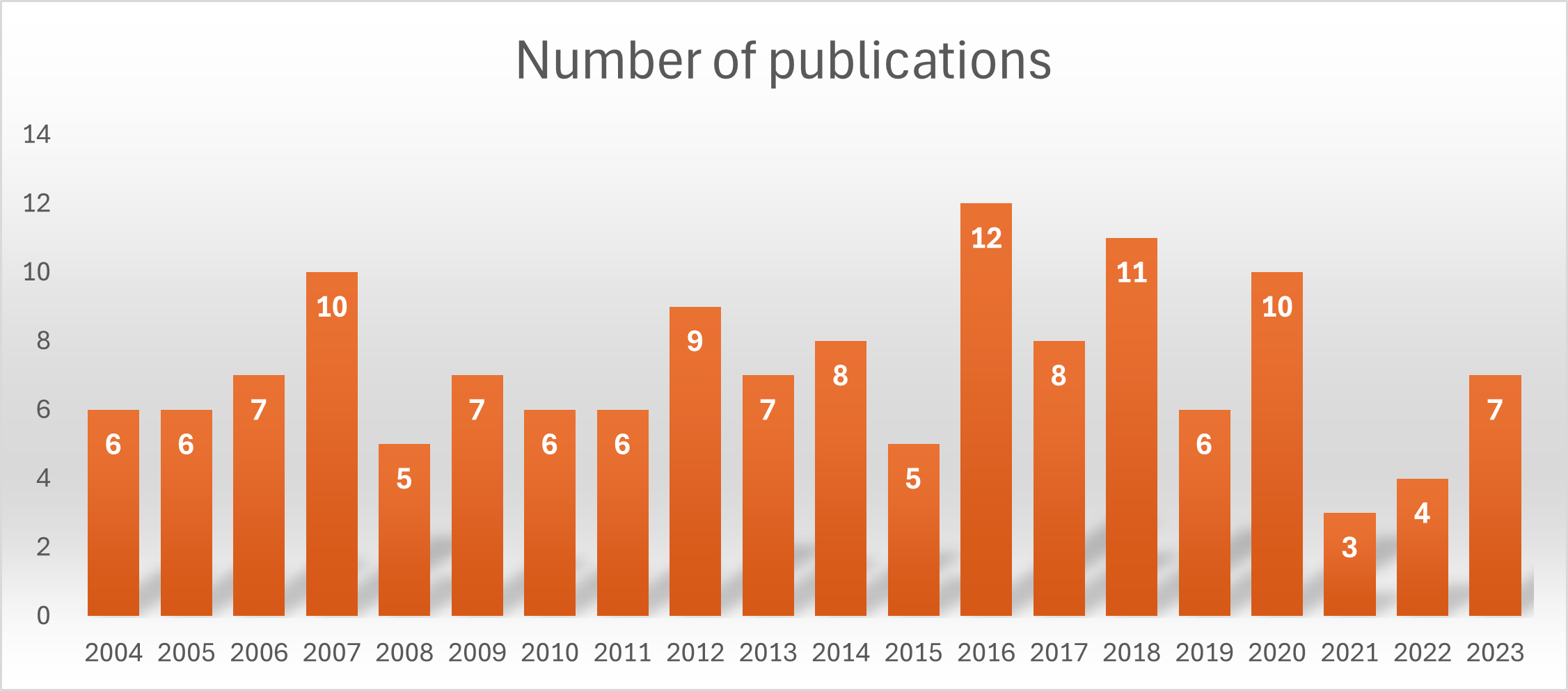}
    \end{center}
    \caption{Number of publications in scientific journals over the past 20 years, containing the keyword \revision{\emph{space weather},} which include in the text references to \revision{\emph{supercomputer}} or \revision{\emph{high performance computing}}. Data gathered using the OpenAlex database~\cite{openalex}. The median number of these publications over the past 20 years is seven.}
    \label{fig:swepub}
    \end{figure}

    \revision{Scientific codes in the area of space physics show very similar characteristics as those described in the previous section. However, the domain of space weather is one step closer to real-time operations and require high availability of computer resources, and high reliability on the outcomes of the computations. To forecast the impact of the solar activity on the Earth, several advanced computer models take advantage of \ac{HPC} systems. Multiple models exists still in their research phase while a few have already transitioned to an operational phase and are used to inform the public and industry}. The process of transforming academic codes into consolidated operational software is called \ac{R2O}. Historically, the domain of space weather has been a continuous user of \ac{HPC} resources. \Cref{fig:swepub} shows that over the past 20 years the number of publications in the domain of space weather that makes explicit use of \ac{HPC} systems has been constant. In general the domain of space physics (including solar physics, space weather, heliospheric physics, planetology, and plasma physics) is known to exploit the latest computing technology, including new generations of processors, accelerators, memory, storage, and network interconnection~\cite{Ruzicka2024}~\cite{Romein2021}~\cite{Zhang2022}~\cite{Ren2021}~\cite{Deluzet2023}.

\begin{figure}[!ht]
    \begin{center}
    \includegraphics[width=\linewidth]{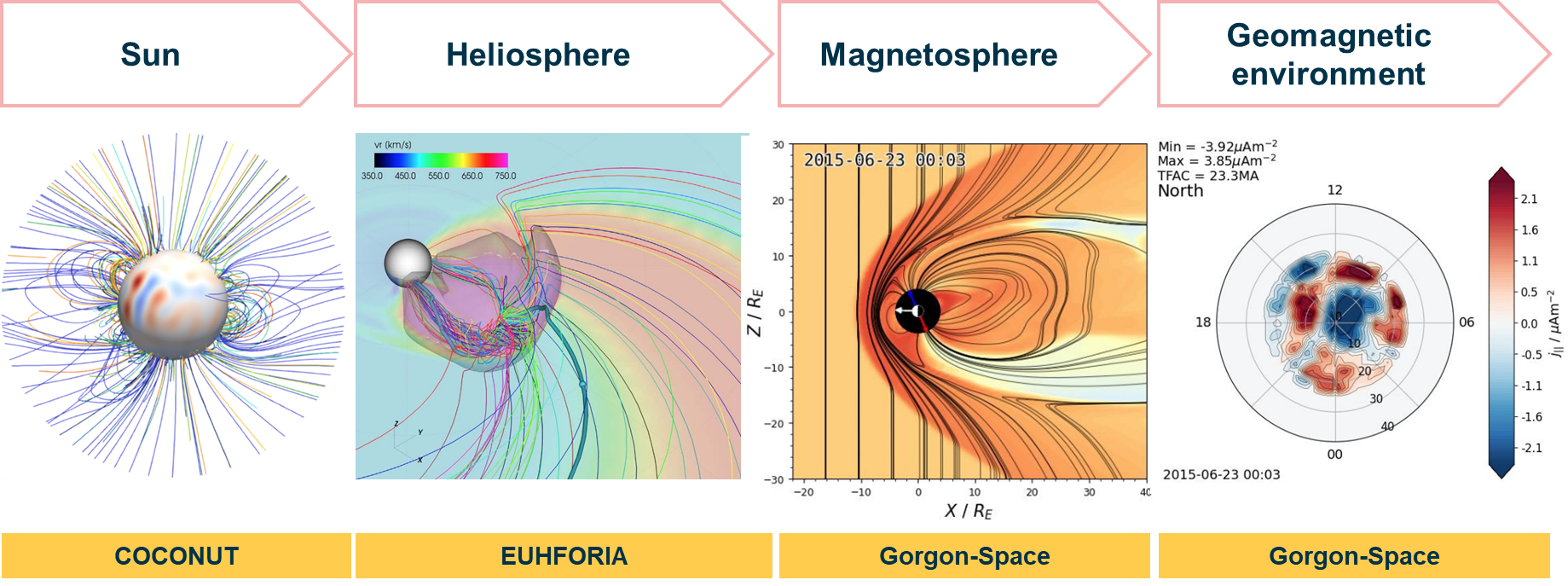}
    \end{center}
    \caption{Three computer models used in the space weather domain to connect solar energetic events on the Sun with its impacts on Earth. COCONUT: modelling the state of the lower solar corona. EUHFORIA: transports of solar plasma towards the planets of the Solar System. Gorgon-Space: calculates the geomagnetic environment around our planet.}
    \label{fig:swechain}
    \end{figure}

    The physics of the space environment cover a large range of scales in time and space. It is common practice to study the phenomena observed at each characteristic scale by independent models. This translates in general into computer models that calculate numerical approximations, covering different segments of the full space environment. To capture as much detail as possible, these numerical models can take advantage of very large computer resources to extend their applicability to a large range of characteristic scales. \Cref{fig:swechain} shows a very small selection of individual computer models studying different segments of the space weather domain that connects the solar activity with its effects on Earth. While most of the physics in these codes are very similar, the numerical methods and characteristic scales processed by each code are different.

    In the domain of space physics and space weather the three most common approximations to model the corresponding physics are: \revision{\emph{(i)}} the particle kinetic scale~\cite{Bacchini2023-sl, Lapenta2017-cu, Chen2019-ns, Chen2012-vy}, \revision{\emph{(ii)}} the fluid continuous scale~\cite{Keppens2021-et, Hinterreiter2019-st, Eggington2018-hf}, and \revision{\emph{(iii)}} the hybrid scale, where particles and fluids coexist~\cite{Le2016-qy, Lottermoser1998-qj, Swift1996-ev, Shi2021-pq}. The equations solved by these numerical approximations can be very diverse, and different computer hardware might be better adapted to each of them. It is reasonable to solve the dynamics of billions of particles in a group of accelerators, while solving the memory intensive resolution of a continuum using a network of general purpose \acp{CPU}~\cite{Kreuzer2018}. To make a selection, a careful knowledge of the algorithms, the hardware, and how to connect the two of them is necessary. 
    
    \revision{\ac{HPC} is used in space applications} to produce data. With high value data, scientific discoveries can be published, and mitigating actions can take place to reduce the impacts of the space environment on our technology. As the \ac{HPC} systems \revision{become larger, models grow to use them, generating much more data per run. This makes it more challenging to process and translate the outputs of these models into information that can be used in the daily operations of the end-users of space weather forecasts}. High precision numerical models producing data at a high cadence can lead to \revision{gigabyte} or even \revision{terabyte} of data in a single execution. \revision{Therefore, there} is a growing need to process and visualise data in the same location where it is produced, \emph{on the fly}, at the same time as the code is executed. In modern \revision{\ac{HPC}} systems \ac{GPU} and visualisation nodes are also available to the end user. This allows to perform data gathering, analysis, processing, and visualisation in the same system where the execution of the code took place. This also requires the end user to provide all the data pipelines necessary to extract valuable data from the raw outputs, which will then be discarded by the end of the execution.

    \revision{We encourage application developers to also evaluate the potential use of high-order numerical methods. It is a common misconception to think that high-order numerical methods are more resource intensive. It is clear that if a problem uses the same time and space discretisation, high-order methods will take much longer to converge. However, this is an inadequate way to measure performance~\cite{Wang2013-ub}. To obtain the same error low-order methods require much finer discretisation and longer convergence times. The use of high-order methods is then critical for applications that require fine precision in their resolution, as for example in the case of sharp shocks, energy dissipation, or multi-scale effects. In an optimal code, the order of the methods shall be adaptive to optimise the use of resources.}
    
    \revision{Application developers should also consider implementing high-order methods in cases where performance analysis show that the software is memory bound. This approach may help the software to switch the balance towards higher computing intensity. This is of course dependent on the specific numerical method used, as memory stencils might also increase to accommodate the larger data size required by these methods. The application developer should keep in mind how the different cache and memory levels are used in the computationally intensive segments of their software \citep{Yotov2007,Williams2024,guillet:tel-04277746}. Reducing the data movement among the different memory levels, and performing vector operations, can further help transforming memory bound into compute bound codes.}

\section{Data processing in experimental High-Energy Physics}
\label{sec:hep}


Energy efficiency has become a pressing challenge in \ac{HEP} computing as the field grapples with the ever-growing demand to manage and process massive datasets. With experiments at the \ac{LHC} already having generated over an exabyte of data, and the forthcoming \ac{HL-LHC} expected to add another exabyte per experiment annually, the infrastructure is facing unprecedented challenges. This rapid expansion underscores the need for innovative approaches to improve energy efficiency across data processing, storage, and distribution.

\ac{HEP} computing relies on highly complex workflows, particularly for reconstructing and simulating particle collisions. Many of these workflows have been developed over decades, with vast, intricate code-bases optimised for traditional computing architectures. However, this reliance on legacy poses significant challenges for modernisation. 

Distributed computing, exemplified by the \ac{WLCG}, has been instrumental in enabling \ac{HEP}’s global research efforts. This extensive network connects hundreds of computing \revision{centre}s worldwide and has successfully supported the immense computational needs of the field for years. However, adapting such a vast and distributed infrastructure to modern, energy-efficient architectures presents another layer of complexity. The extensive volume of legacy code, coupled with the need for algorithmic and structural redesigns, complicates the process of integrating new hardware technologies. 

Recognising the urgency of these challenges, the \ac{HEP} community has launched robust research and development initiatives to enhance energy efficiency and modernise its computational ecosystem. These efforts include optimising workflows, reducing unnecessary data movement, and developing software tailored to more efficient hardware platforms. CERN Openlab~\cite{cernOpenlab}, a collaborative partnership between \ac{HEP} scientists and industry leaders, has helped drive this progress. Over two decades, it has guided the transition to modern computing paradigms, from x86 commodity clusters to accelerated hardware architectures like \acp{GPU}. More recently, it has facilitated explorations of tools such as SYCL~\cite{sycl}, Intel OneAPI~\cite{oneAPI}, and other portability libraries, which facilitate the migration of legacy applications to heterogeneous systems. These innovations have not only improved efficiency but have also increased the adaptability and maintainability of \ac{HEP} software, ensuring its compatibility with future hardware advancements.


Experimental \ac{HEP} computing is increasingly challenged by the need to manage and process vast and ever-growing datasets while improving energy efficiency. As the field pushes the limits of data-intensive science with experiments at the \ac{LHC} and prepares for the High Luminosity upgrade, \revision{innovative solutions are needed to address the very large energy demands}. 

\begin{tcolorbox}[colback=green!5!white,colframe=green!75!black]    
    \textbf{Observation 14:} The field of \ac{HEP} faces a significant challenge in achieving energy efficiency due to its reliance on legacy code and traditional architectures. The transition to modern, energy-efficient platforms such as \acp{GPU} and \acp{FPGA} demands not only rewriting code but also fundamentally redesigning algorithms to align with the parallel processing and memory hierarchies of these platforms. \revision{Adopting programming models such as SYCL~\cite{sycl}, OneAPI~\cite{oneAPI}, and other portability libraries increases the adaptability and maintainability of \ac{HEP} software, facilitating compatibility with novel energy efficient hardware technologies and meeting future computational demands.}
\end{tcolorbox}

\subsection{Data Analysis in Collider Physics}
\label{sec:hep:collider}


\begin{figure}[!ht]
    \includegraphics[width=\textwidth]{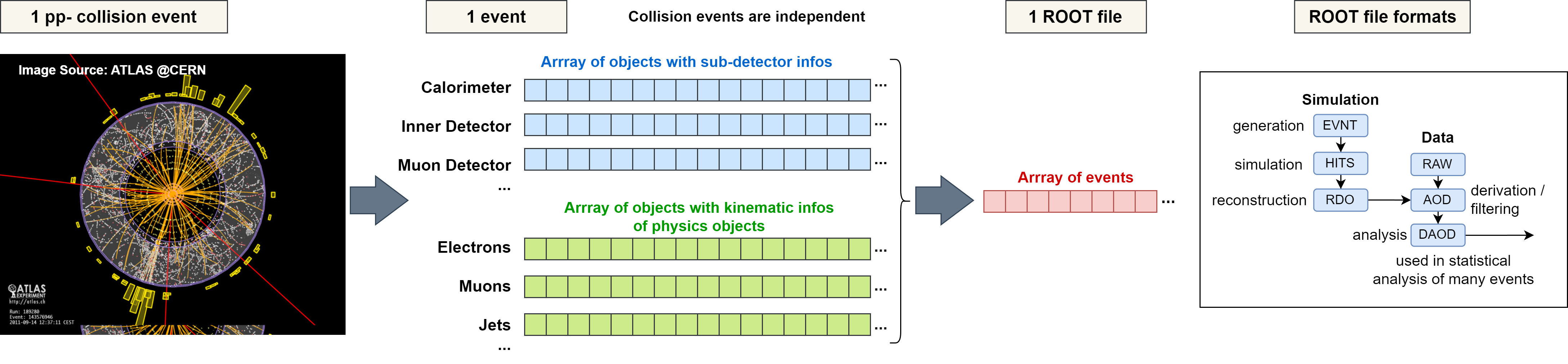}
    \caption{Stages in the analysis in High Energy Physics for the example of the ATLAS experiment. Both simulation and real data processing produce different levels of data files, which are stored in the common ROOT~\cite{rene_brun_2020_3895860} file format. The \ac{DAOD} \revision{files} containing structured object data are commonly used as input for end user analysis.~\cite{Elmsheuser2020}.}\label{fig:atlas_datafiles}
\end{figure}

The computing model in \ac{HEP} involves several different approaches on how code is developed and how resources are used. The core part of pre-processing all the data collected at the experiments and generating Monte Carlo simulation data is performed using the main analysis frameworks of the experiments, usually maintained by experienced developers as performance is critical. Programming languages such as C++ remain prevalent and in general changes undergo a review procedure before being put into production. Examples for these core components are Athena~\cite{atlas_collaboration_2019_2641997} for the ATLAS experiment, the CMS offline software (CMS-SW~\cite{Bayatian:922757}) for CMS, or the Belle II software basf2~\cite{Kuhr_2018}. At this point of the analysis chain, raw data is converted into highly structured object data, and finally a reduced set can be derived as filtered structured object data, \revision{see \cref{fig:atlas_datafiles}}. 

\begin{figure}[!ht]
    \includegraphics[width=0.95\textwidth]{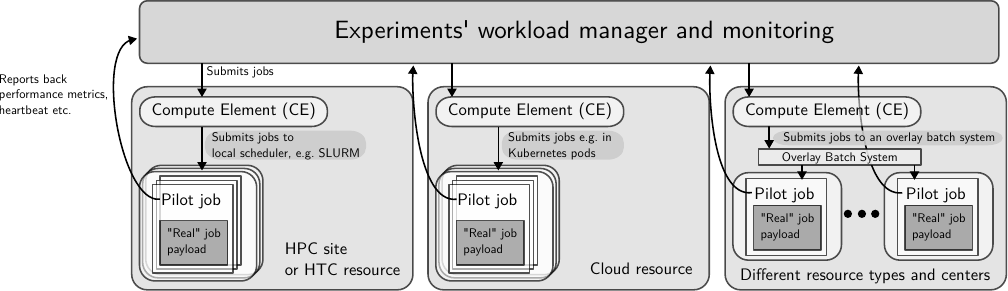}
    \caption{Simplified and generalised schematic of job execution on the \ac{WLCG}. Experiments operate their own workload managers, which submit pilot jobs to compute elements at various resource \revision{centre}s. \revision{These} in turn submit jobs to local schedulers such as Slurm or HTCondor, cloud resources, or even overlay batch systems grouping various resources into another batch system. The pilots launch the actual payload job and report performance metrics and heartbeats back to the experiments' central monitoring infrastructure. \revision{Note that most workflows in \ac{HEP} rather fit a \ac{HTC} model as analyses are event-based, i.e, focus is put on the overall throughput aiming to maximise the usage of available resources. Individual compute jobs are commonly limited to single nodes and can use dedicated \ac{HTC} resources, which might be built from very heterogeneous systems operated by members of the community. The overall model of operation bundles together such dedicated resources, \ac{HPC} resources and cloud resources into one overlay batch system.}%
    }\label{fig:grid_execution}
\end{figure}

These tasks are executed on the \ac{WLCG}, which leverages both dedicated resources and an increasing amount of resources \revision{that} can be used opportunistically. Dedicated workload managers operated by the experiments such as PanDA~\cite{Maeno2024}, the CMS workload manager~\cite{Cinquilli_2012} or DIRAC~\cite{Adrian_Casajus_2010} are used on top of the resource's own workload managers such as Slurm~\cite{slurm} or HTCondor~\cite{htcondor}. The experiment-specific workload managers wrap the actual computing payload within so-called pilots, which monitor the job execution and report back on resource usage and potential problems to the experiment-specific workload managers and monitoring systems as illustrated in \cref{fig:grid_execution}. The highly different error conditions of such a large, distributed systems require a significant amount of manpower to hunt down any issues and to intervene if a drop in efficiency is detected. Efforts to automate the most common cases are continuously undertaken~\cite{Girolamo:OpInt}.

\begin{figure}[!ht]
    \includegraphics[width=0.8\textwidth]{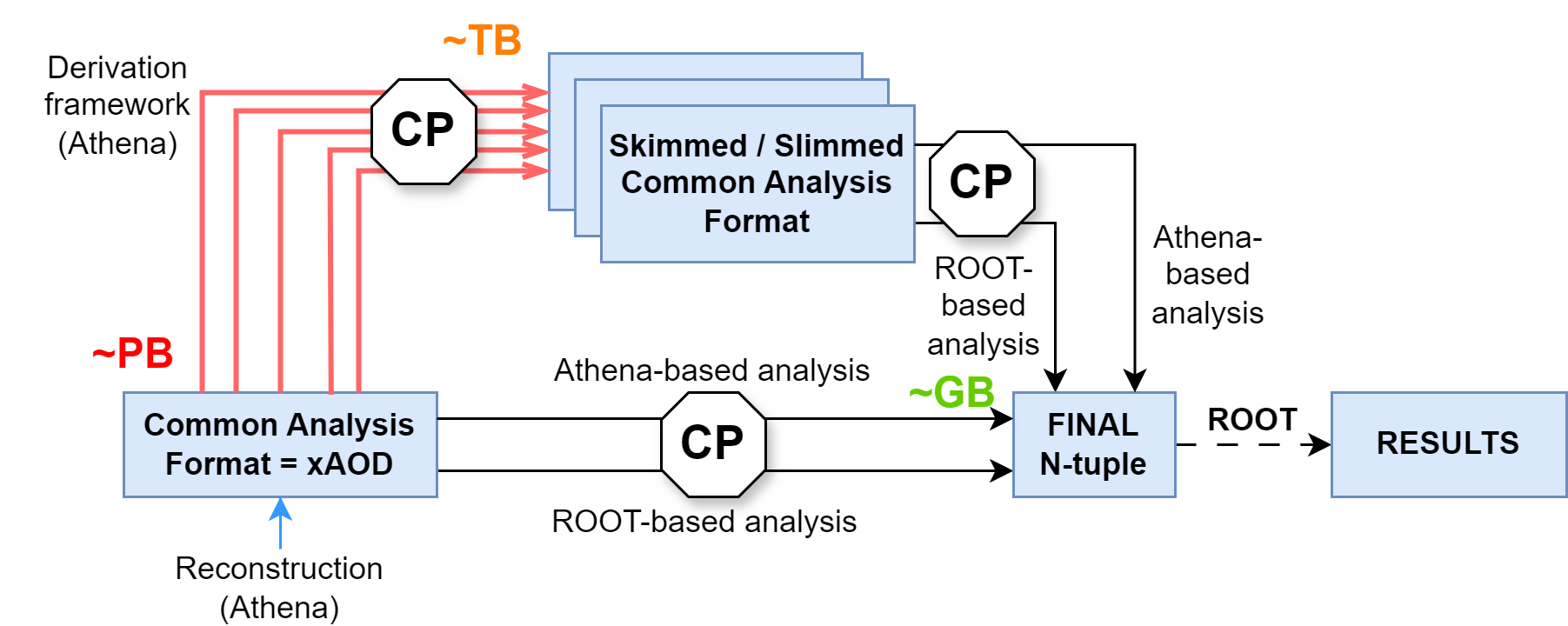}
    \caption{Data reduction for end user analysis for the example of the ATLAS experiment. \ac{AOD} files with structured object data after reconstruction are still sizable and \revision{derived data sets are produced for end users}. \emph{CP} relates to common calibration and object selections, and in the final stage, columnar n-tuple files are produced for consumption by end user analysis tools~\cite{Elmsheuser2020}.}\label{fig:atlas_filtering}
\end{figure}

The physicists performing actual data analysis are commonly leveraging such pre-processed data and pre-generated Monte Carlo samples in their analysis tasks. \revision{They} may be using the core frameworks to further filter the data or develop extensions as part of their work, and can also execute these on the \ac{WLCG}. \revision{However}, they usually do not work with the raw data directly, but with data filtering and histogramming tools operating on these pre-filtered data samples stored in a columnar data format as shown in \cref{fig:atlas_filtering}. 
Both the centrally organised \emph{production usage} and the end user analysis are subject to growing computing demands, due to an actual increase in the amount and complexity of the data, and to more complex algorithms employing \ac{ML} techniques or interactive data filtering with a short feedback loop.
    

 Most production workflows churn through data event by event, producing filtered and highly structured data and / or statistically summing up the individual results. By design, there is no interference between the individual events during analysis (assuming effects such as pile-up have \revision{been accounted for}). For that reason, the \revision{workflows could adapt to dynamic availability of resources if eviction of the jobs without loss of computation time was possible. To accommodate this, frameworks need to} be extended to checkpoint either regularly or on demand, as only the current event number and other status data (such as random seeds) and the intermediate output need to be stored. \revision{This} would allow to suspend the execution without loss of computation time or migrate the computation to a location with \revision{underutilised resources, ideally} close-by. In the current computing model, though, the processes are executed within a pilot environment that monitors the execution, reports back resource usage, and sends regular heartbeats to the central experiment workload management and monitoring systems. These, in turn, disallow the underlying schedulers of the compute site to checkpoint and restore running jobs or migrate them to another node as the meta-schedulers of the experiments would declare such a job as failed. It should be noted that this can also cause problems with \ac{DVFS} due to the expected wallclock time being calculated from a previously reported benchmarking value. \revision{Removing this limitation would increase the overall efficiency due to better usage of existing resources, for example by backfilling empty slots opportunistically, without blocking main users of these resources by enabling eviction without loss of computation time.}




It has become common in \ac{HEP} to wrap both the production workflows and the end user analysis in containers with a well-defined environment and software packages maintained by the corresponding communities. While this certainly helps with analysis reproducibility and eases the time spent until a new student can execute code for the first time, it is often also used as a reason not to port the code to a newer base operating system or new releases of the base libraries used within the community. \revision{However, porting code is preferred when possible, as it} can have positive effects both on analysis correctness, performance, and energy efficiency. 


An ongoing trend especially in end-user analysis is a large push for interactive data analysis, i.e, fast access to columnar data formats and dynamic filtering for statistical analysis. This requires not only high bandwidth to the actual storage and access to compute resources, which are either dedicated to the use case or can be provisioned on-demand, but also a change to the underlying storage formats for increased performance when changing access patterns quickly between different analyses. RNTuple, which is effectively developed to become the successor of the classic TTree columnar data storage format for arbitrary C++ types and collections, has proven to surpass the existing format in terms of performance and space efficiency. It even outperforms industry standards such as HDF5 and Apache Parquet in access patterns common to \ac{HEP} analyses~\cite{Lopez-Gomez:2808833}. However, adoption of new formats proves to be a slow process due to the plethora of different analysis frameworks built on top of the common ROOT framework~\cite{rene_brun_2020_3895860} and reliance of existing code on specific behaviour of TTree, as backwards compatibility had to be broken. 
    

Some of the user and production workflows in \ac{HEP} rely on large local scratch space. This is \revision{the consequence of modular tooling relying, e.g., on exchange data formats between event generators and analysis tools, such as the verbose \emph{Les Houches} event file format~\cite{Alwall2007}}. In terms of performance \revision{and} efficiency, it would be worthwhile to investigate a more direct way to connect \revision{event generators and analysis tools} to each other, avoiding on-disk data exchange formats. \revision{But this is hindered by} large code bases and barriers between different developer subcommunities and programming languages. 
    

It must be noted that most workflows operate in a data streaming model, i.e, data is read and written out event-by-event with the exception of smaller statistical summary output such as histograms. While there is a requirement for high data throughput especially for the input files, there is no actual requirement of POSIX semantics such as locking, directory structures or parallel exclusive file access. \revision{Data} is accessed through the common base framework ROOT, which can deal well with streaming mode or data access through caches~\cite{Boccali2019}. Still, the operational model usually implies data replication to the site \revision{on} which workflows are executed. \revision{Some} of the used tools, especially in end user analysis, also expect POSIX semantics to be available. A stronger decoupling would allow to switch to other storage systems such as object stores as already operated in few large \ac{HEP} resource centres~\cite{Ellis2020}, which would help to reduce the complexity of the necessary storage systems and hence also improve \revision{energy efficiency}. 
    

\begin{tcolorbox}[colback=green!5!white,colframe=green!75!black]
    \textbf{Observation 15:} For production usage, \revision{the experiment's workload managers should be improved} to grant checkpointability and eviction functionality, as well as \revision{to better handle different operational configurations of the compute resources (e.g., \ac{DVFS})}.

\end{tcolorbox}


\section{Lattice Field Theory and Lattice Quantum Chromodynamics}
\label{sec:lqcd}

Over the past two decades the \ac{SM} of particle physics has been largely confirmed as a correct effective description of the interactions of all known fundamental particles.
The discovery of the Higgs boson~\cite{ATLAS:2012yve,CMS:2012qbp} can be argued to mark the transition from an era of experimentally mapping the \ac{SM} to the so-called \emph{precision frontier}~\cite{Heinrich:2020ybq} era.
In the search for physics beyond the \ac{SM}, high-energy experiments at the \ac{LHC} are complemented by B-factories such as LHCb~\cite{LHCbCollaboration:2806113} and Belle-2~\cite{BERTACCHI2023107} and by high-intensity experiments in Mainz~\cite{Schlimme:2024eky}, at Jefferson Lab~\cite{JeffersonLabSoLID:2022iod} and at Fermilab~\cite{shiltsev2017fermilab}.
The latter also hosts the Muon $g-2$~\cite{Keshavarzi:2019bjn} experiment aiming at a substantial improvement of the experimental determination of the anomalous magnetic moment of the muon, $a_\mu = (g-2)_\mu$, a key quantity for which the present experimental value disagrees with the data-driven theoretical determination based on dispersion relations~\cite{KESHAVARZI2022115675}.

Many of these experimental efforts aim to identify very small deviations from the \ac{SM} at high energy scales through imprints at much lower energy scales.
Their expected smallness necessitates experimental and theoretical precision at the sub-percent level and below and the study of phenomena at low energies in turn implies that hadronic and non-perturbative effects play a significant role.
As a result, the achievable theoretical precision depends crucially on non-perturbative calculations using \ac{LFT} methods, chief of which is \ac{LQCD}.
Simulations in \ac{LFT} in general and \ac{LQCD} in particular consume substantial fractions of the available compute budgets on the largest supercomputers and thus contribute significantly to the overall energy consumed by \ac{HPC} systems \revision(see \cref{fig:usestats}).

The past twenty years have also witnessed first results of \ac{LQCD} simulations at physical light quark masses~\cite{BMW:2008jgk,MILC:2013ffd,Abdel-Rehim:2015owa,ETM:2015ned}, removing one of the largest sources of systematic uncertainty.
In order to achieve the required sub-percent precision with reliable uncertainties, however, such calculations must fully account for strong and electromagnetic isospin breaking effects and be performed in large physical volumes, at fine lattice resolutions and employing very high statistics.
Based on current state-of-the-art algorithms, one can easily estimate~\cite{boyle2022latticeqcdcomputationalfrontier} that this necessitates several Exaflop-years of computing time, showing not only that machines beyond the current generation of Exascale supercomputers are required, but that \ac{LFT} practitioners must take special care to make efficient use of these resources if the overall energy efficiency of \ac{HPC} is to be improved.
This implies dedicated performance-engineering efforts to ensure that all employed algorithms are implemented in such a way as to achieve performance as close to optimal as possible, subject to machine limitations and the arithmetic intensity of the algorithms in question.

Members of the \ac{LFT} community have been trail-blazers in the \ac{HPC} field, having contributed to the development of various architectures over the past forty years~\cite{Cabibbo:1984zp,Bodin:2004brr,Mawhinney:1999rji,Chen:2000bu,Pleiter2010,haring2011ibm} as well as the corresponding algorithms to make use of these machines.
\ac{LFT} practitioners were also amongst the earliest adopters of \ac{GPU} acceleration~\cite{Egri:2006zm,Barros:2008rd} for key kernels and today, most state-of-the-art calculations would not be possible without \ac{GPU} offloading.
Despite the relatively high level of expertise, the maintenance, improvement and adaptation of the many software frameworks used for \ac{LFT} calculations in lock-step with current hardware trends has become a substantial \ac{RSE} challenge also for this community.

Software frameworks in \ac{LFT} span many hundreds of thousands of lines of code and have accumulated substantial technical debt through their evolution.
At the same time many of the algorithms used for \ac{LFT} calculations have become relatively complex and the hardware landscape has been subject to significant technological diversification and an increased pace of change compared to even just a decade ago (see \cref{sec:hpc:hw}).
As a consequence, \ac{LFT} software frameworks now have to target multiple memory and execution spaces, multiple communication \acp{API}, and various programming models in order to ensure efficiency and performance-portability.

While historically large parts of these frameworks were written by doctoral students or early-career postdoctoral researchers addressing particular research questions, it is very likely that current challenges can only be met by more dedicated long-term investment into software engineering efforts.
This is demonstrated most prominently by the Grid~\cite{Yamaguchi:2022feu,Boyle:2016lbp} and QUDA~\cite{Clark:2009wm,Babich:2011np,Clark:2016rdz} libraries, which use abstraction through C++ features and, in the case of QUDA, autotuning of kernel launch parameters and communication policies, to provide portable and highly optimised implementations of efficient data structures, computational kernels and algorithms for \ac{LFT}.

More generally, performance-portability approaches such as Kokkos and SYCL have been explored~\cite{joo2019performance,simon_kokkos} by the \ac{LQCD} community and could serve as a relatively easy-to-learn basis for the design of \emph{mostly} performance-portable frameworks to be linked against libraries like Grid or QUDA for particularly demanding kernels or algorithms.
Programmer-productivity layers such as Grid Python Toolkit~(GPT)~\cite{gpt}, Lyncs~\cite{Bacchio:2022bjk} or PyQUDA~\cite{PyQUDA}, which provide Python interfaces to Grid and QUDA, respectively, can instead be used by less experienced students or for exploratory work.

\begin{tcolorbox}[colback=green!5!white,colframe=green!75!black]
    \textbf{Observation 16:} Optimised performance-portable libraries such as Grid and QUDA give access to highly efficient implementations of various kernels and algorithms and should be used, if possible, to benefit from the many years of development and performance-engineering invested in them. They can be combined with programmer-productivity interfaces like Grid Python Toolkit, Lyncs or PyQUDA to make Grid and QUDA accessible to those without the necessary background to use them directly.
\end{tcolorbox}

A calculation in \ac{LFT} usually proceeds through two stages requiring \ac{HPC} resources.
Generally, the first stage consists in the generation of sets of ensembles of representative field configurations sampled from a probability distribution given by the action of the underlying theory using \ac{MCMC} methods, usually variants of the \ac{HMC} algorithm~\cite{Duane:1987de}.
By their nature, the samples in a Markov chain have a sequential dependency and the ensemble generation stage in \ac{LFT} is thus a capability-class computational problem run at the edge of the strong scaling window in order to minimise the real time required for generating independent samples.
Energy efficiency improvements at this stage thus correspond to either significantly improving or even replacing the \ac{HMC} as a sampling algorithm or improving the efficiency of the various kernels and iterative algorithms that are used in such a simulation.  

At the second stage the path integral for observables of interest is evaluated using the representative field configurations generated in the first stage.
Most observables in \ac{LFT} are $n$-point correlation functions, usually in the time-momentum representation, between operators carrying different quantum numbers as relevant for the problem in question.
These give access to the masses and other properties of hadrons as well as matrix elements of operators of interest between these hadronic states.
The evaluation of these correlation functions involves the computation of fermion propagators and their combination with projection tensors through so-called \emph{Wick contractions} as well as the injection of momentum through Fourier transforms.
These calculations straddle the boundary between capability and capacity problem classes and are generally run on as few compute nodes as possible based on their memory footprint.


Generally speaking both stages are dominated by the repeated solution of linear systems of the type $Dx = b$, where $D$ is a very large, very regular and very sparse and potentially very poorly-conditioned complex-valued matrix called the \emph{Dirac} operator.
In state-of-the-art calculations with $128^3\cdot256$ lattice points, the vector $x$ represents close to $10^{10}$ unknowns with near-term future calculations set to increase this by at least one order of magnitude.
\revision{Depending on the type of fermion discretisation, the application of $D$ can have an arithmetic intensity as low as $0.3$ in double precision and is thus strongly memory-bandwidth-bound.}

\begin{figure}
    \includegraphics[width=\linewidth]{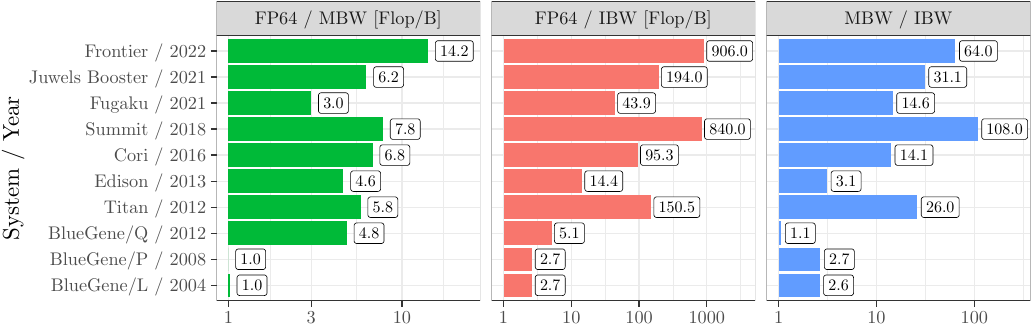}
    \caption{Per-node ratios of peak \ac{FP64}, peak memory bandwidth (MBW) and peak bidirectional interconnect bandwidth (IBW) for representative \ac{HPC} systems used for calculations in lattice field theory. Data from~\cite{boyle2022latticeqcdcomputationalfrontier}, \revision{and available in~\cite{plots_data_zenodo}}.}
    \label{fig:perf_ratios}
\end{figure}

The development of \ac{HPC} systems since the beginning of the millennium has been marked by a substantial increase in the available peak \ac{FP64}, a commensurate but slower increase in peak memory bandwidth, and a very slow increase in peak interconnect bandwidth~\cite{McCalpin2022, Khan2021}.
This is illustrated in \cref{fig:perf_ratios} which shows per-node ratios of these three metrics for a selection of representative systems employed for \ac{LFT} calculations (data from~\cite{boyle2022latticeqcdcomputationalfrontier}).
The first two iterations of the BlueGene line of supercomputers, for example, had memory and interconnect bandwidths that were well-balanced for sparse stencil type problems, such that \ac{LQCD} kernels could achieve in excess of 50\% of the peak \ac{FP64} rate even when running on many nodes.
By contrast, systems which rely on \acp{GPU} for their raw computational power can only exploit this potential for problems with high arithmetic intensity such as dense matrix multiplication.

\begin{figure}
  \includegraphics[width=0.5\linewidth]{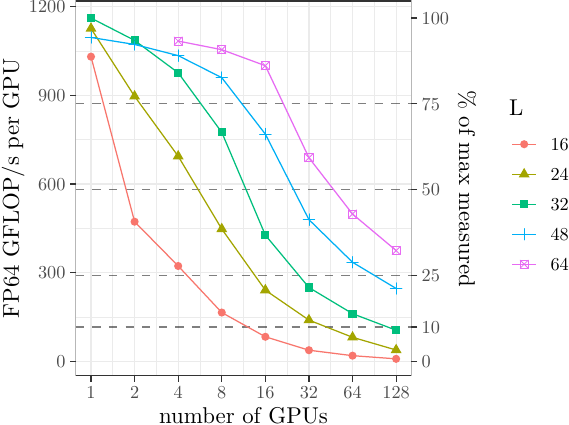}
  \caption{Strong scaling study of the double-precision Wilson-clover twisted mass Dirac operator as implemented in the QUDA~\cite{Clark:2009wm,Babich:2011np,Clark:2016rdz} library for different lattice volumes $V = 2 \cdot L^4$ on JUWELS Booster~\cite{Alvarez2021}. The points indicate the performance per \ac{GPU} in GFLOP/s (higher is better) for the different cases with lines added to guide the eye. The right-hand axis indicates the performance relative to the fastest case in percent. The $L=64$ problem size does not fit on less than 4~\acp{GPU}. NVIDIA specifies a peak \ac{FP64} rate of 9.7~TFLOP/s for the A100 \ac{GPU}. \revision{Data available in~\cite{plots_data_zenodo}}.}
  \label{fig:dslash_strong}
\end{figure}

Even more challenging is the fact that interconnect bandwidth has not increased to the same extent as memory bandwidth and floating point performance as it is easily the most significant bottleneck to scalability on current machines.
This is somewhat offset by the fact that in order to efficiently use the \acp{GPU} on these systems, per-node computational volumes need to be rather large, resulting in surface-to-volume ratios that allow at least for some overlap of communication and computation.
Even so, these imbalances negatively affect efficiency such that, on a machine like JUWELS Booster~\citep{Alvarez2021}, highly optimised kernels only achieve around 10\% of the peak FP64 rate on a single node, dropping quickly as the problem is strong-scaled. 
This is shown in \cref{fig:dslash_strong} for a double-precision Wilson-clover twisted mass Dirac operator using different lattice volumes $V = 2 \cdot L^4$ as implemented in the QUDA~\cite{Clark:2009wm,Babich:2011np,Clark:2016rdz} library.
As a result, even though full system peak performance has increased by about a factor of 100 between BlueGene/L and JUWELS Booster, for example, the effective improvement for \ac{LFT} calculations has been quite a bit lower.

The moderate growth in the computational power effectively available to \ac{LFT} calculations has been accompanied by the development of highly algorithmically efficient solvers based on the adaptive \ac{MG} preconditioning of flexible algorithms such as \ac{FGMRES} or \ac{GCR}, at least for a subset of lattice formulations employing Wilson and Wilson-clover~\cite{Brannick:2007ue,Frommer:2013fsa} or twisted mass~\cite{Alexandrou:2016izb} fermions.
By design, \ac{MG} algorithms exhibit a low degree of data parallelism at the coarsest level of the aggregation hierarchy. 
As a result, the efficient implementation of these algorithms on \acp{GPU} is only possible through fine-grained optimisations~\cite{Clark:2016rdz} of the underlying computational kernels.
Once optimised in this way, the coarsest-grid kernels become dominated by a combination of communication \ac{API}  and network latency, as a result of which the fastest implementations have moved away from \ac{MPI} and instead make use of proprietary solutions such as NVSHMEM for significantly improved strong-scaling.

\begin{figure}[ht]
    \includegraphics[width=0.5\linewidth]{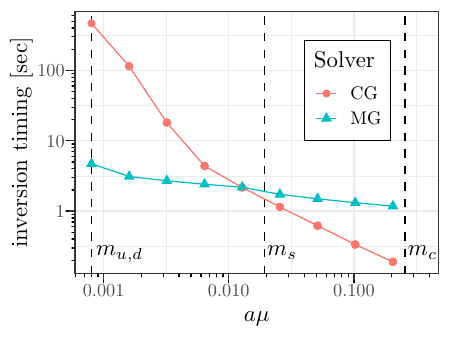}
    \caption{Comparison of time-to-solution (lower is better) for the inversion of the Wilson-clover twisted mass Dirac operator in the calculation of the fermion derivative on a $64^3\cdot128$ lattice on 8~JUWELS~Booster~\cite{Alvarez2021} nodes as a function of the quark mass, $\mu$ using the fastest mixed-precision conjugate gradient algorithm available performing a single solve (red)\revision{,} and multigrid-preconditioned \ac{GCR} performing two solves and including unavoidable overheads (blue). The dashed vertical lines indicate the physical bare average up/down (light), strange and charm quark masses in lattice units. Both algorithms are implemented in the QUDA library~\cite{Clark:2009wm,Babich:2011np,Clark:2016rdz}. \revision{Data available in~\cite{plots_data_zenodo}}}
    \label{fig:mg_vs_cg}
\end{figure}

The efficiency of \ac{MG} algorithms is demonstrated in \cref{fig:mg_vs_cg}, which shows the time required to invert the Wilson-clover twisted mass Dirac operator on a $64^3\cdot128$ lattice on 8~JUWELS Booster nodes in the calculation of the fermion derivative in the \ac{HMC} as a function of the quark mass.
The comparison is between two solvers implemented in the QUDA library~\cite{Clark:2009wm,Babich:2011np,Clark:2016rdz}, the fastest available mixed-precision \ac{CG} performing a single inversion and \ac{MG}-preconditioned \ac{GCR} doing two inversions due to the nature of the problem in question.
Even though the MG solver is running with too small a local computational volume, 
it outperforms \ac{CG} by around a factor of 100 at the physical light quark mass.

Because it is unlikely that the computational imbalance of \ac{HPC} systems will improve, it is important for the \ac{LFT} community to invest in a broad spectrum of algorithmic research, even premising that the situation will worsen further.
This implies that the target should be to increase arithmetic intensity and reduce interconnect requirements whenever possible.
Beyond making use of mixed and adaptive fixed-bit-width~\cite{Clark:2023xnn} representations, arithmetic intensity can also be increased by modifying solvers to operate on \emph{multiple right-hand-sides} where possible~\cite{boyle2022latticeqcdcomputationalfrontier,Boyle:2024pio}, also enabling the usage of vendor library routines for efficient batched matrix-multiplication in certain cases.
Techniques for communication-avoidance can be applied both at the level of solvers~\cite{ESPINOZAVALVERDE2023108869,PhysRevD.108.034503} as well as in sampling algorithms, for instance through domain-decomposition~\cite{Boyle:2022pai}.

At the level of sampling algorithms, the issue of critical slowing down~\cite{Schaefer:2009xx} requires further urgent research to reduce the cost of generating independent samples at fine lattice resolutions.
Promising directions of study include flow-based sampling~\cite{Kanwar:2024ujc}, approaches to Fourier acceleration~\cite{Jung:2024nuv,Abbott:2022zhs,Finkenrath:2022ogg} and parallel tempering~\cite{Bonanno:2024zyn}.
As a complement to improved sampling, variance-reduction techniques for particularly noisy observables, most notably those with quark-line disconnected contributions, can also be effective for improving the overall energy efficiency of calculations in \ac{LFT}.
A recent example is multigrid multilevel Monte Carlo~\cite{frommer2022multilevel,Whyte:2022vrk}.

In the calculation of correlation functions, if the number of particles in the considered state is larger than two, if the number of required momentum combinations is large, or if the number of insertion points, $n$, grows beyond three, overall computational cost may be dominated by tensor contractions rather than linear system solves.
Sub-expression elimination and the caching of common intermediate results can thus lead to significant efficiency improvements compared to the na{\"i}ve sequential calculation of a large collection of such correlation functions.
It has been shown~\cite{Chen:2023zyy} that \revision{very good use can be made of accelerators} once memory management~\cite{10.1145/3506705}, as well as dependency management and scheduling~\cite{9820666} have been addressed, and once the many small-matrix-multiplications are streamed in batches~\cite{10.1007/978-3-319-41321-1_2} to profit from \ac{GPU} stream parallelism.

In conclusion it can be said that in \ac{LFT} in general and \ac{LQCD} in particular energy efficiency is almost synonymous with performance optimisation and algorithmic research.
In light of current hardware trends, this implies on the one hand that performance-portability and dedicated performance-engineering are central pillars \revision{that} allow to maximise machine utilisation and therefore energy efficiency.
On the other hand, the problem of critical slowing down and the expected further growth of the imbalance between computational power and memory, as well as interconnect bandwidth, necessitate a broad spectrum of algorithmic study aimed at improved sampling algorithms, variance reduction techniques, \revision{and} communication-avoiding linear solvers.


\begin{tcolorbox}[colback=green!5!white,colframe=green!75!black]
    \textbf{Observation 17:} Continued algorithmic research is essential to tackle variance reduction for noisy observables as well as the issue of critical slowing down through improved or accelerated sampling. In addition, the growing imbalance between computational power and memory as well as interconnect bandwidth must be addressed through communication-avoidance approaches.    
\end{tcolorbox}

\section{Conclusions}
\label{sec:concl}

\ac{HPC} systems are amongst the scientific instruments with the highest energy consumption. To improve their energy efficiency, a combined effort from technology providers, system operators, software developers, and users is required. In particular, application developers and users need to be conscious of energy efficiency from two perspectives: \emph{(i)} they need to know that energy optimisation techniques applied at hardware and system management levels do have an impact on how they experience \ac{HPC} systems, and \emph{(ii)} they need to learn how to optimise their codes and execution scripts to minimise the energy consumed by their own jobs. This is especially important in the fields of astrophysics, high-energy physics, and lattice field theory, which consume large volumes of compute hours on \ac{HPC} resources.

From the system design and operations, the growing hardware heterogeneity (\acp{CPU}, \acp{GPU} and other accelerators) observed in the \revision{recent} years is a direct consequence of trying to improve performance per Watt on \ac{HPC} systems. Heterogeneous computers are  more difficult to program, which hinders application portability across systems. High-level programming models, frameworks, and domain-specific languages can help to alleviate the burden, but users should enploy the correct backends and libraries on each platform. Data management is also a main concern for energy efficiency and \ac{HPC} users should strive to keep their data as local as possible, hide communication behind computation tasks, and employ I/O optimised libraries to optimally use the file system. \ac{HPC} users must also be aware that they may easily experience performance \revision{variability even on the same HPC machine} across equal runs, since system management automatically adjusts voltage and frequency of compute devices during operations, and may even reallocate jobs to maximise system throughput for a given energy. Therefore, understanding performance metrics requires consultation of monitoring information and direct contact with \ac{HPC} user support teams. 

What is solely in the hands of the application developers and users is making the application codes themselves more energy efficient, and running them in optimal configurations. \revision{Based on our experience, we assume in this paper that nowadays most application developers start their work building on an existing code, which is already parallelised\footnote{\revision{If the start point would be a serial code that the developer is parallelising, it is important to conduct regression tests to guarantee consistency between the results of serial and parallel versions.}}. Their task is then to either add some new aspect to the code, making the simulation more realistic, or improving the performance and scaling of the existing implementation.} This means first analysing the code with performance analysis tools, detecting inefficiencies, and \revision{making} the necessary modifications to maximise the single-node performance (considering also using reduced or mixed arithmetic precision), to then improve its scaling when running in parallel on a growing number of nodes. Once this is done, it should be explored whether the application is tolerant to reductions \revision{in the} operational frequency of the hardware. Furthermore, application configuration and job settings should always be verified before starting long production runs. The most efficient configuration is likely to be different between \ac{HPC} systems with diverse architectures, and one cannot rely on pre-tuned configurations. Energy is also saved by ensuring long-term sustainability and availability of any code development, since this avoids iterating past efforts and errors over and over again. Professional software development strategies, including version control, automated testing, \ac{CI/CD} pipelines, and applying \ac{FAIR} principles for code and data management are therefore crucial.

It must be stressed that the most important asset in any field is its people. The long-term future of \ac{HPC} depends on the ability of the community to attract, train and keep the engagement of talents from the younger generations. \revision{Therefore, it must be invested} in training future \ac{HPC} experts and providing them with the necessary support and a tolerant and open environment that enable minority groups to enter the field and motivate all to stay. Training courses focused on algorithms, performance analysis and optimisations, data access patterns and frameworks must exist for each user community. It is also crucial to establish career paths for \ac{HPC} experts, in particular for research software engineers (\ac{RSE}), who are necessary to \revision{optimise code}, both from the energy and performance perspective. Unfortunately, the traditional view in the domain sciences determines the academic merits of its members based only on their publication record \revision{in} the highest impact journals from their fields. \revision{This must change}, so that application software releases and publications about code improvements are equally valued, recognising their crucial contributions to the advancement of science.

\section*{Conflict of Interest Statement}

The authors declare that the research was conducted in the absence of any commercial or financial relationships that could be construed as a potential conflict of interest.

\section*{Author Contributions}

This paper is the result of a collaborative effort between experts in \ac{HPC}, Astrophysics, and \ac{HEP}. 
ES has developed the concept, coordinated the overall paper effort, and reviewed the content\revision{. ES} is the main author of \cref{sec:intro} and \cref{sec:concl}. \cref{sec:hpc} has been jointly written by ES (main author of subsection~\ref{sec:hpc:hw}), MF (main author of subsections~\ref{sec:hpc:sw} and \ref{sec:hpc:prog}), and MG (main author of subsection~\ref{sec:hpc:data}. SP and JA are the authors of \cref{sec:astro} and are responsible for subsections~\ref{sec:astro:stellar} and \ref{sec:astro:space}, respectively. MG and OF have jointly written~\cref{sec:hep} as the main authors of \revision{its introduction and subsection} \ref{sec:hep:collider}, respectively. BK has written \cref{sec:lqcd} and contributed to subsections~\ref{sec:hpc:prog} and \ref{sec:hpc:sw}. All authors have reviewed the final manuscript before submission.

\section*{Funding}
\revision{This work was funded in part by the Deutsche Forschungsgemeinschaft (DFG, German Research Foundation) as part of the CRC 1639/1 NuMeriQS – 511713970. SP acknowledges funding by the DFG as part of the PUNCH4NFDI project (project number 460248186) and by the project ‘‘NRW-Cluster for data intensive radio astronomy: Big Bang to Big Data (B3D)’’ funded through the programme ‘‘Profilbildung 2020’’, an initiative of the Ministry of Culture and Science of the State of North Rhine-Westphalia.}

\section*{Acknowledgments}
\revision{The authors thank Florian Janetzko (JSC) for providing access to the user statistics represented in \cref{fig:usestats}.}

\section*{Supplemental Data}

\section*{Data Availability Statement}
\revision{The datasets for \cref{fig:usestats,fig:perf_ratios,fig:dslash_strong,fig:mg_vs_cg} are available via Zenodo~\cite{plots_data_zenodo}.}


\bibliographystyle{Frontiers-Harvard}
\bibliography{ref}

\begin{thebibliography}{210}
\providecommand{\natexlab}[1]{#1}
\expandafter\ifx\csname urlstyle\endcsname\relax
  \providecommand{\doi}[1]{doi:\discretionary{}{}{}#1}\else
  \providecommand{\doi}{doi:\discretionary{}{}{}\begingroup
  \urlstyle{rm}\Url}\fi
\providecommand{\selectlanguage}[1]{\relax}
\providecommand{\bibAnnoteFile}[1]{%
  \IfFileExists{#1}{\begin{quotation}\noindent\textsc{Key:} #1\\
  \textsc{Annotation:}\ \input{#1}\end{quotation}}{}}
\providecommand{\bibAnnote}[2]{%
  \begin{quotation}\noindent\textsc{Key:} #1\\
  \textsc{Annotation:}\ #2\end{quotation}}

\bibitem[{Aad et~al.(2012)}]{ATLAS:2012yve}
Aad, G. et~al. (2012).
\newblock {Observation of a new particle in the search for the Standard Model
  Higgs boson with the ATLAS detector at the LHC}.
\newblock \emph{Phys. Lett. B} 716, 1--29.
\newblock
  \href{https://www.sciencedirect.com/science/article/pii/S037026931200857X?via%3Dihub}{DOI:
  10.1016/j.physletb.2012.08.020}
\bibAnnoteFile{ATLAS:2012yve}

\bibitem[{Abbott et~al.(2022)}]{Abbott:2022zhs}
Abbott, R. et~al. (2022).
\newblock {Gauge-equivariant flow models for sampling in lattice field theories
  with pseudofermions}.
\newblock \emph{Phys. Rev. D} 106, 074506.
\newblock \href{https://doi.org/10.1103/PhysRevD.106.074506}{DOI:
  10.1103/PhysRevD.106.074506}
\bibAnnoteFile{Abbott:2022zhs}

\bibitem[{Abdel-Rehim et~al.(2015)}]{Abdel-Rehim:2015owa}
Abdel-Rehim, A. et~al. (2015).
\newblock {Nucleon and pion structure with lattice QCD simulations at physical
  value of the pion mass}.
\newblock \emph{Phys. Rev. D} 92, 114513.
\newblock \href{https://doi.org/10.1103/PhysRevD.92.114513}{DOI:
  10.1103/PhysRevD.92.114513}
\bibAnnoteFile{Abdel-Rehim:2015owa}

\bibitem[{Abdel-Rehim et~al.(2017)}]{ETM:2015ned}
Abdel-Rehim, A. et~al. (2017).
\newblock {First physics results at the physical pion mass from $N_f=2$ Wilson
  twisted mass fermions at maximal twist}.
\newblock \emph{Phys. Rev. D} 95, 094515.
\newblock \href{https://doi.org/10.1103/PhysRevD.95.094515}{DOI:
  10.1103/PhysRevD.95.094515}
\bibAnnoteFile{ETM:2015ned}

\bibitem[{Abdelfattah et~al.(2016)Abdelfattah, Haidar, Tomov, and
  Dongarra}]{10.1007/978-3-319-41321-1_2}
Abdelfattah, A., Haidar, A., Tomov, S., and Dongarra, J. (2016).
\newblock Performance, design, and autotuning of batched {GEMM} for {GPUs}.
\newblock In \emph{High Performance Computing}, eds. J.~M. Kunkel, P.~Balaji,
  and J.~Dongarra (Springer International Publishing), 21--38.
\newblock
  \href{https://link.springer.com/chapter/10.1007/978-3-319-41321-1\_2}{DOI:
  10.1007/978-3-319-41321-1\_2}
\bibAnnoteFile{10.1007/978-3-319-41321-1_2}

\bibitem[{Alexandrou et~al.(2016)Alexandrou, Bacchio, Finkenrath, Frommer,
  Kahl, and Rottmann}]{Alexandrou:2016izb}
Alexandrou, C., Bacchio, S., Finkenrath, J., Frommer, A., Kahl, K., and
  Rottmann, M. (2016).
\newblock {Adaptive Aggregation-based Domain Decomposition Multigrid for
  Twisted Mass Fermions}.
\newblock \emph{Phys. Rev. D} 94, 114509.
\newblock \href{https://doi.org/10.1103/PhysRevD.94.114509}{DOI:
  10.1103/PhysRevD.94.114509}
\bibAnnoteFile{Alexandrou:2016izb}

\bibitem[{Alvarez(2021)}]{Alvarez2021}
Alvarez, D. (2021).
\newblock {JUWELS} {C}luster and {B}ooster: {E}xascale {P}athfinder with
  {M}odular {S}upercomputing {A}rchitecture at {J}uelich {S}upercomputing
  {C}entre.
\newblock \emph{Journal of large-scale research facilities}
  \href{https://jlsrf.org/index.php/lsf/article/view/183}{DOI:
  10.17815/jlsrf-7-183}
\bibAnnoteFile{Alvarez2021}

\bibitem[{Alwall et~al.(2007)Alwall, Ballestrero, Bartalini, Belov, Boos,
  Buckley et~al.}]{Alwall2007}
Alwall, J., Ballestrero, A., Bartalini, P., Belov, S., Boos, E., Buckley, A.,
  et~al. (2007).
\newblock A standard format for les houches event files.
\newblock \emph{Computer Physics Communications} 176, 300–304.
\newblock \href{http://dx.doi.org/10.1016/j.cpc.2006.11.010}{DOI:
  10.1016/j.cpc.2006.11.010}
\bibAnnoteFile{Alwall2007}

\bibitem[{{AMD}(2024{\natexlab{a}})}]{mi300A}
[Dataset] {AMD} (2024{\natexlab{a}}).
\newblock {AMD CDNA 3 architecture}.
\newblock
  \url{https://www.amd.com/content/dam/amd/en/documents/instinct-tech-docs/white-papers/amd-cdna-3-white-paper.pdf}.
\newblock Accessed: 2024-12-08
\bibAnnoteFile{mi300A}

\bibitem[{{AMD}(2024{\natexlab{b}})}]{hip}
[Dataset] {AMD} (2024{\natexlab{b}}).
\newblock {HIP}.
\newblock \url{https://github.com/ROCm/HIP}.
\newblock Accessed: 2024-10-12
\bibAnnoteFile{hip}

\bibitem[{Arrington et~al.(2023)}]{JeffersonLabSoLID:2022iod}
Arrington, J. et~al. (2023).
\newblock {The solenoidal large intensity device (SoLID) for JLab 12 GeV}.
\newblock \emph{J. Phys. G} 50, 110501.
\newblock \href{http://dx.doi.org/10.1088/1361-6471/acda21}{DOI:
  10.1088/1361-6471/acda21}
\bibAnnoteFile{JeffersonLabSoLID:2022iod}

\bibitem[{{ASCL at Michigan Technological University}(2025)}]{ascl}
[Dataset] {ASCL at Michigan Technological University} (2025).
\newblock {Astrophysics Source Code Library}.
\newblock \url{https://ascl.net}.
\newblock Accessed: 2025-01-28
\bibAnnoteFile{ascl}

\bibitem[{{ATLAS Collaboration}(2019)}]{atlas_collaboration_2019_2641997}
{ATLAS Collaboration} (2019).
\newblock Athena \href{https://doi.org/10.5281/zenodo.2641997}{DOI:
  10.5281/zenodo.2641997}
\bibAnnoteFile{atlas_collaboration_2019_2641997}

\bibitem[{Auweter et~al.(2014)Auweter, Bode, Brehm, Brochard, Hammer, Huber
  et~al.}]{Auweter14}
Auweter, A., Bode, A., Brehm, M., Brochard, L., Hammer, N., Huber, H., et~al.
  (2014).
\newblock A case study of energy aware scheduling on supermuc.
\newblock In \emph{Supercomputing}, eds. J.~M. Kunkel, T.~Ludwig, and H.~W.
  Meuer (Springer International Publishing), 394--409.
\newblock \href{https://doi.org/10.1007/978-3-319-07518-1\_25}{DOI:
  10.1007/978-3-319-07518-1\_25}
\bibAnnoteFile{Auweter14}

\bibitem[{Babich et~al.(2011)Babich, Clark, Joo, Shi, Brower, and
  Gottlieb}]{Babich:2011np}
Babich, R., Clark, M.~A., Joo, B., Shi, G., Brower, R.~C., and Gottlieb, S.
  (2011).
\newblock {Scaling Lattice QCD beyond 100 GPUs}.
\newblock In \emph{{SC11 International Conference for High Performance
  Computing, Networking, Storage and Analysis Seattle, Washington, November
  12-18, 2011}}.
\newblock \doi{10.1145/2063384.2063478}.
\newblock \href{https://doi.org/10.1145/2063384.2063478}{DOI:
  10.1145/2063384.2063478}
\bibAnnoteFile{Babich:2011np}

\bibitem[{Bacchini(2023)}]{Bacchini2023-sl}
Bacchini, F. (2023).
\newblock {RelSIM}: A relativistic semi-implicit method for particle-in-cell
  simulations.
\newblock \emph{Astrophys. J. Suppl. Ser.} 268, 60.
\newblock \href{https://doi.org/10.3847/1538-4365/acefba}{DOI:
  10.3847/1538-4365/acefba}
\bibAnnoteFile{Bacchini2023-sl}

\bibitem[{Bacchio et~al.(2022)Bacchio, Finkenrath, and
  Stylianou}]{Bacchio:2022bjk}
Bacchio, S., Finkenrath, J., and Stylianou, C. (2022).
\newblock {Lyncs-API: a Python API for Lattice QCD applications}.
\newblock \emph{PoS} LATTICE2021, 542.
\newblock \href{https://doi.org/10.22323/1.396.0542}{DOI: 10.22323/1.396.0542}
\bibAnnoteFile{Bacchio:2022bjk}

\bibitem[{Barros et~al.(2008)Barros, Babich, Brower, Clark, and
  Rebbi}]{Barros:2008rd}
Barros, K., Babich, R., Brower, R., Clark, M.~A., and Rebbi, C. (2008).
\newblock {Blasting through lattice calculations using CUDA}.
\newblock \emph{PoS} LATTICE2008, 045.
\newblock \href{http://dx.doi.org/10.22323/1.066.0045}{DOI:
  10.22323/1.066.0045}
\bibAnnoteFile{Barros:2008rd}

\bibitem[{{Bate} et~al.(2003){Bate}, {Bonnell}, and {Bromm}}]{Bate:2003}
{Bate}, M.~R., {Bonnell}, I.~A., and {Bromm}, V. (2003).
\newblock {The formation of a star cluster: predicting the properties of stars
  and brown dwarfs}.
\newblock \emph{MNRAS} 339, 577--599.
\newblock \href{https://academic.oup.com/mnras/article/339/3/577/970800}{DOI:
  10.1046/j.1365-8711.2003.06210.x}
\bibAnnoteFile{Bate:2003}

\bibitem[{Bauer et~al.(2012)Bauer, Treichler, Slaughter, and Aiken}]{legion}
Bauer, M., Treichler, S., Slaughter, E., and Aiken, A. (2012).
\newblock Legion: Expressing locality and independence with logical regions.
\newblock In \emph{SC'12: Proceedings of the International Conference on High
  Performance Computing, Networking, Storage and Analysis} (IEEE), 1--11.
\newblock \href{https://doi.org/10.1109/SC.2012.71}{DOI: 10.1109/SC.2012.71}
\bibAnnoteFile{legion}

\bibitem[{Bazavov et~al.(2013)}]{MILC:2013ffd}
Bazavov, A. et~al. (2013).
\newblock {Leptonic Decay-Constant Ratio $f_{K^+}/f_{\pi^+}$ from Lattice QCD
  with Physical Light Quarks}.
\newblock \emph{Phys. Rev. Lett.} 110, 172003.
\newblock \href{https://doi.org/10.1103/PhysRevLett.110.172003}{DOI:
  10.1103/PhysRevLett.110.172003}
\bibAnnoteFile{MILC:2013ffd}

\bibitem[{Beckingsale et~al.(2019)Beckingsale, Burmark, Hornung, Jones,
  Killian, Kunen et~al.}]{RAJA}
Beckingsale, D.~A., Burmark, J., Hornung, R., Jones, H., Killian, W., Kunen,
  A.~J., et~al. (2019).
\newblock Raja: Portable performance for large-scale scientific applications.
\newblock In \emph{2019 IEEE/ACM International Workshop on Performance,
  Portability and Productivity in HPC (P3HPC)}. 71--81.
\newblock \href{https://ieeexplore.ieee.org/document/8945721}{DOIL
  10.1109/P3HPC49587.2019.00012}
\bibAnnoteFile{RAJA}

\bibitem[{{Bedorf} and {Portegies Zwart}(2020)}]{Bedorf:2020}
{Bedorf}, J. and {Portegies Zwart}, S. (2020).
\newblock {Bonsai-SPH: A GPU accelerated astrophysical Smoothed Particle
  Hydrodynamics code}.
\newblock \emph{SciPost Astronomy} 1, 001.
\newblock \href{https://scipost.org/SciPostAstro.1.1.001}{DOI:
  10.21468/SciPostAstro.1.1.001}
\bibAnnoteFile{Bedorf:2020}

\bibitem[{Ben-Nun et~al.(2019)Ben-Nun, de~Fine~Licht, Ziogas, Schneider, and
  Hoefler}]{DaCe}
Ben-Nun, T., de~Fine~Licht, J., Ziogas, A.~N., Schneider, T., and Hoefler, T.
  (2019).
\newblock Stateful dataflow multigraphs: a data-centric model for performance
  portability on heterogeneous architectures.
\newblock In \emph{Proceedings of the International Conference for High
  Performance Computing, Networking, Storage and Analysis} (New York, NY, USA:
  Association for Computing Machinery), SC '19.
\newblock \href{https://doi.org/10.1145/3295500.3356173}{DOI:
  10.1145/3295500.3356173}
\bibAnnoteFile{DaCe}

\bibitem[{Bertacchi(2023)}]{BERTACCHI2023107}
Bertacchi, V. (2023).
\newblock Recent results from belle ii.
\newblock \emph{Nuclear and Particle Physics Proceedings} 324-329, 107--112.
\newblock \href{https://doi.org/10.1016/j.nuclphysbps.2023.01.022}{DOI:
  10.1016/j.nuclphysbps.2023.01.022}
\bibAnnoteFile{BERTACCHI2023107}

\bibitem[{Bhalachandra et~al.(2017)Bhalachandra, Porterfield, Olivier, Prins,
  and Fowler}]{Bhalachandra2017}
Bhalachandra, S., Porterfield, A., Olivier, S.~L., Prins, J.~F., and Fowler,
  R.~J. (2017).
\newblock Improving energy efficiency in memory-constrained applications using
  core-specific power control.
\newblock In \emph{Proceedings of the 5th International Workshop on Energy
  Efficient Supercomputing} (Association for Computing Machinery), E2SC'17.
\newblock \href{https://doi.org/10.1145/3149412.3149418}{DOI:
  10.1145/3149412.3149418}
\bibAnnoteFile{Bhalachandra2017}

\bibitem[{Bhalachandra et~al.(2015)Bhalachandra, Porterfield, and
  Prins}]{Bhalachandra2015}
Bhalachandra, S., Porterfield, A., and Prins, J.~F. (2015).
\newblock Using dynamic duty cycle modulation to improve energy efficiency in
  high performance computing.
\newblock In \emph{Proceedings of the 2015 IEEE International Parallel and
  Distributed Processing Symposium Workshop} (USA: IEEE Computer Society),
  {IPDPSW '15}, 911–918.
\newblock \href{https://doi.org/10.1109/IPDPSW.2015.144}{DOI:
  10.1109/IPDPSW.2015.144}
\bibAnnoteFile{Bhalachandra2015}

\bibitem[{Boccali(2019)}]{Boccali2019}
Boccali, T. (2019).
\newblock Computing models in high energy physics.
\newblock \emph{Reviews in Physics} 4, 100034.
\newblock \href{http://dx.doi.org/10.1016/j.revip.2019.100034}{DOI:
  10.1016/j.revip.2019.100034}
\bibAnnoteFile{Boccali2019}

\bibitem[{Bodin et~al.(2004)}]{Bodin:2004brr}
Bodin, F. et~al. (2004).
\newblock {apeNEXT: A Multi-TFlops computer for elementary particle physics}.
\newblock \emph{Adv. Parallel Comput.} 13, 355--362.
\newblock
  \href{https://www.sciencedirect.com/science/article/abs/pii/S0927545204800479}{DOI:
  10.1016/S0927-5452(04)80047-9}
\bibAnnoteFile{Bodin:2004brr}

\bibitem[{Bonanno et~al.(2024)Bonanno, Clemente, D'Elia, Maio, and
  Parente}]{Bonanno:2024zyn}
Bonanno, C., Clemente, G., D'Elia, M., Maio, L., and Parente, L. (2024).
\newblock {Full QCD with milder topological freezing}.
\newblock \emph{JHEP} 08, 236.
\newblock
  \href{https://doi.org/10.1007/JHEP08(2024)236}{10.1007/JHEP08(2024)236}
\bibAnnoteFile{Bonanno:2024zyn}

\bibitem[{Bosilca et~al.(2013)Bosilca, Bouteiller, Danalis, Faverge,
  H{\'e}rault, and Dongarra}]{parsec}
Bosilca, G., Bouteiller, A., Danalis, A., Faverge, M., H{\'e}rault, T., and
  Dongarra, J.~J. (2013).
\newblock Parsec: Exploiting heterogeneity to enhance scalability.
\newblock \emph{Computing in Science \& Engineering} 15, 36--45.
\newblock \href{https://doi.org/10.1109/MCSE.2013.98}{DOI:
  10.1109/MCSE.2013.98}
\bibAnnoteFile{parsec}

\bibitem[{Boyle et~al.(2022{\natexlab{a}})Boyle, Bollweg, Brower, Christ,
  DeTar, Edwards et~al.}]{boyle2022latticeqcdcomputationalfrontier}
Boyle, P., Bollweg, D., Brower, R., Christ, N., DeTar, C., Edwards, R., et~al.
  (2022{\natexlab{a}}).
\newblock Lattice qcd and the computational frontier
  \href{https://arxiv.org/abs/2204.00039}{arXiv:2204.00039}
\bibAnnoteFile{boyle2022latticeqcdcomputationalfrontier}

\bibitem[{Boyle(2024)}]{Boyle:2024pio}
Boyle, P.~A. (2024).
\newblock {Multiple right hand side multigrid for domain wall fermions with a
  multigrid preconditioned block conjugate gradient algorithm}
  \href{https://arxiv.org/abs/2409.03904}{arXiv:2409.03904}
\bibAnnoteFile{Boyle:2024pio}

\bibitem[{Boyle et~al.(2022{\natexlab{b}})Boyle, Bollweg, Kelly, and
  Yamaguchi}]{Boyle:2022pai}
Boyle, P.~A., Bollweg, D., Kelly, C., and Yamaguchi, A. (2022{\natexlab{b}}).
\newblock {Algorithms for domain wall Fermions}.
\newblock \emph{PoS} LATTICE2021, 470.
\newblock \href{http://dx.doi.org/10.22323/1.396.0470}{DOI:
  10.22323/1.396.0470}
\bibAnnoteFile{Boyle:2022pai}

\bibitem[{Boyle et~al.(2016)Boyle, Cossu, Yamaguchi, and
  Portelli}]{Boyle:2016lbp}
Boyle, P.~A., Cossu, G., Yamaguchi, A., and Portelli, A. (2016).
\newblock {Grid: A next generation data parallel C++ QCD library}.
\newblock \emph{PoS} LATTICE2015, 023.
\newblock \href{http://dx.doi.org/10.22323/1.251.0023}{DOI:
  10.22323/1.251.0023}
\bibAnnoteFile{Boyle:2016lbp}

\bibitem[{Brannick et~al.(2008)Brannick, Brower, Clark, Osborn, and
  Rebbi}]{Brannick:2007ue}
Brannick, J., Brower, R.~C., Clark, M.~A., Osborn, J.~C., and Rebbi, C. (2008).
\newblock {Adaptive Multigrid Algorithm for Lattice QCD}.
\newblock \emph{Phys. Rev. Lett.} 100, 041601.
\newblock \href{https://doi.org/10.1103/PhysRevLett.100.041601}{DOI::
  10.1103/PhysRevLett.100.041601}
\bibAnnoteFile{Brannick:2007ue}

\bibitem[{Brun et~al.(2020)Brun, Rademakers, Canal, Naumann, Couet, Moneta
  et~al.}]{rene_brun_2020_3895860}
Brun, R., Rademakers, F., Canal, P., Naumann, A., Couet, O., Moneta, L., et~al.
  (2020).
\newblock root-project/root: v6.18/02
  \href{https://doi.org/10.5281/zenodo.3895860}{DOI: 10.5281/zenodo.3895860}
\bibAnnoteFile{rene_brun_2020_3895860}

\bibitem[{{BSC}(2019)}]{dimemas}
[Dataset] {BSC} (2019).
\newblock {Dimemas: predict parallel performance using a single {CPU} machine}.
\newblock \url{https://tools.bsc.es/dimemas}.
\newblock Accessed: 2024-12-08
\bibAnnoteFile{dimemas}

\bibitem[{{BSC}(2024)}]{paraver}
[Dataset] {BSC} (2024).
\newblock {Paraver: a flexible performance analysis tool}.
\newblock \url{https://tools.bsc.es/paraver}.
\newblock Accessed: 2024-12-08
\bibAnnoteFile{paraver}

\bibitem[{Cabibbo(1984)}]{Cabibbo:1984zp}
Cabibbo, N. (1984).
\newblock {APE: A High Performance Processor for Lattice QCD}.
\newblock In \emph{{Symposium on Old and New Problems in Fundamental Physics,
  held in Honor of G.C. Wick}}. 137--144.
\newblock \href{https://inspirehep.net/literature/211710}{inspireHEP:211710}
\bibAnnoteFile{Cabibbo:1984zp}

\bibitem[{Carlson et~al.(2000)Carlson, Draper, Culler, Yelick, Brooks, and
  Warren}]{upc}
Carlson, W.~W., Draper, J.~M., Culler, D.~E., Yelick, K.~A., Brooks, E.~D., and
  Warren, K.~H. (2000).
\newblock Introduction to upc and language specification.
\newblock
  \href{https://api.semanticscholar.org/CorpusID:59868665}{CorpusID:59868665}
\bibAnnoteFile{upc}

\bibitem[{Casajus et~al.(2010)Casajus, Graciani, Paterson, Tsaregorodtsev, and
  (on behalf~ofthe Lhcb Dirac~Team)}]{Adrian_Casajus_2010}
Casajus, A., Graciani, R., Paterson, S., Tsaregorodtsev, A., and (on
  behalf~ofthe Lhcb Dirac~Team) (2010).
\newblock {DIRAC pilot framework and the DIRAC Workload Management System}.
\newblock \emph{Journal of Physics: Conference Series} 219, 062049.
\newblock \href{https://dx.doi.org/10.1088/1742-6596/219/6/062049}{DOI:
  10.1088/1742-6596/219/6/062049}
\bibAnnoteFile{Adrian_Casajus_2010}

\bibitem[{{CERN}(2024)}]{cernOpenlab}
[Dataset] {CERN} (2024).
\newblock {CERN openlab}.
\newblock \url{https://openlab.cern}.
\newblock Accessed: 2024-11-27
\bibAnnoteFile{cernOpenlab}

\bibitem[{{CERN openlab}(2024)}]{Cerabyte}
[Dataset] {CERN openlab} (2024).
\newblock {CERN openlab project with Cerabyte}.
\newblock \url{https://openlab.cern/cerabyte}.
\newblock Accessed: 2024-12-01
\bibAnnoteFile{Cerabyte}

\bibitem[{Chatrchyan et~al.(2012)}]{CMS:2012qbp}
Chatrchyan, S. et~al. (2012).
\newblock {Observation of a New Boson at a Mass of 125 GeV with the CMS
  Experiment at the LHC}.
\newblock \emph{Phys. Lett. B} 716, 30--61.
\newblock
  \href{https://www.sciencedirect.com/science/article/pii/S0370269312008581}{DOI:
  10.1016/j.physletb.2012.08.021}
\bibAnnoteFile{CMS:2012qbp}

\bibitem[{Chen et~al.(2001)}]{Chen:2000bu}
Chen, D. et~al. (2001).
\newblock {QCDOC: A 10-teraflops scale computer for lattice QCD}.
\newblock \emph{Nucl. Phys. B Proc. Suppl.} 94, 825--832.
\newblock
  \href{https://www.sciencedirect.com/science/article/abs/pii/S0920563201010143}{DOI:
  10.1016/S0920-5632(01)01014-3}
\bibAnnoteFile{Chen:2000bu}

\bibitem[{Chen et~al.(2012)Chen, Chac{\'o}n, and Barnes}]{Chen2012-vy}
Chen, G., Chac{\'o}n, L., and Barnes, D.~C. (2012).
\newblock An efficient mixed-precision, hybrid {CPU--GPU} implementation of a
  nonlinearly implicit one-dimensional particle-in-cell algorithm.
\newblock \emph{J. Comput. Phys.} 231, 5374--5388.
\newblock \href{https://doi.org/10.1016/j.jcp.2012.04.040}{DOI:
  10.1016/j.jcp.2012.04.040}
\bibAnnoteFile{Chen2012-vy}

\bibitem[{Chen et~al.(2023)Chen, Edwards, and Mao}]{Chen:2023zyy}
Chen, J., Edwards, R.~G., and Mao, W. (2023).
\newblock {Graph Contractions for Calculating Correlation Functions in Lattice
  QCD}.
\newblock In \emph{{Platform for Advanced Scientific Computing}}.
\newblock \href{https://dl.acm.org/doi/10.1145/3592979.3593409}{DOI:
  10.1145/3592979.3593409}
\bibAnnoteFile{Chen:2023zyy}

\bibitem[{Chen and T{\'o}th(2019)}]{Chen2019-ns}
Chen, Y. and T{\'o}th, G. (2019).
\newblock Gauss's law satisfying {Energy-Conserving} {Semi-Implicit}
  {Particle-in-Cell} method.
\newblock \emph{J. Comput. Phys.} 386, 632--652.
\newblock \href{https://doi.org/10.1016/j.jcp.2019.02.032}{DOI:
  10.1016/j.jcp.2019.02.032}
\bibAnnoteFile{Chen2019-ns}

\bibitem[{{CINECA}(2024)}]{leonardo}
[Dataset] {CINECA} (2024).
\newblock {Leonardo: Pre-Exascale Supercomputer}.
\newblock \url{https://leonardo-supercomputer.cineca.eu/}.
\newblock Accessed: 2024-12-08
\bibAnnoteFile{leonardo}

\bibitem[{Cinquilli et~al.(2012)Cinquilli, Evans, Foulkes, Hufnagel,
  Mascheroni, Norman et~al.}]{Cinquilli_2012}
Cinquilli, M., Evans, D., Foulkes, S., Hufnagel, D., Mascheroni, M., Norman,
  M., et~al. (2012).
\newblock {The CMS workload management system}.
\newblock \emph{Journal of Physics: Conference Series} 396, 032113.
\newblock \href{https://dx.doi.org/10.1088/1742-6596/396/3/032113}{DOI:
  10.1088/1742-6596/396/3/032113}
\bibAnnoteFile{Cinquilli_2012}

\bibitem[{Clark et~al.(2010)Clark, Babich, Barros, Brower, and
  Rebbi}]{Clark:2009wm}
Clark, M.~A., Babich, R., Barros, K., Brower, R.~C., and Rebbi, C. (2010).
\newblock {Solving Lattice QCD systems of equations using mixed precision
  solvers on GPUs}.
\newblock \emph{Comput. Phys. Commun.} 181, 1517--1528.
\newblock \doi{10.1016/j.cpc.2010.05.002}.
\newblock \href{https://doi.org/10.1016/j.cpc.2010.05.002}{DOI:
  10.1016/j.cpc.2010.05.002}
\bibAnnoteFile{Clark:2009wm}

\bibitem[{Clark et~al.(2023)Clark, Howarth, Tu, Wagner, and
  Weinberg}]{Clark:2023xnn}
Clark, M.~A., Howarth, D., Tu, J., Wagner, M., and Weinberg, E. (2023).
\newblock {Maximizing the Bang Per Bit}.
\newblock \emph{PoS} LATTICE2022, 338.
\newblock \href{http://dx.doi.org/10.22323/1.430.0338}{DOI:
  10.22323/1.430.0338}
\bibAnnoteFile{Clark:2023xnn}

\bibitem[{Clark et~al.(2016)Clark, Jo{\'o}, Strelchenko, Cheng, Gambhir, and
  Brower}]{Clark:2016rdz}
Clark, M.~A., Jo{\'o}, B., Strelchenko, A., Cheng, M., Gambhir, A., and Brower,
  R.~C. (2016).
\newblock {Accelerating Lattice QCD Multigrid on GPUs Using Fine-Grained
  Parallelization}.
\newblock In \emph{SC '16: Proceedings of the International Conference for High
  Performance Computing, Networking, Storage and Analysis}. 795--806.
\newblock \doi{10.1109/SC.2016.67}.
\newblock \href{https://doi.org/10.1109/SC.2016.67}{DOI: 10.1109/SC.2016.67}
\bibAnnoteFile{Clark:2016rdz}

\bibitem[{{CMS Collaboration}(2006)}]{Bayatian:922757}
{CMS Collaboration} (2006).
\newblock \emph{{CMS Physics: Technical Design Report Volume 1: Detector
  Performance and Software}}.
\newblock Technical design report. CMS (Geneva: CERN).
\newblock \href{https://cds.cern.ch/record/922757}{CERN CDS Record: 922757}
\bibAnnoteFile{Bayatian:922757}

\bibitem[{Corbal{\'a}n and Brochard(2018)}]{EAR}
Corbal{\'a}n, J. and Brochard, L. (2018).
\newblock Ear: Energy management framework for supercomputers.
\newblock \href{https://api.semanticscholar.org/CorpusID:252911624}{Link}
\bibAnnoteFile{EAR}

\bibitem[{{CSCS}(2023)}]{cscs_stats:2023}
[Dataset] {CSCS} (2023).
\newblock {Annual Report 2023 from the Swiss National Supercomputing Centre}.
\newblock \url{https://report2023.cscs.ch/facts-and-figures}.
\newblock Accessed: 2025-01-23
\bibAnnoteFile{cscs_stats:2023}

\bibitem[{Daiß et~al.(2021)Daiß, Simberg, Reverdell, Biddiscombe, Pollinger,
  Kaiser et~al.}]{9460406}
Daiß, G., Simberg, M., Reverdell, A., Biddiscombe, J., Pollinger, T., Kaiser,
  H., et~al. (2021).
\newblock Beyond fork-join: Integration of performance portable kokkos kernels
  with hpx.
\newblock In \emph{2021 IEEE International Parallel and Distributed Processing
  Symposium Workshops (IPDPSW)}. 377--386.
\newblock \href{https://doi.org/10.1109/IPDPSW52791.2021.00066}{DOI:
  10.1109/IPDPSW52791.2021.00066}
\bibAnnoteFile{9460406}

\bibitem[{Daiß et~al.(2022)Daiß, Singanaboina, Diehl, Kaiser, and
  Pflüger}]{10029450}
Daiß, G., Singanaboina, S.~Y., Diehl, P., Kaiser, H., and Pflüger, D. (2022).
\newblock From merging frameworks to merging stars: Experiences using hpx,
  kokkos and simd types.
\newblock In \emph{2022 IEEE/ACM 7th International Workshop on Extreme Scale
  Programming Models and Middleware (ESPM2)}. 10--19.
\newblock \href{https://doi.org/10.1109/ESPM256814.2022.00007}{DOI:
  10.1109/ESPM256814.2022.00007}
\bibAnnoteFile{10029450}

\bibitem[{Deluzet et~al.(2023)Deluzet, Fubiani, Garrigues, Guillet, and
  Narski}]{Deluzet2023}
Deluzet, F., Fubiani, G., Garrigues, L., Guillet, C., and Narski, J. (2023).
\newblock Efficient parallelization for 3d-3v sparse grid particle-in-cell:
  Shared memory architectures.
\newblock \emph{Journal of Computational Physics} 480, 112022.
\newblock \doi{10.1016/j.jcp.2023.112022}.
\newblock \href{https://doi.org/10.1016/j.jcp.2023.112022}{DOI:
  10.1016/j.jcp.2023.112022}
\bibAnnoteFile{Deluzet2023}

\bibitem[{Di~Girolamo et~al.(2022)Di~Girolamo, Legger, Paparrigopoulos,
  Schovancová, Beermann, Boehler et~al.}]{Girolamo:OpInt}
Di~Girolamo, A., Legger, F., Paparrigopoulos, P., Schovancová, J., Beermann,
  T., Boehler, M., et~al. (2022).
\newblock Preparing distributed computing operations for the hl-lhc era with
  operational intelligence.
\newblock \emph{Frontiers in Big Data} 4.
\newblock
  \href{https://www.frontiersin.org/journals/big-data/articles/10.3389/fdata.2021.753409}{DOI:
  10.3389/fdata.2021.753409}
\bibAnnoteFile{Girolamo:OpInt}

\bibitem[{Diehl et~al.(2024)Diehl, Brandt, and Kaiser}]{hpx2}
Diehl, P., Brandt, S.~R., and Kaiser, H. (2024).
\newblock Shared memory parallelism in modern c++ and hpx.
\newblock \emph{SN Computer Science} 5, 459.
\newblock \href{https://doi.org/10.1007/s42979-024-02769-6}{DOI:
  10.1007/s42979-024-02769-6}
\bibAnnoteFile{hpx2}

\bibitem[{Duane et~al.(1987)Duane, Kennedy, Pendleton, and
  Roweth}]{Duane:1987de}
Duane, S., Kennedy, A.~D., Pendleton, B.~J., and Roweth, D. (1987).
\newblock {Hybrid Monte Carlo}.
\newblock \emph{Phys. Lett. B} 195, 216--222.
\newblock
  \href{https://www.sciencedirect.com/science/article/abs/pii/037026938791197X}{DOI:
  10.1016/0370-2693(87)91197-X}
\bibAnnoteFile{Duane:1987de}

\bibitem[{Durr et~al.(2008)}]{BMW:2008jgk}
Durr, S. et~al. (2008).
\newblock {Ab-Initio Determination of Light Hadron Masses}.
\newblock \emph{Science} 322, 1224--1227.
\newblock \href{https://doi.org/10.1126/science.1163233}{DOI:
  10.1126/science.1163233}
\bibAnnoteFile{BMW:2008jgk}

\bibitem[{Dykes et~al.(2021)Dykes, Foyer, Richardson, Svedin, Podobas, Jansson
  et~al.}]{MAMBA}
Dykes, T., Foyer, C., Richardson, H., Svedin, M., Podobas, A., Jansson, N.,
  et~al. (2021).
\newblock Mamba: Portable array-based abstractions for heterogeneous
  high-performance systems.
\newblock In \emph{2021 International Workshop on Performance, Portability and
  Productivity in HPC (P3HPC)}. 10--21.
\newblock \href{https://ieeexplore.ieee.org/document/9652860}{DOI:
  10.1109/P3HPC54578.2021.00005}
\bibAnnoteFile{MAMBA}

\bibitem[{Dörrich et~al.(2023)Dörrich, Fan, and Kist}]{Doerrich2023}
Dörrich, M., Fan, M., and Kist, A.~M. (2023).
\newblock Impact of mixed precision techniques on training and inference
  efficiency of deep neural networks.
\newblock \emph{IEEE Access} 11, 57627--57634.
\newblock \href{https://ieeexplore.ieee.org/abstract/document/10146255}{DOI:
  10.1109/ACCESS.2023.3284388}
\bibAnnoteFile{Doerrich2023}

\bibitem[{Edwards and Joo(2005)}]{chroma}
Edwards, R.~G. and Joo, B. (2005).
\newblock {The Chroma software system for lattice QCD}.
\newblock \emph{Nucl. Phys. B Proc. Suppl.} 140, 832.
\newblock \href{https://doi.org/10.1016/j.nuclphysbps.2004.11.254}{DOI:
  10.1016/j.nuclphysbps.2004.11.254}
\bibAnnoteFile{chroma}

\bibitem[{Eggington et~al.(2018)Eggington, Mejnertsen, Desai, Eastwood, and
  Chittenden}]{Eggington2018-hf}
Eggington, J. W.~B., Mejnertsen, L., Desai, R.~T., Eastwood, J.~P., and
  Chittenden, J.~P. (2018).
\newblock Forging links in earth's plasma environment.
\newblock \emph{Astron. Geophys.} 59, 6.26--6.28.
\newblock \href{https://doi.org/10.1093/astrogeo/aty275}{DOI:
  10.1093/astrogeo/aty275}
\bibAnnoteFile{Eggington2018-hf}

\bibitem[{Egri et~al.(2007)Egri, Fodor, Hoelbling, Katz, Nogradi, and
  Szabo}]{Egri:2006zm}
Egri, G.~I., Fodor, Z., Hoelbling, C., Katz, S.~D., Nogradi, D., and Szabo,
  K.~K. (2007).
\newblock {Lattice QCD as a video game}.
\newblock \emph{Comput. Phys. Commun.} 177, 631--639.
\newblock \href{http://dx.doi.org/10.1016/j.cpc.2007.06.005}{DOI:
  10.1016/j.cpc.2007.06.005}
\bibAnnoteFile{Egri:2006zm}

\bibitem[{Ellis et~al.(2020)Ellis, Brew, Patargias, Adye, Appleyard, Dewhurst
  et~al.}]{Ellis2020}
Ellis, K., Brew, C., Patargias, G., Adye, T., Appleyard, R., Dewhurst, A.,
  et~al. (2020).
\newblock Xrootd and object store: A new paradigm.
\newblock \emph{EPJ Web of Conferences} 245, 04006.
\newblock \href{http://dx.doi.org/10.1051/epjconf/202024504006}{DOI:
  10.1051/epjconf/202024504006}
\bibAnnoteFile{Ellis2020}

\bibitem[{Elmsheuser et~al.(2020)Elmsheuser, Anastopoulos, Boyd, Catmore, Gray,
  Krasznahorkay et~al.}]{Elmsheuser2020}
Elmsheuser, J., Anastopoulos, C., Boyd, J., Catmore, J., Gray, H.,
  Krasznahorkay, A., et~al. (2020).
\newblock Evolution of the atlas analysis model for run-3 and prospects for
  hl-lhc.
\newblock \emph{EPJ Web of Conferences} 245, 06014.
\newblock \href{http://dx.doi.org/10.1051/epjconf/202024506014}{DOI:
  10.1051/epjconf/202024506014}
\bibAnnoteFile{Elmsheuser2020}

\bibitem[{Ene and Anireh(2022)}]{Ene2022}
Ene, D. and Anireh, V.~I. (2022).
\newblock Performance evaluation of parallel algorithms.
\newblock \emph{International Journal of Computer Science and Engineering} 9,
  10–14.
\newblock \doi{10.14445/23488387/ijcse-v9i6p102}.
\newblock \href{https://doi.org/10.1016/j.jcp.2023.112022}{DOI:
  10.14445/23488387/ijcse-v9i6p102}
\bibAnnoteFile{Ene2022}

\bibitem[{Espinoza-Valverde et~al.(2023)Espinoza-Valverde, Frommer,
  Ramirez-Hidalgo, and Rottmann}]{ESPINOZAVALVERDE2023108869}
Espinoza-Valverde, J., Frommer, A., Ramirez-Hidalgo, G., and Rottmann, M.
  (2023).
\newblock Coarsest-level improvements in multigrid for lattice qcd on
  large-scale computers.
\newblock \emph{Computer Physics Communications} 292, 108869.
\newblock \href{https://doi.org/10.1016/j.cpc.2023.108869}{DOI:
  10.1016/j.cpc.2023.108869}
\bibAnnoteFile{ESPINOZAVALVERDE2023108869}

\bibitem[{{EuroHPC Joint Undertaking}(2022)}]{eurohpc_stats:2022}
[Dataset] {EuroHPC Joint Undertaking} (2022).
\newblock {EuroHPC JU 2022 Consolidated Annual Activity Report}.
\newblock
  \url{https://eurohpc-ju.europa.eu/document/download/c7a5fc77-4236-41d8-979d-56af07607c25_en?filename=Annex%20to%20Decision%2012.2023-EuroHPC%20JU%20Consolidated%20Annual%20Activity%20Report%202022_0.pdf}.
\newblock Accessed: 2025-01-23
\bibAnnoteFile{eurohpc_stats:2022}

\bibitem[{{European Commission}(2024)}]{EUenergyefficiency}
[Dataset] {European Commission} (2024).
\newblock {EU Energy Efficiency Directive}.
\newblock
  \url{https://energy.ec.europa.eu/topics/energy-efficiency/energy-efficiency-targets-directive-and-rules/energy-efficiency-directive\_en}.
\newblock Accessed: 2024-10-14
\bibAnnoteFile{EUenergyefficiency}

\bibitem[{Finkenrath(2022)}]{Finkenrath:2022ogg}
Finkenrath, J. (2022).
\newblock {Tackling critical slowing down using global correction steps with
  equivariant flows: the case of the Schwinger model}
  \href{https://arxiv.org/pdf/2201.02216}{arXiv:2201.02216}
\bibAnnoteFile{Finkenrath:2022ogg}

\bibitem[{Frommer et~al.(2014)Frommer, Kahl, Krieg, Leder, and
  Rottmann}]{Frommer:2013fsa}
Frommer, A., Kahl, K., Krieg, S., Leder, B., and Rottmann, M. (2014).
\newblock {Adaptive Aggregation-Based Domain Decomposition Multigrid for the
  Lattice Wilson--Dirac Operator}.
\newblock \emph{SIAM J. Sci. Comput.} 36, A1581--A1608.
\newblock \href{https://doi.org/10.1137/130919507}{DOI: 10.1137/130919507}
\bibAnnoteFile{Frommer:2013fsa}

\bibitem[{Frommer et~al.(2022)Frommer, Khalil, and
  Ramirez-Hidalgo}]{frommer2022multilevel}
Frommer, A., Khalil, M.~N., and Ramirez-Hidalgo, G. (2022).
\newblock A multilevel approach to variance reduction in the stochastic
  estimation of the trace of a matrix.
\newblock \emph{SIAM Journal on Scientific Computing} 44, A2536--A2556.
\newblock \href{https://doi.org/10.1137/21M1441894}{DOI: 10.1137/21M1441894}
\bibAnnoteFile{frommer2022multilevel}

\bibitem[{{Fryxell} et~al.(2000){Fryxell}, {Olson}, {Ricker}, {Timmes},
  {Zingale}, {Lamb} et~al.}]{Fryxell:2000}
{Fryxell}, B., {Olson}, K., {Ricker}, P., {Timmes}, F.~X., {Zingale}, M.,
  {Lamb}, D.~Q., et~al. (2000).
\newblock {FLASH: An Adaptive Mesh Hydrodynamics Code for Modeling
  Astrophysical Thermonuclear Flashes}.
\newblock \emph{ApJS} 131, 273--334.
\newblock \href{https://iopscience.iop.org/article/10.1086/317361}{DOI:
  10.1086/317361}
\bibAnnoteFile{Fryxell:2000}

\bibitem[{Geimer et~al.(2010)Geimer, Wolf, Wylie, Ábrahám, Becker, and
  Mohr}]{Geimer2010}
Geimer, M., Wolf, F., Wylie, B. J.~N., Ábrahám, E., Becker, D., and Mohr, B.
  (2010).
\newblock The scalasca performance toolset architecture.
\newblock \emph{Concurrency and Computation: Practice and Experience} 22,
  702--719.
\newblock \href{https://doi.org/10.1002/cpe.1556}{DOI: 10.1002/cpe.1556}
\bibAnnoteFile{Geimer2010}

\bibitem[{Germain et~al.(2000)Germain, McCorquodale, Parker, and
  Johnson}]{uintah}
Germain, J. D. d.~S., McCorquodale, J., Parker, S.~G., and Johnson, C.~R.
  (2000).
\newblock Uintah: A massively parallel problem solving environment.
\newblock In \emph{Proceedings the Ninth International Symposium on
  High-Performance Distributed Computing} (IEEE), 33--41.
\newblock \href{https://doi.org/10.1109/HPDC.2000.868632}{DOI:
  10.1109/HPDC.2000.868632}
\bibAnnoteFile{uintah}

\bibitem[{Godoy et~al.(2023{\natexlab{a}})Godoy, Valero-Lara, Teranishi,
  Balaprakash, and Vetter}]{10.1145/3605731.3605886}
Godoy, W., Valero-Lara, P., Teranishi, K., Balaprakash, P., and Vetter, J.
  (2023{\natexlab{a}}).
\newblock Evaluation of openai codex for hpc parallel programming models kernel
  generation.
\newblock In \emph{Proceedings of the 52nd International Conference on Parallel
  Processing Workshops} (New York, NY, USA: Association for Computing
  Machinery), ICPP Workshops '23, 136–144.
\newblock \href{https://doi.org/10.1145/3605731.3605886}{DOI:
  10.1145/3605731.3605886}
\bibAnnoteFile{10.1145/3605731.3605886}

\bibitem[{Godoy et~al.(2023{\natexlab{b}})Godoy, Valero-Lara, Dettling,
  Trefftz, Jorquera, Sheehy et~al.}]{Godoy:2023}
Godoy, W.~F., Valero-Lara, P., Dettling, T.~E., Trefftz, C., Jorquera, I.,
  Sheehy, T., et~al. (2023{\natexlab{b}}).
\newblock Evaluating performance and portability of high-level programming
  models: Julia, python/numba, and kokkos on exascale nodes.
\newblock In \emph{2023 IEEE International Parallel and Distributed Processing
  Symposium Workshops (IPDPSW)}. 373--382.
\newblock \href{https://doi.org/10.1109/IPDPSW59300.2023.00068}{DOI:
  10.1109/IPDPSW59300.2023.00068}
\bibAnnoteFile{Godoy:2023}

\bibitem[{Gombosi et~al.(2023)Gombosi, Chen, Huang, Manchester, Sokolov, Toth
  et~al.}]{Gombosi2023-rb}
Gombosi, T.~I., Chen, Y., Huang, Z., Manchester, W.~B., Sokolov, I., Toth, G.,
  et~al. (2023).
\newblock Adaptive global magnetohydrodynamic simulations.
\newblock In \emph{Space and Astrophysical Plasma Simulation} (Cham: Springer
  International Publishing). 211--253.
\newblock \href{https://doi.org/10.1007/978-3-031-11870-8\_7}{DOI:
  10.1007/978-3-031-11870-8\_7}
\bibAnnoteFile{Gombosi2023-rb}

\bibitem[{{Google}(2024)}]{TPU}
[Dataset] {Google} (2024).
\newblock {TPU (Cloud)}.
\newblock \url{https://cloud.google.com/tpu/docs}.
\newblock Accessed: 2024-10-15
\bibAnnoteFile{TPU}

\bibitem[{{Graphcore}(2024)}]{graphcore}
[Dataset] {Graphcore} (2024).
\newblock {Graphcore}.
\newblock \url{https://www.graphcore.ai/}.
\newblock Accessed: 2024-10-15
\bibAnnoteFile{graphcore}

\bibitem[{{Grimm} and {Stadel}(2014)}]{Grimm:2014}
{Grimm}, S.~L. and {Stadel}, J.~G. (2014).
\newblock {The GENGA Code: Gravitational Encounters in N-body Simulations with
  GPU Acceleration}.
\newblock \emph{The Astrophysical Journal} 796, 23.
\newblock
  \href{https://iopscience.iop.org/article/10.1088/0004-637X/796/1/23}{DOIL
  10.1088/0004-637X/796/1/23}
\bibAnnoteFile{Grimm:2014}

\bibitem[{{Groq Inc.}(2024)}]{groq}
[Dataset] {Groq Inc.} (2024).
\newblock {Groq}.
\newblock \url{https://groq.com/}.
\newblock Accessed: 2024-10-15
\bibAnnoteFile{groq}

\bibitem[{Guillet(2023)}]{guillet:tel-04277746}
Guillet, C. (2023).
\newblock \emph{{Sparse approach to accelerate Particle-In-Cell method in 3D}}.
\newblock Theses, {Universit{\'e} Paul Sabatier - Toulouse III}
\bibAnnoteFile{guillet:tel-04277746}

\bibitem[{{GWT-TUD GmbH}(2024)}]{vampir}
[Dataset] {GWT-TUD GmbH} (2024).
\newblock {Vampir - Performance Optimization}.
\newblock \url{https://vampir.eu/}.
\newblock Accessed: 2024-12-08
\bibAnnoteFile{vampir}

\bibitem[{Haring et~al.(2011)Haring, Ohmacht, Fox, Gschwind, Satterfield,
  Sugavanam et~al.}]{haring2011ibm}
Haring, R., Ohmacht, M., Fox, T., Gschwind, M., Satterfield, D., Sugavanam, K.,
  et~al. (2011).
\newblock The ibm blue gene/q compute chip.
\newblock \emph{Ieee Micro} 32, 48--60.
\newblock \href{https://ieeexplore.ieee.org/document/6109225}{DOI:
  10.1109/MM.2011.108}
\bibAnnoteFile{haring2011ibm}

\bibitem[{Heinrich(2021)}]{Heinrich:2020ybq}
Heinrich, G. (2021).
\newblock {Collider Physics at the Precision Frontier}.
\newblock \emph{Phys. Rept.} 922, 1--69.
\newblock
  \href{https://www.sciencedirect.com/science/article/abs/pii/S0370157321001198}{10.1016/j.physrep.2021.03.006}
\bibAnnoteFile{Heinrich:2020ybq}

\bibitem[{Herten et~al.(2024)Herten, Achilles, Alvarez, Badwaik, Behle, Bode
  et~al.}]{herten2024}
Herten, A., Achilles, S., Alvarez, D., Badwaik, J., Behle, E., Bode, M., et~al.
  (2024).
\newblock Application-driven exascale: The jupiter benchmark suite.
\newblock In \emph{Proceedings of the International Conference for High
  Performance Computing, Networking, Storage, and Analysis} (IEEE Press), SC
  '24.
\newblock
  \href{https://doi.org/10.1109/SC41406.2024.00038}{DOI:10.1109/SC41406.2024.00038}
\bibAnnoteFile{herten2024}

\bibitem[{Hinterreiter et~al.(2019)Hinterreiter, Magdalenic, Temmer, Verbeke,
  Jebaraj, Samara et~al.}]{Hinterreiter2019-st}
Hinterreiter, J., Magdalenic, J., Temmer, M., Verbeke, C., Jebaraj, I.~C.,
  Samara, E., et~al. (2019).
\newblock Assessing the performance of {EUHFORIA} modeling the background solar
  wind.
\newblock \emph{Sol. Phys.} 294, 170.
\newblock \href{https://doi.org/10.1007/s11207-019-1558-8}{DOI:
  10.1007/s11207-019-1558-8}
\bibAnnoteFile{Hinterreiter2019-st}

\bibitem[{{HLRS}(2023)}]{hlrs_stats:2023}
[Dataset] {HLRS} (2023).
\newblock {HLRS Annual Report 2023}.
\newblock \url{https://www.hlrs.de/about/profile/annual-report}.
\newblock Accessed: 2025-01-23
\bibAnnoteFile{hlrs_stats:2023}

\bibitem[{Hotta and Kusano(2021)}]{Hotta2021-xi}
Hotta, H. and Kusano, K. (2021).
\newblock Solar differential rotation reproduced with high-resolution
  simulation.
\newblock \emph{Nat. Astron.} 5, 1100--1102.
\newblock \href{https://doi.org/10.1038/s41550-021-01459-0}{DOI:
  10.1038/s41550-021-01459-0}
\bibAnnoteFile{Hotta2021-xi}

\bibitem[{{HTCondor Software Suite}(2024)}]{htcondor}
[Dataset] {HTCondor Software Suite} (2024).
\newblock {HTcondor collaboration}.
\newblock \url{https://htcondor.org/}.
\newblock Accessed: 2024-12-08
\bibAnnoteFile{htcondor}

\bibitem[{Hunold and Steiner(2020)}]{hunold2020benchmarking}
Hunold, S. and Steiner, S. (2020).
\newblock Benchmarking julia’s communication performance: Is julia hpc ready
  or full hpc?
\newblock In \emph{IEEE/ACM Performance Modeling, Benchmarking and Simulation
  of High Performance Computer Systems (PMBS)}.
\newblock \href{https://doi.org/10.1109/PMBS51919.2020.00008}{DOI:
  10.1109/PMBS51919.2020.00008}
\bibAnnoteFile{hunold2020benchmarking}

\bibitem[{{Ian Cutress and Anton Shilov}(2019)}]{Cutress2019}
[Dataset] {Ian Cutress and Anton Shilov} (2019).
\newblock {The Larrabee Chapter Closes: Intel's Final Xeon Phi Processors Now
  in EOL}.
\newblock
  \url{https://www.anandtech.com/show/14305/intel-xeon-phi-knights-mill-now-eol}.
\newblock Accessed: 2024-12-08
\bibAnnoteFile{Cutress2019}

\bibitem[{Ilsche et~al.(2024)Ilsche, Schrader, and Schöne}]{ilsche2024}
Ilsche, T., Schrader, S., and Schöne, R. (2024).
\newblock Optimizing idle power of hpc systems: Practical insights and methods.
\newblock In \emph{2024 IEEE International Conference on Cluster Computing
  Workshops (CLUSTER Workshops)}. 19--25.
\newblock
  \href{https://ieeexplore.ieee.org/document/10740847}{10.1109/CLUSTERWorkshops61563.2024.00014}
\bibAnnoteFile{ilsche2024}

\bibitem[{{Intel}(2024{\natexlab{a}})}]{oneAPI}
[Dataset] {Intel} (2024{\natexlab{a}}).
\newblock {OneAPI}.
\newblock
  \url{https://www.intel.com/content/www/us/en/developer/tools/oneapi/overview.htm}.
\newblock Accessed: 2024-11-27
\bibAnnoteFile{oneAPI}

\bibitem[{{Intel}(2024{\natexlab{b}})}]{vtune}
[Dataset] {Intel} (2024{\natexlab{b}}).
\newblock {VTune Profiler: Performance Analysis for Applications \& Systems }.
\newblock
  \url{https://www.intel.com/content/www/us/en/developer/tools/oneapi/vtune-profiler.html}.
\newblock Accessed: 2024-12-08
\bibAnnoteFile{vtune}

\bibitem[{Ishiyama(2016)}]{Ishiyama2016-qq}
Ishiyama, T. (2016).
\newblock Supercomputer simulations of structure formation in the universe.
\newblock \emph{Proc. Int. Astron. Union} 12, 10--16.
\newblock \href{https://doi.org/10.1016/j.cpc.2008.12.031}{DOI:
  10.1016/j.cpc.2008.12.031}
\bibAnnoteFile{Ishiyama2016-qq}

\bibitem[{Janetzko et~al.(2025)Janetzko, Kostrzewa, and
  Suarez}]{plots_data_zenodo}
[Dataset] Janetzko, F., Kostrzewa, B., and Suarez, E. (2025).
\newblock Data, build and plot scripts accompanying "energy efficiency trends
  in hpc: what high-energy and astrophycisists need to know".
\newblock \doi{10.5281/ZENODO.14790357}.
\newblock \href{https://doi.org/10.5281/zenodo.14790357}{DOI:
  10.5281/zenodo.14790357}
\bibAnnoteFile{plots_data_zenodo}

\bibitem[{Jevons(1865)}]{Jevons:1865}
Jevons, W. (1865).
\newblock \emph{The Coal Question: An Enquiry Concerning the Progress of the
  Nation, and the Probable Exhaustion of Our Coal-mines}.
\newblock Making Of The Modern World. Part 2 (Macmillan)
\bibAnnoteFile{Jevons:1865}

\bibitem[{Jiang et~al.(2024)Jiang, Shi, Chen, Gong, and Yang}]{PyQUDA}
[Dataset] Jiang, X., Shi, C., Chen, Y., Gong, M., and Yang, Y.-B. (2024).
\newblock Use quda for lattice qcd calculation with python.
\newblock \href{https://arxiv.org/abs/2411.08461}{arXiv:2411.08461}
\bibAnnoteFile{PyQUDA}

\bibitem[{Jo{\'o} et~al.(2019)Jo{\'o}, Kurth, Clark, Kim, Trott, Ibanez
  et~al.}]{joo2019performance}
Jo{\'o}, B., Kurth, T., Clark, M.~A., Kim, J., Trott, C.~R., Ibanez, D., et~al.
  (2019).
\newblock Performance portability of a wilson dslash stencil operator mini-app
  using kokkos and sycl.
\newblock In \emph{2019 IEEE/ACM International Workshop on Performance,
  Portability and Productivity in HPC (P3HPC)} (IEEE), 14--25.
\newblock \href{https://doi.org/10.1109/P3HPC49587.2019.00007}{DOI:
  10.1109/P3HPC49587.2019.00007}
\bibAnnoteFile{joo2019performance}

\bibitem[{{Juelich}(2024)}]{SIONLib}
[Dataset] {Juelich} (2024).
\newblock {SIONLib}.
\newblock \url{https://apps.fz-juelich.de/jsc/sionlib/docu/index.html}.
\newblock Accessed: 2024-12-01
\bibAnnoteFile{SIONLib}

\bibitem[{{J\"{u}lich Supercomputing Centre}(2024{\natexlab{a}})}]{jupiter}
[Dataset] {J\"{u}lich Supercomputing Centre} (2024{\natexlab{a}}).
\newblock {JUPITER: The arrival of Exascale in Europe}.
\newblock \url{https://www.fz-juelich.de/en/ias/jsc/jupiter}.
\newblock Accessed: 2024-12-08
\bibAnnoteFile{jupiter}

\bibitem[{{J\"{u}lich Supercomputing Centre}(2024{\natexlab{b}})}]{llview}
[Dataset] {J\"{u}lich Supercomputing Centre} (2024{\natexlab{b}}).
\newblock {LLview job monitoring}.
\newblock
  \url{https://www.fz-juelich.de/en/ias/jsc/services/user-support/software-tools/llview}.
\newblock Accessed: 2024-10-20
\bibAnnoteFile{llview}

\bibitem[{{Julita Corbalan and Andreas Smolenko and M. D’Amico and German
  Llort and E. Mercadal and Judit Gimeenez and Carmen
  Navarrete}(2021)}]{deepest:2021}
[Dataset] {Julita Corbalan and Andreas Smolenko and M. D’Amico and German
  Llort and E. Mercadal and Judit Gimeenez and Carmen Navarrete} (2021).
\newblock {Deliverable D2.3: Benchmarking, evaluation and prediction report}.
\newblock
  \url{https://deep-projects.eu/wp-content/uploads/2023/09/DEEP-EST_D23_Benchmarking_evaluation_and_prediction_report_v10.pdf}.
\newblock Accessed: 2025-01-27
\bibAnnoteFile{deepest:2021}

\bibitem[{Jung and Christ(2024)}]{Jung:2024nuv}
Jung, C. and Christ, N.~H. (2024).
\newblock {Riemannian manifold HMC with fermions}.
\newblock \emph{PoS} LATTICE2023, 009.
\newblock \href{https://doi.org/10.22323/1.453.0009}{DOI: 10.22323/1.453.0009}
\bibAnnoteFile{Jung:2024nuv}

\bibitem[{Kaiser et~al.(2014)Kaiser, Heller, Adelstein-Lelbach, Serio, and
  Fey}]{hpx}
Kaiser, H., Heller, T., Adelstein-Lelbach, B., Serio, A., and Fey, D. (2014).
\newblock Hpx: A task based programming model in a global address space.
\newblock In \emph{Proceedings of the 8th International Conference on
  Partitioned Global Address Space Programming Models}. 1--11.
\newblock \href{https://doi.org/10.1145/2676870.267688}{DOI:
  10.1145/2676870.267688}
\bibAnnoteFile{hpx}

\bibitem[{Kale and Krishnan(1993)}]{charm++}
Kale, L.~V. and Krishnan, S. (1993).
\newblock Charm++: a portable concurrent object oriented system based on c++.
\newblock \emph{SIGPLAN Not.} 28, 91–108.
\newblock \href{https://doi.org/10.1145/167962.165874}{DOI:
  10.1145/167962.165874}
\bibAnnoteFile{charm++}

\bibitem[{Kanwar(2024)}]{Kanwar:2024ujc}
Kanwar, G. (2024).
\newblock {Flow-based sampling for lattice field theories}.
\newblock In \emph{{40th International Symposium on Lattice Field Theory}}.
\newblock \href{https://arxiv.org/abs/2401.01297}{arXiv:2401.01297}
\bibAnnoteFile{Kanwar:2024ujc}

\bibitem[{Kawazura and Kimura(2024)}]{Kawazura2024-tw}
Kawazura, Y. and Kimura, S.~S. (2024).
\newblock Inertial range of magnetorotational turbulence.
\newblock \emph{Sci. Adv.} 10, eadp4965.
\newblock \href{https://doi.org/10.1126/sciadv.adp4965}{DOI:
  10.1126/sciadv.adp4965}
\bibAnnoteFile{Kawazura2024-tw}

\bibitem[{Keppens et~al.(2021)Keppens, Teunissen, Xia, and
  Porth}]{Keppens2021-et}
Keppens, R., Teunissen, J., Xia, C., and Porth, O. (2021).
\newblock {MPI-AMRVAC}: A parallel, grid-adaptive {PDE} toolkit.
\newblock \emph{Comput. Math. Appl.} 81, 316--333.
\newblock \href{https://doi.org/10.1016/j.camwa.2020.03.023}{DOI:
  10.1016/j.camwa.2020.03.023}
\bibAnnoteFile{Keppens2021-et}

\bibitem[{Keshavarzi(2019)}]{Keshavarzi:2019bjn}
Keshavarzi, A. (2019).
\newblock The muon $g-2$ experiment at fermilab.
\newblock \emph{EPJ Web Conf.}
  \href{https://www.epj-conferences.org/articles/epjconf/abs/2019/17/epjconf\_phipsi18\_05003}{DOI:
  10.1051/epjconf/201921205003}
\bibAnnoteFile{Keshavarzi:2019bjn}

\bibitem[{Keshavarzi et~al.(2022)Keshavarzi, Khaw, and
  Yoshioka}]{KESHAVARZI2022115675}
Keshavarzi, A., Khaw, K.~S., and Yoshioka, T. (2022).
\newblock Muon $g-2$: A review.
\newblock \emph{Nuclear Physics B}
  \href{https://doi.org/10.1016/j.nuclphysb.2022.115675}{DOI:
  10.1016/j.nuclphysb.2022.115675}
\bibAnnoteFile{KESHAVARZI2022115675}

\bibitem[{Khan et~al.(2021)Khan, Sim, Vazhkudai, Butt, and Kim}]{Khan2021}
Khan, A., Sim, H., Vazhkudai, S.~S., Butt, A.~R., and Kim, Y. (2021).
\newblock An analysis of system balance and architectural trends based on
  top500 supercomputers.
\newblock In \emph{The International Conference on High Performance Computing
  in Asia-Pacific Region} (Association for Computing Machinery), HPCAsia '21,
  11–22.
\newblock \href{https://doi.org/10.1145/3432261.3432263}{DOI:
  10.1145/3432261.3432263}
\bibAnnoteFile{Khan2021}

\bibitem[{{Khronos}(2024)}]{sycl}
[Dataset] {Khronos} (2024).
\newblock {SYCL}.
\newblock \url{https://www.khronos.org/sycl/}.
\newblock Accessed: 2024-10-10
\bibAnnoteFile{sycl}

\bibitem[{Kodama et~al.(2020)Kodama, Odajima, Arima, and Sato}]{kodama2020}
Kodama, Y., Odajima, T., Arima, E., and Sato, M. (2020).
\newblock Evaluation of power management control on the supercomputer fugaku.
\newblock In \emph{2020 IEEE International Conference on Cluster Computing
  (CLUSTER)}. 484--493.
\newblock \href{https://ieeexplore.ieee.org/document/9229623}{DOI:
  10.1109/CLUSTER49012.2020.00069}
\bibAnnoteFile{kodama2020}

\bibitem[{Kreuzer et~al.(2018)Kreuzer, Eicker, Amaya, and Suarez}]{Kreuzer2018}
Kreuzer, A., Eicker, N., Amaya, J., and Suarez, E. (2018).
\newblock Application performance on a cluster-booster system.
\newblock In \emph{2018 IEEE International Parallel and Distributed Processing
  Symposium Workshops (IPDPSW)}. 69--78.
\newblock \href{https://ieeexplore.ieee.org/document/8425386}{DOI:
  10.1109/IPDPSW.2018.00019}
\bibAnnoteFile{Kreuzer2018}

\bibitem[{Kreuzer et~al.(2021)Kreuzer, Kreutz, Steinbusch, Huda, Llort,
  Corbalan et~al.}]{Kreuzer2021}
Kreuzer, A., Kreutz, J., Steinbusch, B., Huda, Z.~U., Llort, G., Corbalan, J.,
  et~al. (2021).
\newblock \emph{{B}est {P}ractices {G}uide} (Forschungszentrum Jülich GmbH
  Zentralbibliothek, Verlag Jülich), vol.~48 of \emph{Schriften des
  Forschungszentrums Jülich IAS Series}.
\newblock 187--232.
\newblock \href{http://hdl.handle.net/2128/30533}{{FZJ} Record:905853}
\bibAnnoteFile{Kreuzer2021}

\bibitem[{Kuhr et~al.(2018)Kuhr, Pulvermacher, Ritter, Hauth, and
  Braun}]{Kuhr_2018}
Kuhr, T., Pulvermacher, C., Ritter, M., Hauth, T., and Braun, N. (2018).
\newblock {The Belle II Core Software: Belle II Framework Software Group}.
\newblock \emph{Computing and Software for Big Science} 3.
\newblock
  \href{http://dx.doi.org/10.1007/s41781-018-0017-9}{10.1007/s41781-018-0017-9}
\bibAnnoteFile{Kuhr_2018}

\bibitem[{Lani et~al.(2013)Lani, Villedie, Bensassi, Koloszar, Vymazal, Yalim
  et~al.}]{Lani2013-jp}
Lani, A., Villedie, N., Bensassi, K., Koloszar, L., Vymazal, M., Yalim, S.~M.,
  et~al. (2013).
\newblock {COOLFluiD}: an open computational platform for multi-physics
  simulation and research.
\newblock In \emph{21st {AIAA} Computational Fluid Dynamics Conference}
  (Reston, Virginia: American Institute of Aeronautics and Astronautics).
\newblock \href{https://doi.org/10.2514/6.2013-2589}{DOI: 10.2514/6.2013-2589}
\bibAnnoteFile{Lani2013-jp}

\bibitem[{Lapenta(2017)}]{Lapenta2017-cu}
Lapenta, G. (2017).
\newblock Exactly energy conserving semi-implicit particle in cell formulation.
\newblock \emph{J. Comput. Phys.} 334, 349--366
\bibAnnoteFile{Lapenta2017-cu}

\bibitem[{Le et~al.(2016)Le, Daughton, Karimabadi, and Egedal}]{Le2016-qy}
Le, A., Daughton, W., Karimabadi, H., and Egedal, J. (2016).
\newblock Hybrid simulations of magnetic reconnection with kinetic ions and
  fluid electron pressure anisotropy.
\newblock \emph{Phys. Plasmas} 23, 032114.
\newblock \href{https://doi.org/10.1063/1.4943893}{DOI: 10.1063/1.4943893}
\bibAnnoteFile{Le2016-qy}

\bibitem[{Lehner and Wettig(2023)}]{PhysRevD.108.034503}
Lehner, C. and Wettig, T. (2023).
\newblock Gauge-equivariant neural networks as preconditioners in lattice qcd.
\newblock \emph{Phys. Rev. D} 108, 034503.
\newblock \href{https://doi.org/10.1103/PhysRevD.108.034503}{DOI:
  10.1103/PhysRevD.108.034503}
\bibAnnoteFile{PhysRevD.108.034503}

\bibitem[{Lehner et~al.(2024)}]{gpt}
[Dataset] Lehner, C. et~al. (2024).
\newblock {Grid Python Toolkit (GPT)}.
\newblock \url{https://github.com/lehner/gpt}.
\newblock Accessed: 2024-12-03
\bibAnnoteFile{gpt}

\bibitem[{LHCb~Collaboration(2022)}]{LHCbCollaboration:2806113}
LHCb~Collaboration, L.~E. (2022).
\newblock \emph{{Future physics potential of LHCb}}.
\newblock Tech. rep., CERN, Geneva.
\newblock \href{https://cds.cern.ch/record/2806113}{CERN Record:2806113}
\bibAnnoteFile{LHCbCollaboration:2806113}

\bibitem[{Li et~al.(2023)Li, Michelogiannakis, Cook, Cooray, and Chen}]{Li2023}
Li, J., Michelogiannakis, G., Cook, B., Cooray, D., and Chen, Y. (2023).
\newblock Analyzing resource utilization in an {HPC} system: A case study of
  {NERSC’s} perlmutter.
\newblock In \emph{International Conference on High Performance Computing
  (ISC)}. 297–316.
\newblock \href{https://doi.org/10.1007/978-3-031-32041-5\_16}{DOI:
  10.1007/978-3-031-32041-5\_16}
\bibAnnoteFile{Li2023}

\bibitem[{Lopez-Gomez and Blomer(2023)}]{Lopez-Gomez:2808833}
Lopez-Gomez, J. and Blomer, J. (2023).
\newblock {RNTuple performance: Status and Outlook}.
\newblock \emph{J. Phys. : Conf. Ser.} 2438, 012118.
\newblock
  \href{https://iopscience.iop.org/article/10.1088/1742-6596/2438/1/012118}{DOI:
  10.1088/1742-6596/2438/1/012118}
\bibAnnoteFile{Lopez-Gomez:2808833}

\bibitem[{Lottermoser et~al.(1998)Lottermoser, Scholer, and
  Matthews}]{Lottermoser1998-qj}
Lottermoser, R.-F., Scholer, M., and Matthews, A.~P. (1998).
\newblock Ion kinetic effects in magnetic reconnection: Hybrid simulations.
\newblock \emph{J. Geophys. Res.} 103, 4547--4559.
\newblock \href{ https://doi.org/10.1029/97JA01872}{DOI: 10.1029/97JA01872}
\bibAnnoteFile{Lottermoser1998-qj}

\bibitem[{{LuxProvide}(2024)}]{meluxina}
[Dataset] {LuxProvide} (2024).
\newblock {MeluXina}.
\newblock \url{https://www.luxprovide.lu/meluxina/}.
\newblock Accessed: 2024-12-08
\bibAnnoteFile{meluxina}

\bibitem[{Maeno et~al.(2024)Maeno, Alekseev, Barreiro~Megino, De, Guan,
  Karavakis et~al.}]{Maeno2024}
Maeno, T., Alekseev, A., Barreiro~Megino, F.~H., De, K., Guan, W., Karavakis,
  E., et~al. (2024).
\newblock Panda: Production and distributed analysis system.
\newblock \emph{Computing and Software for Big Science} 8.
\newblock \href{http://dx.doi.org/10.1007/s41781-024-00114-3}{DOI:
  10.1007/s41781-024-00114-3}
\bibAnnoteFile{Maeno2024}

\bibitem[{Maloney et~al.(2024)Maloney, Suarez, Eicker, Guimarães, and
  Frings}]{Maloney2024}
Maloney, S., Suarez, E., Eicker, N., Guimarães, F., and Frings, W. (2024).
\newblock Analyzing hpc monitoring data with a view towards efficient resource
  utilization.
\newblock In \emph{2024 IEEE 36th International Symposium on Computer
  Architecture and High Performance Computing (SBAC-PAD)}. 170--181.
\newblock \href{https://ieeexplore.ieee.org/document/10763585}{DOI:
  10.1109/SBAC-PAD63648.2024.00023}
\bibAnnoteFile{Maloney2024}

\bibitem[{Mawhinney(1999)}]{Mawhinney:1999rji}
Mawhinney, R.~D. (1999).
\newblock {The One teraflops QCDSP computer}.
\newblock \emph{Parallel Comput.} 25, 1281.
\newblock \href{http://dx.doi.org/10.1016/S0167-8191(99)00051-4}{DOI:
  10.1016/S0167-8191(99)00051-4}
\bibAnnoteFile{Mawhinney:1999rji}

\bibitem[{McAlpine et~al.(2022)McAlpine, Helly, Schaller, Sawala, Lavaux,
  Jasche et~al.}]{McAlpine2022-uv}
McAlpine, S., Helly, J.~C., Schaller, M., Sawala, T., Lavaux, G., Jasche, J.,
  et~al. (2022).
\newblock {SIBELIUS-DARK}: a galaxy catalogue of the local volume from a
  constrained realisation simulation.
\newblock \emph{Mon. Not. R. Astron. Soc.}
  \href{https://doi.org/10.1093/mnras/stac295}{DOI: 10.1093/mnras/stac295}
\bibAnnoteFile{McAlpine2022-uv}

\bibitem[{{McCalpin}(2022)}]{McCalpin2022}
[Dataset] {McCalpin}, J. (2022).
\newblock {Memory Bandwidth and system balance in HPC systems}.
\newblock
  \url{https://sites.utexas.edu/jdm4372/2016/11/22/sc16-invited-talk-memory-bandwidth-and-system-balance-in-hpc-systems/}.
\newblock Accessed: 2024-12-08
\bibAnnoteFile{McCalpin2022}

\bibitem[{Moore(2006)}]{moore2006}
Moore, G.~E. (2006).
\newblock Cramming more components onto integrated circuits.
\newblock \emph{IEEE Solid-State Circuits Society Newsletter [Reprinted from
  Electronics (1965)]} ,
  33--35\href{https://ieeexplore.ieee.org/document/4785860}{DOI:
  10.1109/N-SSC.2006.4785860}
\bibAnnoteFile{moore2006}

\bibitem[{{MPI Forum}(2024)}]{mpi}
[Dataset] {MPI Forum} (2024).
\newblock {Message Passing Interface documentation}.
\newblock \url{https://www.mpi-forum.org/docs/}.
\newblock Accessed: 2024-11-30
\bibAnnoteFile{mpi}

\bibitem[{{NERSC}(2023)}]{nersc_stats:2023}
[Dataset] {NERSC} (2023).
\newblock {National Energy Research Scientific Computing CenterTable of
  Contents, 2023 Annual Report}.
\newblock
  \url{https://www.nersc.gov/assets/Annual-Reports/2023-NERSC-Annual-Report-compressed.pdf}.
\newblock Accessed: 2025-01-23
\bibAnnoteFile{nersc_stats:2023}

\bibitem[{Nichols et~al.(2024)Nichols, Davis, Xie, Rajaram, and
  Bhatele}]{10.1145/3625549.3658689}
Nichols, D., Davis, J.~H., Xie, Z., Rajaram, A., and Bhatele, A. (2024).
\newblock Can large language models write parallel code? (New York, NY, USA:
  Association for Computing Machinery), HPDC '24, 281–294.
\newblock
  \href{https://doi.org/10.1145/3625549.3658689}{DOI:10.1145/3625549.3658689}
\bibAnnoteFile{10.1145/3625549.3658689}

\bibitem[{Nieplocha et~al.(2006)Nieplocha, Palmer, Tipparaju, Krishnan, Trease,
  and Aprà}]{globalarrays}
Nieplocha, J., Palmer, B., Tipparaju, V., Krishnan, M., Trease, H., and Aprà,
  E. (2006).
\newblock Advances, applications and performance of the global arrays shared
  memory programming toolkit.
\newblock \emph{The International Journal of High Performance Computing
  Applications} 20, 203--231.
\newblock \href{https://doi.org/10.1177/1094342006064503}{DOI:
  10.1177/1094342006064503}
\bibAnnoteFile{globalarrays}

\bibitem[{{Nitadori} and {Aarseth}(2012)}]{Nitadori:2012}
{Nitadori}, K. and {Aarseth}, S.~J. (2012).
\newblock {Accelerating NBODY6 with graphics processing units}.
\newblock \emph{MNRAS} 424, 545--552.
\newblock \href{https://academic.oup.com/mnras/article/424/1/545/1010486}{DOI:
  10.1111/j.1365-2966.2012.21227.x}
\bibAnnoteFile{Nitadori:2012}

\bibitem[{Nordlund et~al.(2018)Nordlund, Ramsey, Popovas, and
  K{\"u}ffmeier}]{Nordlund2018-se}
Nordlund, {\AA}., Ramsey, J.~P., Popovas, A., and K{\"u}ffmeier, M. (2018).
\newblock dispatch: a numerical simulation framework for the exa-scale era --
  i. fundamentals.
\newblock \emph{Mon. Not. R. Astron. Soc.} 477, 624--638.
\newblock \href{https://doi.org/10.1093/mnras/sty599}{DOI:
  10.1093/mnras/sty599}
\bibAnnoteFile{Nordlund2018-se}

\bibitem[{Numrich and Reid(1998)}]{coarray}
Numrich, R.~W. and Reid, J. (1998).
\newblock Co-array fortran for parallel programming.
\newblock \emph{SIGPLAN Fortran Forum} 17, 1–31.
\newblock \href{https://doi.org/10.1145/289918.289920}{DOI:
  10.1145/289918.289920}
\bibAnnoteFile{coarray}

\bibitem[{{NVIDIA}(2024{\natexlab{a}})}]{cuda}
[Dataset] {NVIDIA} (2024{\natexlab{a}}).
\newblock {CUDA Toolkit}.
\newblock \url{https://developer.nvidia.com/cuda-toolkit}.
\newblock Accessed: 2024-10-10
\bibAnnoteFile{cuda}

\bibitem[{{NVIDIA}(2024{\natexlab{b}})}]{gh200}
[Dataset] {NVIDIA} (2024{\natexlab{b}}).
\newblock {Grace-Hopper architecture}.
\newblock
  \url{https://resources.nvidia.com/en-us-grace-cpu/nvidia-grace-hopper}.
\newblock Accessed: 2024-12-08
\bibAnnoteFile{gh200}

\bibitem[{{OpenMP Architecture Review Board}(2024)}]{openmp}
[Dataset] {OpenMP Architecture Review Board} (2024).
\newblock {OpenMP}.
\newblock \url{https://www.openmp.org/}.
\newblock Accessed: 2024-10-15
\bibAnnoteFile{openmp}

\bibitem[{{OurResearch}(2025)}]{openalex}
[Dataset] {OurResearch} (2025).
\newblock {OpenAlex: The open catalog to the global research system}.
\newblock \url{https://openalex.org/}.
\newblock Accessed: 2025-01-28
\bibAnnoteFile{openalex}

\bibitem[{Pereira et~al.(2017)Pereira, Couto, Ribeiro, Rua, Cunha, Fernandes
  et~al.}]{Pereira2017}
Pereira, R., Couto, M., Ribeiro, F., Rua, R., Cunha, J., Fernandes, J.~a.~P.,
  et~al. (2017).
\newblock Energy efficiency across programming languages: how do energy, time,
  and memory relate?
\newblock In \emph{Proceedings of the 10th ACM SIGPLAN International Conference
  on Software Language Engineering} (New York, NY, USA: Association for
  Computing Machinery), SLE 2017, 256–267.
\newblock \href{https://doi.org/10.1145/3136014.3136031}{DOI:
  10.1145/3136014.3136031}
\bibAnnoteFile{Pereira2017}

\bibitem[{Pleiter and Maurer(2010)}]{Pleiter2010}
Pleiter, D. and Maurer, T. (2010).
\newblock Qpace -- a qcd parallel computer based on cell processors.
\newblock In \emph{Proceedings of The XXVII International Symposium on Lattice
  Field Theory — PoS(LAT2009)} (Sissa Medialab), LAT2009, 001.
\newblock \href{http://dx.doi.org/10.22323/1.091.0001}{DOI:
  10.22323/1.091.0001}
\bibAnnoteFile{Pleiter2010}

\bibitem[{Portegies~Zwart and McMillan(2018)}]{Portegies:2018}
Portegies~Zwart, S. and McMillan, S. (2018).
\newblock \emph{Astrophysical Recipes; The art of {AMUSE}} (IOP Publishing).
\newblock \href{https://dx.doi.org/10.1088/978-0-7503-1320-9}{DOI:
  10.1088/978-0-7503-1320-9}
\bibAnnoteFile{Portegies:2018}

\bibitem[{{Python Software Foundation}(2025)}]{python}
[Dataset] {Python Software Foundation} (2025).
\newblock {Python}.
\newblock \url{https://www.python.org}.
\newblock Accessed: 2025-02-08
\bibAnnoteFile{python}

\bibitem[{{Rein} and {Liu}(2012)}]{Rein:2012}
{Rein}, H. and {Liu}, S.~F. (2012).
\newblock {REBOUND: an open-source multi-purpose N-body code for collisional
  dynamics}.
\newblock \emph{astro-ph} 537, A128.
\newblock
  \href{https://www.aanda.org/articles/aa/full_html/2012/01/aa18085-11/aa18085-11.html}{DOI:
  10.1051/0004-6361/201118085}
\bibAnnoteFile{Rein:2012}

\bibitem[{Ren et~al.(2021)Ren, Luo, Peng, Wu, and Li}]{Ren2021}
Ren, J., Luo, J., Peng, I., Wu, K., and Li, D. (2021).
\newblock Optimizing large-scale plasma simulations on persistent memory-based
  heterogeneous memory with effective data placement across memory hierarchy.
\newblock In \emph{Proceedings of the ACM International Conference on
  Supercomputing} (ACM), ICS ’21, 203–214.
\newblock \href{https://doi.org/10.1145/3447818.3460356}{DOI:
  10.1145/3447818.3460356}
\bibAnnoteFile{Ren2021}

\bibitem[{Riley et~al.(2022)Riley, Caplan, Downs, Linker, and
  Lionello}]{Riley2022-rl}
Riley, P., Caplan, R.~M., Downs, C., Linker, J.~A., and Lionello, R. (2022).
\newblock Comparing and contrasting the properties of the inner heliosphere for
  the three most recent solar minima.
\newblock \emph{J. Geophys. Res. Space Phys.} 127, e2022JA030261.
\newblock \href{https://doi.org/10.1029/2022JA030261}{DOI:
  10.1029/2022JA030261}
\bibAnnoteFile{Riley2022-rl}

\bibitem[{Romein(2021)}]{Romein2021}
Romein, J.~W. (2021).
\newblock The tensor-core correlator.
\newblock \emph{Astronomy \& Astrophysics} 656, A52.
\newblock \doi{10.1051/0004-6361/202141896}.
\newblock \href{https://doi.org/10.1051/0004-6361/202141896}{DOI:
  10.1051/0004-6361/202141896}
\bibAnnoteFile{Romein2021}

\bibitem[{Ruzicka et~al.(2024)Ruzicka, Asch, Meneses, Rampp, and
  Laure}]{Ruzicka2024}
[Dataset] Ruzicka, J., Asch, C., Meneses, E., Rampp, M., and Laure, E. (2024).
\newblock A study of performance portability in plasma physics simulations.
\newblock \doi{10.48550/ARXIV.2411.05009}.
\newblock \href{https://doi.org/10.48550/arXiv.2411.05009}{DOI:
  10.48550/arXiv.2411.05009}
\bibAnnoteFile{Ruzicka2024}

\bibitem[{Schaefer et~al.(2009)Schaefer, Sommer, and Virotta}]{Schaefer:2009xx}
Schaefer, S., Sommer, R., and Virotta, F. (2009).
\newblock {Investigating the critical slowing down of QCD simulations}.
\newblock \emph{PoS} LAT2009, 032.
\newblock \href{https://doi.org/10.22323/1.091.0032}{DOI: 10.22323/1.091.0032}
\bibAnnoteFile{Schaefer:2009xx}

\bibitem[{Schaye et~al.(2023)Schaye, Kugel, Schaller, Helly, Braspenning,
  Elbers et~al.}]{Schaye2023-oy}
Schaye, J., Kugel, R., Schaller, M., Helly, J.~C., Braspenning, J., Elbers, W.,
  et~al. (2023).
\newblock The {FLAMINGO} project: cosmological hydrodynamical simulations for
  large-scale structure and galaxy cluster surveys.
\newblock \emph{Mon. Not. R. Astron. Soc.} 526, 4978--5020.
\newblock \href{https://doi.org/10.1093/mnras/stad2419}{DOI:
  10.1093/mnras/stad2419}
\bibAnnoteFile{Schaye2023-oy}

\bibitem[{{SchedMD}(2024)}]{slurm}
[Dataset] {SchedMD} (2024).
\newblock {Slurm scheduler}.
\newblock \url{https://slurm.schedmd.com/}.
\newblock Accessed: 2024-09-12
\bibAnnoteFile{slurm}

\bibitem[{Schlepphorst and Krieg(2023)}]{simon_kokkos}
Schlepphorst, S. and Krieg, S. (2023).
\newblock Benchmarking a portable lattice quantum chromodynamics kernel written
  in kokkos and mpi.
\newblock In \emph{Proceedings of the SC '23 Workshops of The International
  Conference on High Performance Computing, Network, Storage, and Analysis}
  (New York, NY, USA: Association for Computing Machinery), SC-W '23,
  1027–1037.
\newblock \href{https://doi.org/10.1145/3624062.3624179}{DOI:
  10.1145/3624062.3624179}
\bibAnnoteFile{simon_kokkos}

\bibitem[{Schlimme et~al.(2024)}]{Schlimme:2024eky}
Schlimme, S. et~al. (2024).
\newblock {The MESA physics program}.
\newblock \emph{EPJ Web Conf.} 303, 06002.
\newblock \href{http://dx.doi.org/10.1051/epjconf/202430306002}{DOI:
  10.1051/epjconf/202430306002}
\bibAnnoteFile{Schlimme:2024eky}

\bibitem[{Sevilla et~al.(2022)Sevilla, Heim, Ho, Besiroglu, Hobbhahn, and
  Villalobos}]{sevilla2022}
Sevilla, J., Heim, L., Ho, A., Besiroglu, T., Hobbhahn, M., and Villalobos, P.
  (2022).
\newblock Compute trends across three eras of machine learning.
\newblock In \emph{2022 International Joint Conference on Neural Networks
  (IJCNN)}. 1--8.
\newblock \href{https://ieeexplore.ieee.org/document/9891914}{DOI:
  10.1109/IJCNN55064.2022.9891914}
\bibAnnoteFile{sevilla2022}

\bibitem[{Shi et~al.(2021)Shi, Lin, Wang, Wang, and Nishimura}]{Shi2021-pq}
Shi, F., Lin, Y., Wang, X., Wang, B., and Nishimura, Y. (2021).
\newblock {3-D} global hybrid simulations of magnetospheric response to
  foreshock processes.
\newblock \emph{Earth Planets Space} 73.
\newblock \href{https://doi.org/10.1186/s40623-021-01469-2}{DOI:
  10.1186/s40623-021-01469-2}
\bibAnnoteFile{Shi2021-pq}

\bibitem[{Shiltsev(2017)}]{shiltsev2017fermilab}
Shiltsev, V. (2017).
\newblock Fermilab proton accelerator complex status and improvement plans.
\newblock \emph{Modern Physics Letters A} 32, 1730012.
\newblock \href{https://arxiv.org/abs/1705.03075}{arXiv:1705.03075}
\bibAnnoteFile{shiltsev2017fermilab}

\bibitem[{Simmendinger et~al.(2024)Simmendinger, Marquardt, Mäder, and
  Schneider}]{powersched}
Simmendinger, C., Marquardt, M., Mäder, J., and Schneider, R. (2024).
\newblock Powersched - managing power consumption in overprovisioned systems.
\newblock In \emph{IEEE International Conference on Cluster Computing Workshops
  (CLUSTER Workshops)}.
\newblock \href{https://doi.org/10.1109/CLUSTERWorkshops61563.2024.00012}{DOI:
  10.1109/CLUSTERWorkshops61563.2024.00012}
\bibAnnoteFile{powersched}

\bibitem[{Skuse(2019)}]{Skuse:2019}
Skuse, B. (2019).
\newblock The third pillar.
\newblock \emph{Physics World} 32, 40.
\newblock \href{https://dx.doi.org/10.1088/2058-7058/32/3/33}{DOI:
  10.1088/2058-7058/32/3/33}
\bibAnnoteFile{Skuse:2019}

\bibitem[{Sol\'{o}rzano et~al.(2024)Sol\'{o}rzano, Sato, Yamamoto, Shoji,
  Brandt, Schwaller et~al.}]{Solorzano2024}
Sol\'{o}rzano, A. L.~V., Sato, K., Yamamoto, K., Shoji, F., Brandt, J.~M.,
  Schwaller, B., et~al. (2024).
\newblock { Toward Sustainable HPC: In-Production Deployment of Incentive-Based
  Power Efficiency Mechanism on the Fugaku Supercomputer }.
\newblock In \emph{2024 SC24: International Conference for High Performance
  Computing, Networking, Storage and Analysis SC} (IEEE Computer Society),
  342--357.
\newblock
  \href{https://doi.ieeecomputersociety.org/10.1109/SC41406.2024.00030}{DOI:
  10.1109/SC41406.2024.00030}
\bibAnnoteFile{Solorzano2024}

\bibitem[{{Springel} et~al.(2021){Springel}, {Pakmor}, {Zier}, and
  {Reinecke}}]{Springel:2021}
{Springel}, V., {Pakmor}, R., {Zier}, O., and {Reinecke}, M. (2021).
\newblock {Simulating cosmic structure formation with the GADGET-4 code}.
\newblock \emph{MNRAS}
  \href{https://ui.adsabs.harvard.edu/abs/2021MNRAS.506.2871S}{DOI:
  10.1093/mnras/stab1855}
\bibAnnoteFile{Springel:2021}

\bibitem[{{Springel} et~al.(2005){Springel}, {White}, {Jenkins}, {Frenk},
  {Yoshida}, {Gao} et~al.}]{Springel:2005}
{Springel}, V., {White}, S. D.~M., {Jenkins}, A., {Frenk}, C.~S., {Yoshida},
  N., {Gao}, L., et~al. (2005).
\newblock {Simulations of the formation, evolution and clustering of galaxies
  and quasars}.
\newblock \emph{Nature} 435, 629--636.
\newblock
  \href{https://www.nature.com/articles/nature03597}{DOI:10.1038/nature03597}
\bibAnnoteFile{Springel:2005}

\bibitem[{Suarez et~al.(2019)Suarez, Eicker, and Lippert}]{Suarez:2019}
Suarez, E., Eicker, N., and Lippert, T. (2019).
\newblock \emph{{M}odular {S}upercomputing {A}rchitecture: from {I}dea to
  {P}roduction; 3rd} (CRC Press), vol.~3, chap.~9.
\newblock 223--251.
\newblock \doi{10.1201/9781351036863}.
\newblock \href{https://juser.fz-juelich.de/record/862856}{FZJ record: 862856}
\bibAnnoteFile{Suarez:2019}

\bibitem[{Suarez et~al.(2022)Suarez, Eicker, Moschny, Pickartz, Clauss, Plugaru
  et~al.}]{Suarez:2022}
[Dataset] Suarez, E., Eicker, N., Moschny, T., Pickartz, S., Clauss, C.,
  Plugaru, V., et~al. (2022).
\newblock Modular supercomputing architecture.
\newblock \href{https://doi.org/10.5281/zenodo.6508394}{DOI:
  10.5281/zenodo.6508394}
\bibAnnoteFile{Suarez:2022}

\bibitem[{Sundriyal and Sosonkina(2017)}]{Sundriyal2017}
Sundriyal, V. and Sosonkina, M. (2017).
\newblock Reducing idle power consumption in high performance systems.
\newblock In \emph{2017 International Conference on Computational Science and
  Computational Intelligence (CSCI)}. 1629--1632.
\newblock \href{https://ieeexplore.ieee.org/document/8561048}{DOI:
  10.1109/CSCI.2017.283}
\bibAnnoteFile{Sundriyal2017}

\bibitem[{Swift(1996)}]{Swift1996-ev}
Swift, D.~W. (1996).
\newblock Use of a hybrid code for global-scale plasma simulation.
\newblock \emph{J. Comput. Phys.} 126, 109--121.
\newblock \href{https://doi.org/10.1006/jcph.1996.0124}{DOI:
  10.1006/jcph.1996.0124}
\bibAnnoteFile{Swift1996-ev}

\bibitem[{Tarraf et~al.(2024)Tarraf, Schreiber, Cascajo, Besnard, Vef, Huber
  et~al.}]{Tarraf2024}
Tarraf, A., Schreiber, M., Cascajo, A., Besnard, J.-B., Vef, M.-A., Huber, D.,
  et~al. (2024).
\newblock Malleability in modern hpc systems: Current experiences, challenges,
  and future opportunities.
\newblock \emph{IEEE Trans. Parallel Distrib. Syst.} ,
  1551–1564\href{https://doi.org/10.1109/TPDS.2024.3406764}{DOI:
  10.1109/TPDS.2024.3406764}
\bibAnnoteFile{Tarraf2024}

\bibitem[{Teichgräber(2022)}]{juliateich}
Teichgräber, J. M.~R. (2022).
\newblock Julia: A competitive high-level choice for performance portability in
  hpc?
\newblock In \emph{MPCDF Seminar (Performance) Portable Programming of HPC
  Applications}.
\newblock
  \href{https://events.gwdg.de/event/243/contributions/509/attachments/145/186/julia-performance-portability-in-hpc-final.pdf}{Link}
\bibAnnoteFile{juliateich}

\bibitem[{{The ClusterCockpit Project}(2024)}]{ClusterCockpit}
[Dataset] {The ClusterCockpit Project} (2024).
\newblock {ClusterCockpit Generic Datastructure Specification}.
\newblock
  \url{https://github.com/ClusterCockpit/cc-specifications/tree/master/datastructures}.
\newblock Accessed: 2024-10-25
\bibAnnoteFile{ClusterCockpit}

\bibitem[{{The HDF Group}(2025)}]{The_HDF_Group_Hierarchical_Data_Format}
[Dataset] {The HDF Group} (2025).
\newblock {Hierarchical Data Format, version 5}
\bibAnnoteFile{The_HDF_Group_Hierarchical_Data_Format}

\bibitem[{{The Horizon Collaboration}(2024)}]{horizon}
[Dataset] {The Horizon Collaboration} (2024).
\newblock {The Horizon simulation: Modelling galaxy formation in a cosmic
  framework}.
\newblock \url{www.horizon-simulation.org}.
\newblock Accessed: 2024-12-08
\bibAnnoteFile{horizon}

\bibitem[{{The Illustris Collaboration}(2024)}]{illustris}
[Dataset] {The Illustris Collaboration} (2024).
\newblock {The Illustris Project}.
\newblock \url{www.illustris-project.org}.
\newblock Accessed: 2024-12-08
\bibAnnoteFile{illustris}

\bibitem[{{The TNG Collaboration}(2024)}]{tng}
[Dataset] {The TNG Collaboration} (2024).
\newblock {The Illustris TNG Project}.
\newblock \url{www.tng-project.org}.
\newblock Accessed: 2024-12-08
\bibAnnoteFile{tng}

\bibitem[{{The University of Tokyo}(2024)}]{wisteria}
[Dataset] {The University of Tokyo} (2024).
\newblock {Wisteria/BDEC}.
\newblock
  \url{https://www.cc.u-tokyo.ac.jp/en/supercomputer/wisteria/service/}.
\newblock Accessed: 2024-12-08
\bibAnnoteFile{wisteria}

\bibitem[{{The Virgo Collaboration}(2024)}]{eagle}
[Dataset] {The Virgo Collaboration} (2024).
\newblock {The Eagle Project}.
\newblock \url{https://icc.dur.ac.uk/Eagle/}.
\newblock Accessed: 2024-12-08
\bibAnnoteFile{eagle}

\bibitem[{Thoman et~al.(2018)Thoman, Dichev, Heller, Iakymchuk, Aguilar,
  Hasanov et~al.}]{task_parallel_taxonomy}
Thoman, P., Dichev, K., Heller, T., Iakymchuk, R., Aguilar, X., Hasanov, K.,
  et~al. (2018).
\newblock A taxonomy of task-based parallel programming technologies for
  high-performance computing.
\newblock \emph{The Journal of Supercomputing} 74, 1422--1434.
\newblock \href{https://doi.org/10.1007/s11227-018-2238-4}{DOI:
  10.1007/s11227-018-2238-4}
\bibAnnoteFile{task_parallel_taxonomy}

\bibitem[{T{\'o}th et~al.(2012)T{\'o}th, van~der Holst, Sokolov, De~Zeeuw,
  Gombosi, Fang et~al.}]{Toth2012-zn}
T{\'o}th, G., van~der Holst, B., Sokolov, I.~V., De~Zeeuw, D.~L., Gombosi,
  T.~I., Fang, F., et~al. (2012).
\newblock Adaptive numerical algorithms in space weather modeling.
\newblock \emph{J. Comput. Phys.} 231, 870--903.
\newblock \href{https://doi.org/10.1016/j.jcp.2011.02.006}{DOI:
  10.1016/j.jcp.2011.02.006}
\bibAnnoteFile{Toth2012-zn}

\bibitem[{Tr\"{o}pgen et~al.(2024)Tr\"{o}pgen, Sch\"{o}ne, Ilsche, and
  Hackenberg}]{Troepgen_2024_SPEC}
Tr\"{o}pgen, H., Sch\"{o}ne, R., Ilsche, T., and Hackenberg, D. (2024).
\newblock 16 years of spec power: An analysis of x86 energy efficiency trends.
\newblock In \emph{2024 IEEE International Conference on Cluster Computing
  Workshops (CLUSTER Workshops)} (IEEE), 76–80.
\newblock
  \href{http://dx.doi.org/10.1109/CLUSTERWorkshops61563.2024.00020}{DOI:
  10.1109/clusterworkshops61563.2024.00020}
\bibAnnoteFile{Troepgen_2024_SPEC}

\bibitem[{Trott et~al.(2022)Trott, Lebrun-Grandié, Arndt, Ciesko, Dang,
  Ellingwood et~al.}]{kokkos}
Trott, C.~R., Lebrun-Grandié, D., Arndt, D., Ciesko, J., Dang, V., Ellingwood,
  N., et~al. (2022).
\newblock Kokkos 3: Programming model extensions for the exascale era.
\newblock \emph{IEEE Transactions on Parallel and Distributed Systems}
  \href{https://doi.org/10.1109/TPDS.2021.3097283}{DOI:
  10.1109/TPDS.2021.3097283}
\bibAnnoteFile{kokkos}

\bibitem[{Vasil et~al.(2024)Vasil, Lecoanet, Augustson, Burns, Oishi, Brown
  et~al.}]{Vasil2024-zx}
Vasil, G.~M., Lecoanet, D., Augustson, K., Burns, K.~J., Oishi, J.~S., Brown,
  B.~P., et~al. (2024).
\newblock The solar dynamo begins near the surface.
\newblock \emph{Nature} 629, 769--772.
\newblock \href{https://doi.org/10.1038/s41586-024-07315-1}{DOI:
  10.1038/s41586-024-07315-1}
\bibAnnoteFile{Vasil2024-zx}

\bibitem[{Vay et~al.(2021)Vay, Huebl, Almgren, Amorim, Bell, Fedeli
  et~al.}]{Vay2021}
Vay, J.-L., Huebl, A., Almgren, A., Amorim, L.~D., Bell, J., Fedeli, L., et~al.
  (2021).
\newblock Modeling of a chain of three plasma accelerator stages with the warpx
  electromagnetic pic code on gpus.
\newblock \emph{Physics of Plasmas} 28.
\newblock \doi{10.1063/5.0028512}.
\newblock \href{https://doi.org/10.1063/5.0028512}{DOI: 10.1063/5.0028512}
\bibAnnoteFile{Vay2021}

\bibitem[{Verbeke et~al.(2022)Verbeke, Baratashvili, and
  Poedts}]{Verbeke2022-ro}
Verbeke, C., Baratashvili, T., and Poedts, S. (2022).
\newblock {ICARUS}, a new inner heliospheric model with a flexible grid.
\newblock \emph{Astron. Astrophys.} 662, A50.
\newblock \href{ https://doi.org/10.1051/0004-6361/202141981 } {DOI:
  10.1051/0004-6361/202141981}
\bibAnnoteFile{Verbeke2022-ro}

\bibitem[{{Wang} et~al.(2020){Wang}, {Iwasawa}, {Nitadori}, and
  {Makino}}]{Wang:2020}
{Wang}, L., {Iwasawa}, M., {Nitadori}, K., and {Makino}, J. (2020).
\newblock {PETAR: a high-performance N-body code for modelling massive
  collisional stellar systems}.
\newblock \emph{MNRAS} 497, 536--555.
\newblock \href{https://academic.oup.com/mnras/article/497/1/536/5867779}{DOI:
  10.1093/mnras/staa1915}
\bibAnnoteFile{Wang:2020}

\bibitem[{Wang et~al.(2022{\natexlab{a}})Wang, Peng, Ren, Chen, and
  Edwards}]{10.1145/3506705}
Wang, Q., Peng, Z., Ren, B., Chen, J., and Edwards, R.~G. (2022{\natexlab{a}}).
\newblock Memhc: An optimized gpu memory management framework for accelerating
  many-body correlation.
\newblock \emph{ACM Trans. Archit. Code Optim.} 19.
\newblock \href{https://doi.org/10.1145/3506705}{DOI: {10.1145/3506705}}
\bibAnnoteFile{10.1145/3506705}

\bibitem[{Wang et~al.(2022{\natexlab{b}})Wang, Ren, Chen, and
  Edwards}]{9820666}
Wang, Q., Ren, B., Chen, J., and Edwards, R.~G. (2022{\natexlab{b}}).
\newblock Micco: An enhanced multi-gpu scheduling framework for many-body
  correlation functions.
\newblock In \emph{2022 IEEE International Parallel and Distributed Processing
  Symposium (IPDPS)}. 135--145.
\newblock \href{https://ieeexplore.ieee.org/document/9820666}{DOI:
  10.1109/IPDPS53621.2022.00022}
\bibAnnoteFile{9820666}

\bibitem[{Wang et~al.(2013)Wang, Fidkowski, Abgrall, Bassi, Caraeni, Cary
  et~al.}]{Wang2013-ub}
Wang, Z.~J., Fidkowski, K., Abgrall, R., Bassi, F., Caraeni, D., Cary, A.,
  et~al. (2013).
\newblock High-order {CFD} methods: current status and perspective.
\newblock \emph{Int. J. Numer. Methods Fluids} 72, 811--845.
\newblock \href{https://doi.org/10.1002/fld.3767}{DOI: 10.1002/fld.3767}
\bibAnnoteFile{Wang2013-ub}

\bibitem[{Warnecke et~al.(2023)Warnecke, Korpi-Lagg, Gent, and
  Rheinhardt}]{Warnecke2023-uj}
Warnecke, J., Korpi-Lagg, M.~J., Gent, F.~A., and Rheinhardt, M. (2023).
\newblock Numerical evidence for a small-scale dynamo approaching solar
  magnetic prandtl numbers.
\newblock \emph{Nat. Astron.} 7, 662--668.
\newblock \href{https://doi.org/10.1038/s41550-023-01975-1}{DOI:
  10.1038/s41550-023-01975-1}
\bibAnnoteFile{Warnecke2023-uj}

\bibitem[{Weinzierl(2021)}]{Weinzierl:2021}
Weinzierl, T. (2021).
\newblock \emph{The Pillars of Science} (Cham: Springer International
  Publishing).
\newblock 3--9.
\newblock \href{https://doi.org/10.1007/978-3-030-76194-3\_1}{DOI:
  10.1007/978-3-030-76194-3\_1}
\bibAnnoteFile{Weinzierl:2021}

\bibitem[{{Wells} et~al.(1981){Wells}, {Greisen}, and {Harten}}]{Wells:1981}
{Wells}, D.~C., {Greisen}, E.~W., and {Harten}, R.~H. (1981).
\newblock {FITS - a Flexible Image Transport System}.
\newblock \emph{A\&AS}
  \href{https://adsabs.harvard.edu/pdf/1981a&as...44..363w}{SAO/NASA
  Astrophysics Data System}
\bibAnnoteFile{Wells:1981}

\bibitem[{Whyte et~al.(2024)Whyte, Stathopoulos, Romero, and
  Orginos}]{Whyte:2022vrk}
Whyte, T., Stathopoulos, A., Romero, E., and Orginos, K. (2024).
\newblock {Optimizing shift selection in multilevel Monte Carlo for
  disconnected diagrams in lattice QCD}.
\newblock \emph{Comput. Phys. Commun.} 294, 108928.
\newblock \href{https://doi.org/10.1016/j.cpc.2023.108928}{DOI:
  10.1016/j.cpc.2023.108928}
\bibAnnoteFile{Whyte:2022vrk}

\bibitem[{{Wilkinson} et~al.(2016){Wilkinson}, {Dumontier}, {Aalbersberg},
  {Appleton}, {Axton}, {Baak} et~al.}]{Wilkinson:2016}
{Wilkinson}, M.~D., {Dumontier}, M., {Aalbersberg}, I.~J., {Appleton}, G.,
  {Axton}, M., {Baak}, A., et~al. (2016).
\newblock The {FAIR} guiding principles for scientific data management and
  stewardship.
\newblock \emph{Scientific Data}
  \href{https://www.nature.com/articles/sdata201618}{DOI:
  10.1038/sdata.2016.18}
\bibAnnoteFile{Wilkinson:2016}

\bibitem[{Williams et~al.(2024)Williams, Tskhakaya, Costea, Peng,
  Garcia-Gasulla, and Markidis}]{Williams2024}
Williams, J.~J., Tskhakaya, D., Costea, S., Peng, I.~B., Garcia-Gasulla, M.,
  and Markidis, S. (2024).
\newblock \emph{Leveraging HPC Profiling and Tracing Tools to Understand
  the Performance of Particle-in-Cell Monte Carlo Simulations} (Springer
  Nature Switzerland).
\newblock 123–134.
\newblock \href{http://dx.doi.org/10.1007/978-3-031-50684-0\_10}{DOI:
  10.1007/978-3-031-50684-0\_10}
\bibAnnoteFile{Williams2024}

\bibitem[{Williams et~al.(2009)Williams, Waterman, and
  Patterson}]{williams2009}
Williams, S., Waterman, A., and Patterson, D. (2009).
\newblock Roofline: an insightful visual performance model for multicore
  architectures.
\newblock \emph{Commun. ACM} 52, 65–76.
\newblock \href{https://doi.org/10.1145/1498765.1498785}{DOI:
  10.1145/1498765.1498785}
\bibAnnoteFile{williams2009}

\bibitem[{Xia et~al.(2018)Xia, Teunissen, Mellah, Chan{\'e}, and
  Keppens}]{Xia2018-mg}
Xia, C., Teunissen, J., Mellah, I.~E., Chan{\'e}, E., and Keppens, R. (2018).
\newblock {MPI-AMRVAC} 2.0 for solar and astrophysical applications.
\newblock \emph{Astrophys. J. Suppl. Ser.} 234, 30.
\newblock \href{https://doi.org/10.3847/1538-4365/aaa6c8}{DOI:
  10.3847/1538-4365/aaa6c8}
\bibAnnoteFile{Xia2018-mg}

\bibitem[{Yamaguchi et~al.(2022)Yamaguchi, Boyle, Cossu, Filaci, Lehner, and
  Portelli}]{Yamaguchi:2022feu}
Yamaguchi, A., Boyle, P., Cossu, G., Filaci, G., Lehner, C., and Portelli, A.
  (2022).
\newblock {Grid: OneCode and FourAPIs}.
\newblock \emph{PoS} LATTICE2021, 035.
\newblock \href{http://dx.doi.org/10.22323/1.396.0035}{DOI:
  10.22323/1.396.0035}
\bibAnnoteFile{Yamaguchi:2022feu}

\bibitem[{Yotov et~al.(2007)Yotov, Roeder, Pingali, Gunnels, and
  Gustavson}]{Yotov2007}
Yotov, K., Roeder, T., Pingali, K., Gunnels, J., and Gustavson, F. (2007).
\newblock An experimental comparison of cache-oblivious and cache-conscious
  programs.
\newblock In \emph{Proceedings of the nineteenth annual ACM symposium on
  Parallel algorithms and architectures} (ACM), SPAA07, 93–104.
\newblock \doi{10.1145/1248377.1248394}.
\newblock \href{http://dx.doi.org/10.1145/1248377.1248394}{DOI:
  10.1145/1248377.1248394}
\bibAnnoteFile{Yotov2007}

\bibitem[{Zenker et~al.(2016)Zenker, Worpitz, Widera, Huebl, Juckeland,
  Knüpfer et~al.}]{alpaka}
Zenker, E., Worpitz, B., Widera, R., Huebl, A., Juckeland, G., Knüpfer, A.,
  et~al. (2016).
\newblock Alpaka -- an abstraction library for parallel kernel acceleration.
\newblock In \emph{IEEE International Parallel and Distributed Processing
  Symposium Workshops (IPDPSW)}.
\newblock \href{https://doi.org/10.1109/IPDPSW.2016.50}{DOI:
  10.1109/IPDPSW.2016.50}
\bibAnnoteFile{alpaka}

\bibitem[{Zhang and Gladman(2022)}]{Zhang2022}
Zhang, K. and Gladman, B.~J. (2022).
\newblock Glisse: A gpu-optimized planetary system integrator with application
  to orbital stability calculations.
\newblock \emph{New Astronomy} 90, 101659.
\newblock \doi{10.1016/j.newast.2021.101659}.
\newblock \href{https://doi.org/10.1016/j.newast.2021.101659}{DOI:
  10.1016/j.newast.2021.101659}
\bibAnnoteFile{Zhang2022}

\end{thebibliography}


\section*{Figure captions}



\end{document}